\RequirePackage{letltxmacro}
\LetLtxMacro{\LaTeXtextbf}{\textbf}
\documentclass{ieeeaccess}
\LetLtxMacro{\textbf}{\LaTeXtextbf}

\usepackage{textcomp}
\usepackage{makecell}
\usepackage{cite}
\usepackage{amssymb, amsfonts, amsmath, mathtools, amsthm, bbm, bm}

\usepackage{textcomp}
\usepackage{dsfont}

\usepackage{algorithmic}
\usepackage[linesnumbered,ruled,vlined]{algorithm2e}

\usepackage{booktabs, multirow}

\usepackage{graphicx, subcaption, xcolor}
\usepackage{tikz, pgfplots}
\usepackage{caption,setspace}
\NewSpotColorSpace{PANTONE}
\AddSpotColor{PANTONE} {PANTONE3015C} {PANTONE\SpotSpace 3015\SpotSpace C} {1 0.3 0 0.2}
\SetPageColorSpace{PANTONE}
\usetikzlibrary{decorations,decorations.markings, optics}
\usetikzlibrary{arrows,positioning,shapes.geometric,spy}

\def\BibTeX{{\rm B\kern-.05em{\sc i\kern-.025em b}\kern-.08em
		T\kern-.1667em\lower.7ex\hbox{E}\kern-.125emX}}

\DeclareMathOperator*{\argmin}{arg\,min}
\DeclareMathOperator*{\argmax}{arg\,max}
\DeclareMathOperator*{\concat}{concat}
\DeclareMathOperator*{\sgn}{sgn}

\DeclareMathOperator*{\relu}{ReLU}

\newcommand\abs[1]{\left|#1\right|}
\begin{document}
\history{Date of publication xxxx 00, 0000, date of current version xxxx 00, 0000.}
\doi{}

\title{Fast Successive-Cancellation List Flip Decoding of Polar Codes}
\author{\uppercase{Nghia~Doan}\authorrefmark{1}, 
		\uppercase{Seyyed~Ali~Hashemi}\authorrefmark{2}, and
		\uppercase{Warren~J.~Gross}\authorrefmark{1} 
		}
	
\address[1]{Department of Electrical and Computer Engineering, McGill University, Canada (e-mails: nghia.doan@mail.mcgill.ca, warren.gross@mcgill.ca)}
\address[2]{Qualcomm, USA (e-mail: hashemi@qti.qualcomm.com)}

\markboth
{Nghia~Doan \headeretal: Fast Successive-Cancellation List Flip Decoding of Polar Codes}
{Nghia~Doan \headeretal: Fast Successive-Cancellation List Flip Decoding of Polar Codes}

\corresp{Corresponding author: Nghia~Doan (e-mail: nghia.doan@mail.mcgill.ca).}

\begin{abstract}
This work presents a fast successive-cancellation list flip (Fast-SCLF) decoding algorithm for polar codes that addresses the high latency issue associated with the successive-cancellation list flip (SCLF) decoding algorithm. We first propose a bit-flipping strategy tailored to the state-of-the-art fast successive-cancellation list (FSCL) decoding that avoids tree-traversal in the binary tree representation of SCLF, thus reducing the latency of the decoding process. We then derive a parameterized path selection error model to accurately estimate the bit index at which the correct decoding path is eliminated from the initial FSCL decoding. The trainable parameter is optimized online based on an efficient supervised learning framework. Simulation results show that for a polar code of length 512 with 256 information bits, with similar error-correction performance and memory consumption, the proposed Fast-SCLF decoder reduces up to $73.4\%$ of the average decoding latency of the SCLF decoder with the same list size at the frame error rate of $10^{-4}$, while incurring a maximum computational complexity overhead of \textcolor{black}{$27.6\%$}. For the same polar code of length 512 with 256 information bits and at practical signal-to-noise ratios, the proposed decoder with list size 4 reduces \textcolor{black}{$89.3\%$} and $43.7\%$ of the average complexity and decoding latency of the FSCL decoder with list size 32 (FSCL-32), respectively, while also reducing \textcolor{black}{$83.2\%$} of the memory consumption of FSCL-32. The significant improvements of the proposed decoder come at the cost of $0.07$ dB error-correction performance degradation compared with FSCL-32.
\end{abstract}


\titlepgskip=-15pt

\maketitle

\section{Introduction} \label{sec:intro}

\PARstart{P}{olar} codes are the first class of error-correcting codes that is proven to achieve the channel capacity of any binary symmetric channel under the low-complexity successive-cancellation (SC) decoding algorithm \cite{arikan}. Recently, polar codes are selected for use in the enhanced mobile broadband (eMBB) control channel of the fifth generation of cellular communications (5G) standard, where codes with short to moderate block lengths are used \cite{3gpp_report}. The error-correction performance of short to moderate-length polar codes under SC decoding does not satisfy the requirements of the 5G standard. SC list (SCL) decoding was introduced in \cite{tal_list, KaiNiu, Alexios_LLR_SCLD} to improve the error-correction performance of SC decoding by keeping a list of candidate message words at each decoding step. In addition, it was observed that under SCL decoding, the error-correction performance is significantly improved when the polar code is concatenated with a cyclic redundancy check (CRC) \cite{tal_list, KaiNiu}. Furthermore, SC-based decoding of polar code can be represented as a binary tree traversing problem \cite{SimpSCD} and it was shown that the decoders in \cite{arikan, tal_list, KaiNiu, Alexios_LLR_SCLD} experience a high decoding latency as they require a full binary tree traversal. Several fast decoding techniques were introduced to improve the decoding latency of the conventional SC and  SCL decoding algorithms \cite{gabi_fast_pcd, SSCL, Ali_FSSCL, Ardakani_TCOM, Hanif_FastSCL}. In particular, the decoding operations of special constituent codes under the fast SCL (FSCL) decoding algorithms proposed in \cite{SSCL, Ardakani_TCOM, Ali_FSSCL, Hanif_FastSCL} can be carried out at the parent node level, thus reducing the decoding latency caused by the tree traversal.

As the memory requirement of SCL decoding grows linearly with the list size \cite{Ali_mem_effic_PC}, it is of great interest to improve the decoding performance of SCL decoding with a small list size. \textcolor{black}{To address this issue, various bit-flipping algorithms of the \textcolor{black}{SC-based} decoders were proposed to significantly improve the error-correction performance of SC and SCL decoding with the same list size \cite{SCF,DSCF,doan2020neural,9052925, Doan_ICC21, Furkan_TSP20,Ali_DLSC, SCLF, Pan, Yongrun, Lee20}.} In \cite{Ali_DLSC}, given that the initial SCL decoding attempt does not satisfy the CRC verification, the authors proposed an algorithm that flips the first erroneous bit of the best decoding path in the next decoding attempt and the error position is estimated using a correlation matrix. The SCL-Flip (SCLF) decoder proposed in \cite{SCLF} estimates the bit index at which the correct path is discarded from the initial SCL decoding, then in the next decoding attempt all the paths that were discarded at the estimated error position are selected to continue the decoding. It was observed that the decoder in \cite{SCLF} provides a better error-correction performance when compared with the decoder proposed in \cite{Ali_DLSC}. In \cite{Pan}, an improved SCLF decoding algorithm is proposed which addresses the high-order errors of SCL decoding. By utilizing a complex path selection error model and with the same number of additional decoding attempts, the decoder in \cite{Pan} provides a slight error-correction performance improvement when compared with the SCLF decoder \cite{SCLF}. It is worth to note that all the bit-flipping algorithms of SCL decoding introduced in \cite{Ali_DLSC, SCLF, Pan, Yongrun} suffer from a variable decoding latency, which is caused by the sequential nature of the bit-flipping operations. In addition, all the decoders in \cite{Ali_DLSC, SCLF, Pan, Yongrun} fully traverse the polar code decoding tree as required by SCL decoding, thus resulting in a high decoding latency. \textcolor{black}{The authors in \cite{Lee20} proposed a simplified node-based bit-flipping algorithm to improve the decoding latency of SCLF decoding, which is referred as SSCLF decoding in this paper. Specifically, a low-complexity bit-flipping metric based on the path metrics is utilized in \cite{Lee20} to select the first error position of FSCL decoding. However, when applied to an FSCL decoder with the list sizes of 2 and 4, the simplified path-selection scheme of \cite{Lee20} results in a significant FER performance degradation when compared with SCLF decoding \cite{SCLF}.}

In this paper, a fast SCLF (Fast-SCLF) decoding algorithm is proposed to tackle the underlying high-decoding latency of the SCLF decoder \cite{SCLF}. In particular, a bit-flipping strategy tailored to FSCL decoding of polar codes is first introduced. Then, a path selection error metric is derived for the proposed bit-flipping strategy. The proposed path selection error metric utilizes a trainable parameter to improve the estimation accuracy of the error position, which is optimized online using an efficient supervised learning framework. By utilizing online training, the proposed path selection error-model does not require the parameter to be optimized offline at various signal-to-noise ratios (SNRs). Instead, the parameter is automatically optimized at the operating SNR of the decoder, which obviates the need for pilot signals. Our simulation results illustrate that for a polar code of length 512 with 256 information bits at the frame error rate (FER) of $10^{-4}$, with similar error-correction performance and memory consumption, the proposed Fast-SCLF decoder reduces up to $73.4\%$ of the average decoding latency of the SCLF decoder with the same list size, while incurring a maximum computational overhead of \textcolor{black}{$27.6\%$}. For the same polar code of length 512 with 256 information bits and at practical SNR values, the proposed decoder with list size 4 reduces \textcolor{black}{$89.3\%$} and $43.7\%$ of the average complexity and decoding latency of FSCL-32, respectively, while also reducing \textcolor{black}{$83.2\%$} of the memory consumption of FSCL-32. Note that the significant complexity reductions only come at the cost of less than 0.07 dB error-correction performance degradation. \textcolor{black}{For the same polar code of length 512 with 256 information bits, when compared with the SSCLF decoder with list size 4 at the target FER of $10^{-4}$, an FER performance gain of 0.2 dB is obtained for the proposed Fast-SCLF decoder at the cost of $8.3\%$ computational complexity overheads, while the average decoding latency and memory consumption of the Fast-SCLF decoder are relatively similar to those of the SSCLF decoder.}

The remainder of this paper is organized as follows. Section~\ref{sec:polar} provides the background on polar codes and their decoding algorithms. Section~\ref{sec:FSCLF} proposes the Fast-SCLF decoding algorithm. Simulation results are reported in Section~\ref{sec:evaluation}, followed by concluding remarks drawn in Section~\ref{sec:conclude}.

\section{Preliminaries}
\label{sec:polar}

We start this section by first introducing notations. Throughout this paper boldface letters indicate vectors and matrices, while unless otherwise specified non-boldface letters indicate either binary, integer or real numbers. In addition, by $\bm{a}_{i_\text{min}}^{i_\text{max}} = \{a_{i_\text{min}},\ldots, a_{i_\text{max}}\}$ we denote a vector of size $i_\text{max} - i_\text{min} + 1$ containing the $a$ elements from index $i_\text{min}$ to $i_\text{max}$ $(i_\text{min} < i_\text{max})$. Sets are denoted by blackboard bold letters, e.g., $\mathbb{R}$ is the set containing real numbers. Finally, $\mathds{1}_{X}$ is an indicator function where $\mathds{1}_{X}=1$ if the condition $X$ is true, and $\mathds{1}_{X}=0$ otherwise.

\subsection{Polar Encoding} 

A polar code $\mathcal{P}(N,K)$ of length $N$ with $K$ information bits is constructed by applying a linear transformation to the binary message word $\bm{u} = \{u_1,\ldots,u_N\}$ as $\bm{x} = \bm{u}\bm{G}^{\otimes n}$ where $\bm{x} = \{x_1,\ldots,x_N\}$ is the codeword, $\bm{G}^{\otimes m}$ is the $n$-th Kronecker power of the polarizing matrix $\bm{G}=\bigl[\begin{smallmatrix} 1&0\\ 1&1 \end{smallmatrix} \bigr]$, and $n = \log_2 N$. The vector $\bm{u}$ contains a set $\mathbbm{I}$ of $K$ information bit indices and a set $\mathbbm{I}^c$ of $N-K$ frozen bit indices, with $\mathbbm{I}$ and $\mathbbm{I}^c$ are known to the encoder and the decoder. The values of all the frozen bits are set to $0$, while the values of the information bits are independent and identically distributed \cite{arikan}. The codeword $\bm{x}$ is then modulated and sent through the channel. In this paper, binary phase-shift keying (BPSK) modulation and additive white Gaussian noise (AWGN) channel model are considered. Therefore, the soft vector of the transmitted codeword received by the decoder is given as ${\bm{y}=(\mathbf{1}-2\bm{x})+\bm{z}}$, where $\mathbf{1}$ is an all-one vector of size $N$, and $\bm{z}$ is a Gaussian noise vector of size $N$ with variance $\sigma^2$ and zero mean. In the log-likelihood ratio (LLR) domain, the LLR vector of the transmitted codeword is given as
${\bm{\alpha}_n=\ln{\frac{Pr(\bm{x}=0|\bm{y})}{Pr(\bm{x}=1|\bm{y})}}=\frac{2\bm{y}}{\sigma^2}}$.

\subsection{Successive-Cancellation and Successive-Cancellation List Decoding} 

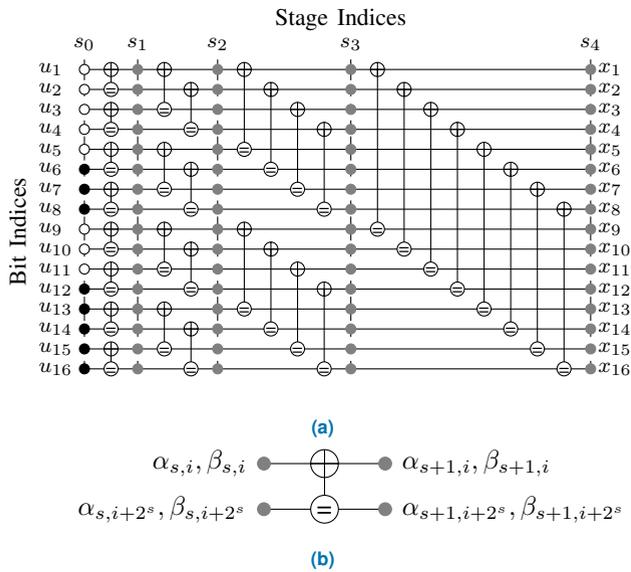
\begin{figure}[t]
	\centering
	\begin{subfigure}{1\columnwidth}
		\centering	
		\begin{tikzpicture}[scale=0.75]

\def\N{16}
\def\xM{7.5}
\def\xss{\xM/\N}
\def\xs{\xss/1}
\def\ys{0.35}
\def\gain{1.0}
\def\markSize{2.5}
\def\PEmarkSize{3.5}

\draw[dashed] (0*\xs,0*\ys) -- (0*\xs,15.5*\ys*\gain) node[above]{\footnotesize{$s_0$}};
\draw[dashed] (2*\xs,0*\ys) -- (2*\xs,15.5*\ys*\gain) node[above]{\footnotesize{$s_1$}};
\draw[dashed] (5*\xs,0*\ys) -- (5*\xs,15.5*\ys*\gain) node[above]{\footnotesize{$s_2$}};
\draw[dashed] (10*\xs,0*\ys) -- (10*\xs,15.5*\ys*\gain) node[above]{\footnotesize{$s_3$}};
\draw[dashed] (19*\xs,0*\ys) -- (19*\xs,15.5*\ys*\gain) node[above]{\footnotesize{$s_4$}};

\foreach \i in{0,...,15}
{
	\pgfmathsetmacro\bIndex{int(16-\i)};
	\draw[] (0,\i*\ys) -- (19*\xs,\i*\ys);
	\ifthenelse{\i > 10}{\draw[] plot[mark=*, mark size = \markSize, mark options={fill=white}] coordinates {(0*\xs,\i*\ys)};}
	
	\ifthenelse{\i < 11 \AND \i > 7}{\draw[] plot[mark=*, mark size = \markSize, mark options={fill=black}] coordinates {(0*\xs,\i*\ys)};}
	
	\ifthenelse{\i < 8 \AND \i > 4}{\draw[] plot[mark=*, mark size = \markSize, mark options={fill=white}] coordinates {(0*\xs,\i*\ys)};}
	
	\ifthenelse{\i < 5}{\draw[] plot[mark=*, mark size = \markSize, mark options={fill=black}] coordinates {(0*\xs,\i*\ys)};}
	
	\draw[] plot[mark=*, mark size = \markSize, mark options={color=gray}] coordinates {(2*\xs,\i*\ys)};
	\draw[] plot[mark=*, mark size = \markSize, mark options={color=gray}] coordinates {(5*\xs,\i*\ys)};
	\draw[] plot[mark=*, mark size = \markSize, mark options={color=gray}] coordinates {(10*\xs,\i*\ys)};
	\draw[] plot[mark=*, mark size = \markSize, mark options={color=gray}] coordinates {(19*\xs,\i*\ys)};
	
	\node[text width=0.5cm] at (-1*\xs,\i*\ys) {\footnotesize{$u_{\bIndex}$}};
	\node[text width=0.5cm] at (20*\xs,\i*\ys) {\footnotesize{$x_{\bIndex}$}};
}

\draw[] plot[mark=oplus, mark size = \PEmarkSize, mark options={fill=white}] coordinates {(\xs,\ys)} --
plot[mark=*, mark size = \PEmarkSize, mark options={fill=white}] coordinates {(\xs,0)} node[above=-0.2]{\scriptsize{=}};

\draw[] plot[mark=oplus, mark size = \PEmarkSize, mark options={fill=white}] coordinates {(\xs,3*\ys)} --
plot[mark=*, mark size = \PEmarkSize, mark options={fill=white}] coordinates {(\xs,2*\ys)} node[above=-0.2]{\scriptsize{=}};

\draw[] plot[mark=oplus, mark size = \PEmarkSize, mark options={fill=white}] coordinates {(\xs,5*\ys)} --
plot[mark=*, mark size = \PEmarkSize, mark options={fill=white}] coordinates {(\xs,4*\ys)} node[above=-0.2]{\scriptsize{=}};

\draw[] plot[mark=oplus, mark size = \PEmarkSize, mark options={fill=white}] coordinates {(\xs,7*\ys)} --
plot[mark=*, mark size = \PEmarkSize, mark options={fill=white}] coordinates {(\xs,6*\ys)} node[above=-0.2]{\scriptsize{=}};

\draw[] plot[mark=oplus, mark size = \PEmarkSize, mark options={fill=white}] coordinates {(\xs,9*\ys)} --
plot[mark=*, mark size = \PEmarkSize, mark options={fill=white}] coordinates {(\xs,8*\ys)} node[above=-0.2]{\scriptsize{=}};

\draw[] plot[mark=oplus, mark size = \PEmarkSize, mark options={fill=white}] coordinates {(\xs,11*\ys)} --
plot[mark=*, mark size = \PEmarkSize, mark options={fill=white}] coordinates {(\xs,10*\ys)} node[above=-0.2]{\scriptsize{=}};

\draw[] plot[mark=oplus, mark size = \PEmarkSize, mark options={fill=white}] coordinates {(\xs,13*\ys)} --
plot[mark=*, mark size = \PEmarkSize, mark options={fill=white}] coordinates {(\xs,12*\ys)} node[above=-0.2]{\scriptsize{=}};

\draw[] plot[mark=oplus, mark size = \PEmarkSize, mark options={fill=white}] coordinates {(\xs,15*\ys)} --
plot[mark=*, mark size = \PEmarkSize, mark options={fill=white}] coordinates {(\xs,14*\ys)} node[above=-0.2]{\scriptsize{=}};

\draw[] plot[mark=oplus, mark size = \PEmarkSize, mark options={fill=white}] coordinates {(3*\xs,3*\ys)} --
plot[mark=*, mark size = \PEmarkSize, mark options={fill=white}] coordinates {(3*\xs,\ys)} node[above=-0.2]{\scriptsize{=}};

\draw[] plot[mark=oplus, mark size = \PEmarkSize, mark options={fill=white}] coordinates {(3*\xs,7*\ys)} --
plot[mark=*, mark size = \PEmarkSize, mark options={fill=white}] coordinates {(3*\xs,5*\ys)} node[above=-0.2]{\scriptsize{=}};

\draw[] plot[mark=oplus, mark size = \PEmarkSize, mark options={fill=white}] coordinates {(4*\xs,2*\ys)} --
plot[mark=*, mark size = \PEmarkSize, mark options={fill=white}] coordinates {(4*\xs,0)} node[above=-0.2]{\scriptsize{=}};

\draw[] plot[mark=oplus, mark size = \PEmarkSize, mark options={fill=white}] coordinates {(4*\xs,6*\ys)} --
plot[mark=*, mark size = \PEmarkSize, mark options={fill=white}] coordinates {(4*\xs,4*\ys)} node[above=-0.2]{\scriptsize{=}};

\draw[] plot[mark=oplus, mark size = \PEmarkSize, mark options={fill=white}] coordinates {(3*\xs,11*\ys)} --
plot[mark=*, mark size = \PEmarkSize, mark options={fill=white}] coordinates {(3*\xs,9*\ys)} node[above=-0.2]{\scriptsize{=}};

\draw[] plot[mark=oplus, mark size = \PEmarkSize, mark options={fill=white}] coordinates {(3*\xs,15*\ys)} --
plot[mark=*, mark size = \PEmarkSize, mark options={fill=white}] coordinates {(3*\xs,13*\ys)} node[above=-0.2]{\scriptsize{=}};

\draw[] plot[mark=oplus, mark size = \PEmarkSize, mark options={fill=white}] coordinates {(4*\xs,10*\ys)} --
plot[mark=*, mark size = \PEmarkSize, mark options={fill=white}] coordinates {(4*\xs,8*\ys)} node[above=-0.2]{\scriptsize{=}};

\draw[] plot[mark=oplus, mark size = \PEmarkSize, mark options={fill=white}] coordinates {(4*\xs,14*\ys)} --
plot[mark=*, mark size = \PEmarkSize, mark options={fill=white}] coordinates {(4*\xs,12*\ys)} node[above=-0.2]{\scriptsize{=}};

\draw[] plot[mark=oplus, mark size = \PEmarkSize, mark options={fill=white}] coordinates {(9*\xs,4*\ys)} --
plot[mark=*, mark size = \PEmarkSize, mark options={fill=white}] coordinates {(9*\xs,0*\ys)} node[above=-0.2]{\scriptsize{=}};

\draw[] plot[mark=oplus, mark size = \PEmarkSize, mark options={fill=white}] coordinates {(8*\xs,5*\ys)} --
plot[mark=*, mark size = \PEmarkSize, mark options={fill=white}] coordinates {(8*\xs,1*\ys)} node[above=-0.2]{\scriptsize{=}};

\draw[] plot[mark=oplus, mark size = \PEmarkSize, mark options={fill=white}] coordinates {(7*\xs,6*\ys)} --
plot[mark=*, mark size = \PEmarkSize, mark options={fill=white}] coordinates {(7*\xs,2*\ys)} node[above=-0.2]{\scriptsize{=}};

\draw[] plot[mark=oplus, mark size = \PEmarkSize, mark options={fill=white}] coordinates {(6*\xs,7*\ys)} --
plot[mark=*, mark size = \PEmarkSize, mark options={fill=white}] coordinates {(6*\xs,3*\ys)} node[above=-0.2]{\scriptsize{=}};

\draw[] plot[mark=oplus, mark size = \PEmarkSize, mark options={fill=white}] coordinates {(9*\xs,12*\ys)} --
plot[mark=*, mark size = \PEmarkSize, mark options={fill=white}] coordinates {(9*\xs,8*\ys)} node[above=-0.2]{\scriptsize{=}};

\draw[] plot[mark=oplus, mark size = \PEmarkSize, mark options={fill=white}] coordinates {(8*\xs,13*\ys)} --
plot[mark=*, mark size = \PEmarkSize, mark options={fill=white}] coordinates {(8*\xs,9*\ys)} node[above=-0.2]{\scriptsize{=}};

\draw[] plot[mark=oplus, mark size = \PEmarkSize, mark options={fill=white}] coordinates {(7*\xs,14*\ys)} --
plot[mark=*, mark size = \PEmarkSize, mark options={fill=white}] coordinates {(7*\xs,10*\ys)} node[above=-0.2]{\scriptsize{=}};

\draw[] plot[mark=oplus, mark size = \PEmarkSize, mark options={fill=white}] coordinates {(6*\xs,15*\ys)} --
plot[mark=*, mark size = \PEmarkSize, mark options={fill=white}] coordinates {(6*\xs,11*\ys)} node[above=-0.2]{\scriptsize{=}};


\draw[] plot[mark=oplus, mark size = \PEmarkSize, mark options={fill=white}] coordinates {(11*\xs,15*\ys)} --
plot[mark=*, mark size = \PEmarkSize, mark options={fill=white}] coordinates {(11*\xs,7*\ys)} node[above=-0.2]{\scriptsize{=}};

\draw[] plot[mark=oplus, mark size = \PEmarkSize, mark options={fill=white}] coordinates {(12*\xs,14*\ys)} --
plot[mark=*, mark size = \PEmarkSize, mark options={fill=white}] coordinates {(12*\xs,6*\ys)} node[above=-0.2]{\scriptsize{=}};

\draw[] plot[mark=oplus, mark size = \PEmarkSize, mark options={fill=white}] coordinates {(13*\xs,13*\ys)} --
plot[mark=*, mark size = \PEmarkSize, mark options={fill=white}] coordinates {(13*\xs,5*\ys)} node[above=-0.2]{\scriptsize{=}};

\draw[] plot[mark=oplus, mark size = \PEmarkSize, mark options={fill=white}] coordinates {(14*\xs,12*\ys)} --
plot[mark=*, mark size = \PEmarkSize, mark options={fill=white}] coordinates {(14*\xs,4*\ys)} node[above=-0.2]{\scriptsize{=}};

\draw[] plot[mark=oplus, mark size = \PEmarkSize, mark options={fill=white}] coordinates {(15*\xs,11*\ys)} --
plot[mark=*, mark size = \PEmarkSize, mark options={fill=white}] coordinates {(15*\xs,3*\ys)} node[above=-0.2]{\scriptsize{=}};

\draw[] plot[mark=oplus, mark size = \PEmarkSize, mark options={fill=white}] coordinates {(16*\xs,10*\ys)} --
plot[mark=*, mark size = \PEmarkSize, mark options={fill=white}] coordinates {(16*\xs,2*\ys)} node[above=-0.2]{\scriptsize{=}};

\draw[] plot[mark=oplus, mark size = \PEmarkSize, mark options={fill=white}] coordinates {(17*\xs,9*\ys)} --
plot[mark=*, mark size = \PEmarkSize, mark options={fill=white}] coordinates {(17*\xs,1*\ys)} node[above=-0.2]{\scriptsize{=}};

\draw[] plot[mark=oplus, mark size = \PEmarkSize, mark options={fill=white}] coordinates {(18*\xs,8*\ys)} --
plot[mark=*, mark size = \PEmarkSize, mark options={fill=white}] coordinates {(18*\xs,0*\ys)} node[above=-0.2]{\scriptsize{=}};

\node[text width=2cm] at (10*\xs,17.5*\ys) {\small{Stage Indices}};
\node[text width=2cm, rotate=90] at (-2.5*\xs,8*\ys) {\small{Bit Indices}};

\end{tikzpicture}
		\vspace*{-0.5\baselineskip}
		\caption{}
	\end{subfigure}
	\begin{subfigure}{1\columnwidth}
		\centering
		{\hspace*{24pt}\begin{tikzpicture}[scale=0.8]
\def\markSize{3}
\draw[] plot[mark=*, mark size = \markSize, mark options={color=gray}] coordinates {(0,0.75)} node[left=0.1] {$\alpha_{s,i},\beta_{s,i}$} -- plot[mark=*, mark size = \markSize, mark options={color=gray}] coordinates {(2,0.75)} node[right=0.1] {$\alpha_{s+1,i},\beta_{s+1,i}$};

\draw[] plot[mark=*, mark size = \markSize, mark options={color=gray}] coordinates {(0,0)} node[left=0.1] {$\alpha_{s,i+2^s},\beta_{s,i+2^s}$} -- plot[mark=*, mark size = \markSize, mark options={color=gray}] coordinates {(2,0)} node[right=0.1] {$\alpha_{s+1,i+2^s},\beta_{s+1,i+2^s}$};

\draw[] plot[mark=oplus, mark size = 6.5, mark options={fill=white}] coordinates {(1,0.75)} --
plot[mark=*, mark size = 6.5, mark options={fill=white}] coordinates {(1,0)} node[above=-0.225]{=};
\end{tikzpicture}}
		\caption{}
	\end{subfigure}	
	\caption{(a) Factor graph representation of $\mathcal{P}(16,8)$, and (b) a processing element (PE) of polar codes.}
	\label{fig:polar:fg}
\end{figure}

An example of a factor-graph representation for $\mathcal{P}(16,8)$ is depicted in Fig.~\ref{fig:polar:fg}(a) with the frozen set $\mathbbm{I}^c=\{1,2,3,4,5,9,10,11\}$. To obtain the message word under SC decoding, the soft LLR values and the hard bit estimations are propagated through all the processing elements (PEs), which are depicted in Fig.~\ref{fig:polar:fg}(b). The PE in Fig.~\ref{fig:polar:fg}(b) performs the computations: $\alpha_{s,i} = f(\alpha_{s+1,i},\alpha_{s+1,i+2^s})$ and $\alpha_{s,i+2^s} = g(\alpha_{s+1,i},\alpha_{s+1,i+2^s},\beta_{s,i})$ \cite{arikan}. Note that $\alpha_{s,i}$ and $\beta_{s,i}$ are the soft LLR value and the hard-bit estimation at the $s$-th stage and the $i$-th bit, respectively, and the min-sum approximation formulations of $f$ and $g$ are $f(a,b) = \min(|a|,|b|)\sgn(a)\sgn(b)$, and $g(a,b,c) = b + (1-2c)a$ \cite{arikan}. The soft LLR values at the $n$-th stage are initialized to $\bm{\alpha}_n$ and the hard-bit estimation of an information bit at the $0$-th stage is obtained as $\hat{u}_{i} = \beta_{0,i}=\frac{1 - \sgn(\alpha_{1,i})}{2}$, $\forall i \in \mathbbm{I}$ \cite{arikan}. Given $\beta_{s,i}$ and $\beta_{s,i+2^s}$, $\beta_{s+1,i}$ and $\beta_{s+1,i+2^s}$ are then computed as $\beta_{s+1,i} = \beta_{s,i} \oplus \beta_{s,i+2^s}$ and $\beta_{s+1,i+2^s} = \beta_{s,i+2^s}$ \cite{arikan}.


SCL decoding was introduced to significantly improve the error-correction performance of SC decoding \cite{tal_list, KaiNiu, Alexios_LLR_SCLD}. Under SCL decoding, the estimation of an information bit $\hat{u}_i$ $(i \in \mathbbm{I})$ is considered to be both $0$ and $1$, causing a path splitting and doubling the number of candidate codewords (decoding paths) after each splitting. To prevent the exponential growth of the number of decoding paths, a path metric is used to select the $L$ most probable paths after each information bit is decoded. The path metric is obtained as \cite{Alexios_LLR_SCLD}
\begin{equation}
\label{equ:polar:PM_LLR}
\text{PM}_l=
\begin{cases}
{\text{PM}}_l + \abs{\alpha_{{0,i}_l}} & \text{ if } \hat{u}_i \neq \frac{1-\sgn(\alpha_{{0,i}_l})}{2},\\
\text{PM}_l & \text{ otherwise,}
\end{cases}
\end{equation}
where $\alpha_{{0,i}_l}$ denotes the soft value of the $i$-th bit $(i \in [1,N])$ at stage $0$ of the $l$-th path. Initially, $\text{PM}_l=0$, $\forall l$. After each information bit is decoded, only the $L$ paths with the smallest path metric values, i.e., the $L$ most-probable decoding paths, are selected from the $2L$ paths to continue the decoding. After the last information bit is decoded, the decoding path that has the smallest path metric is selected as the decoding output. In a practical scenario such as in the 5G standard, the message word is often concatenated with a CRC of size $C$ to allow error detection. In addition, it was observed in \cite{tal_list, KaiNiu} that the error-correction performance of SCL decoding is greatly improved when a CRC is concatenated to polar codes.

\subsection{Fast Successive-Cancellation List Decoding} 
\label{sec:polar:FSCL}

\begin{figure}[t]
	\centering
	\usetikzlibrary{arrows, decorations}

\tikzstyle{vecArrow} = [thick, decoration={markings,mark=at position
	1 with {\arrow[semithick]{open triangle 60}}},
double distance=1.4pt, shorten >= 5.5pt,
preaction = {decorate},
postaction = {draw,line width=1.4pt, white,shorten >= 4.5pt}]
\tikzstyle{innerWhite} = [semithick, white,line width=1.4pt, shorten >= 4.5pt]

\begin{tikzpicture}[scale=0.7]

\def\N{16}
\def\xM{7.5}
\def\xss{\xM/\N}
\def\xs{\xss/1}
\def\ys{0.4}
\def\Ygain{1.075}
\def\Xgain{0.8}
\def\markSize{3}
\def\PEmarkSize{3.8}

\node[text width=2cm] at (3.5*\xs,17.5*\ys) {\small{Stage Indices}};
\node[text width=2cm, rotate=90] at (-2.5*\xs,8*\ys) {\small{Bit Indices}};

\draw[dashed] (0*\xs*\Xgain,-0.25*\ys) -- (0*\xs*\Xgain,14.5*\ys*\Ygain) node[above]{\footnotesize{$s_0$}};
\draw[dashed] (2*\xs*\Xgain,-0.25*\ys) -- (2*\xs*\Xgain,14.5*\ys*\Ygain) node[above]{\footnotesize{$s_1$}};
\draw[dashed] (4*\xs*\Xgain,-0.25*\ys) -- (4*\xs*\Xgain,14.5*\ys*\Ygain) node[above]{\footnotesize{$s_2$}};
\draw[dashed] (6*\xs*\Xgain,-0.25*\ys) -- (6*\xs*\Xgain,14.5*\ys*\Ygain) node[above]{\footnotesize{$s_3$}};
\draw[dashed] (8*\xs*\Xgain,-0.25*\ys) -- (8*\xs*\Xgain,14.5*\ys*\Ygain) node[above]{\footnotesize{$s_4$}};

\draw[] (8*\xs*\Xgain,7.5*\ys) -- (6*\xs*\Xgain,3.5*\ys);
\draw[] (8*\xs*\Xgain,7.5*\ys) -- (6*\xs*\Xgain,11.5*\ys);
\draw[] plot[mark=*, mark size = \markSize, mark options={fill=gray}] coordinates {(8*\xs*\Xgain,7.5*\ys)};

\foreach \i in{0,...,1}
{
	\draw[] (6*\xs*\Xgain,8*\i*\ys+3.5*\ys) -- (4*\xs*\Xgain,8*\i*\ys+1.5*\ys);
	\draw[] (6*\xs*\Xgain,8*\i*\ys+3.5*\ys) -- (4*\xs*\Xgain,8*\i*\ys+5.5*\ys);
	\draw[] plot[mark=*, mark size = \markSize, mark options={fill=gray}] coordinates {(6*\xs*\Xgain,8*\i*\ys+3.5*\ys)};
}

\foreach \i in{0,...,3}
{
	\draw[] (4*\xs*\Xgain,4*\i*\ys+1.5*\ys) -- (2*\xs*\Xgain,4*\i*\ys+0.5*\ys);
	\draw[] (4*\xs*\Xgain,4*\i*\ys+1.5*\ys) -- (2*\xs*\Xgain,4*\i*\ys+2.5*\ys);
	\draw[] plot[mark=*, mark size = \markSize, mark options={fill=gray}] coordinates {(4*\xs*\Xgain,4*\i*\ys+1.5*\ys)};
}

\foreach \i in{0,...,7}
{
	\draw[] (2*\xs*\Xgain,2*\i*\ys+0.5*\ys) -- (0*\xs*\Xgain,2*\i*\ys);
	\draw[] (2*\xs*\Xgain,2*\i*\ys+0.5*\ys) -- (0*\xs*\Xgain,2*\i*\ys+\ys);
	\draw[] plot[mark=*, mark size = \markSize, mark options={fill=gray}] coordinates {(2*\xs*\Xgain,2*\i*\ys+0.5*\ys)};	
}

\foreach \i in{0,...,15}
{
	\pgfmathsetmacro\bIndex{int(16-\i)};
	\ifthenelse{\i > 10}{\draw[] plot[mark=*, mark size = \markSize, mark options={fill=white}] coordinates {(0*\xs,\i*\ys)};}
	
	\ifthenelse{\i < 11 \AND \i > 7}{\draw[] plot[mark=*, mark size = \markSize, mark options={fill=black}] coordinates {(0*\xs,\i*\ys)};}
	
	\ifthenelse{\i < 8 \AND \i > 4}{\draw[] plot[mark=*, mark size = \markSize, mark options={fill=white}] coordinates {(0*\xs,\i*\ys)};}
	
	\ifthenelse{\i < 5}{\draw[] plot[mark=*, mark size = \markSize, mark options={fill=black}] coordinates {(0*\xs,\i*\ys)};}
	
	\node[text width=0.5cm] at (-1*\xs,\i*\ys) {\footnotesize{$u_{\bIndex}$}};
}

\draw[draw=black] (-0.25*\xs,-0.25*\ys) rectangle ++(4.65*\xs*\Xgain,3.5*\ys);
\draw[draw=black] (-0.25*\xs, 3.75*\ys) rectangle ++(4.65*\xs*\Xgain,3.5*\ys);
\draw[draw=black] (-0.25*\xs, 7.75*\ys) rectangle ++(4.65*\xs*\Xgain,3.5*\ys);
\draw[draw=black] (-0.25*\xs, 11.75*\ys) rectangle ++(4.65*\xs*\Xgain,3.5*\ys);

\node[text width=1cm] at (6.8*\xs*\Xgain,\ys) {\footnotesize{Rate-1}};
\node[text width=1cm] at (6.8*\xs*\Xgain,6*\ys) {\footnotesize{REP}};
\node[text width=1cm] at (6.8*\xs*\Xgain,9*\ys) {\footnotesize{SPC}};
\node[text width=1cm] at (6.8*\xs*\Xgain,14*\ys) {\footnotesize{Rate-0}};

\draw[vecArrow] (8*\xs,7.5*\ys) -- (10*\xs,7.5*\ys);
\draw[] (8.8*\xs,8.5*\ys) node[above=0.1]{\footnotesize{Tree}};
\draw[] (8.8*\xs,7.5*\ys) node[above=0.1]{\footnotesize{pruning}};

\node[text width=2cm] at (15.6*\xs,17.5*\ys) {\small{Stage Indices}};

\draw[dashed] (17*\xs*\Xgain,-0.25*\ys) -- (17*\xs*\Xgain,14.25*\ys*\Ygain) node[above]{\footnotesize{$s_2$}};
\draw[dashed] (19*\xs*\Xgain,-0.25*\ys) -- (19*\xs*\Xgain,14.25*\ys*\Ygain) node[above]{\footnotesize{$s_3$}};
\draw[dashed] (21*\xs*\Xgain,-0.25*\ys) -- (21*\xs*\Xgain,14.25*\ys*\Ygain) node[above]{\footnotesize{$s_4$}};

\draw[] (21*\xs*\Xgain,7.5*\ys) -- (19*\xs*\Xgain,3.5*\ys);
\draw[] (21*\xs*\Xgain,7.5*\ys) -- (19*\xs*\Xgain,11.5*\ys);
\draw[] plot[mark=*, mark size = \markSize, mark options={fill=gray}] coordinates {(21*\xs*\Xgain,7.5*\ys)};

\foreach \i in{0,...,1}
{
	\draw[] (19*\xs*\Xgain,8*\i*\ys+3.5*\ys) -- (17*\xs*\Xgain,8*\i*\ys+1.5*\ys);
	\draw[] (19*\xs*\Xgain,8*\i*\ys+3.5*\ys) -- (17*\xs*\Xgain,8*\i*\ys+5.5*\ys);
	\draw[] plot[mark=*, mark size = \markSize, mark options={fill=gray}] coordinates {(19*\xs*\Xgain,8*\i*\ys+3.5*\ys)};
}

\draw[] plot[mark=*, mark size = \markSize, mark options={fill=black}] coordinates {(17*\xs*\Xgain,1.5*\ys)};

\draw[] plot[mark=triangle*, mark size = \markSize*1.25, mark options={fill=black}] coordinates {(17*\xs*\Xgain,4*\ys+1.5*\ys)};

\draw[] plot[mark=square*, mark size = \markSize*1, mark options={fill=black}] coordinates {(17*\xs*\Xgain,8*\ys+1.5*\ys)};

\draw[] plot[mark=*, mark size = \markSize*1, mark options={fill=white}] coordinates {(17*\xs*\Xgain,12*\ys+1.5*\ys)};
\node[text width=1cm] at (15.5*\xs*\Xgain,12*\ys+1.5*\ys) {\footnotesize{Rate-0}};
\node[text width=1cm] at (15.8*\xs*\Xgain,8*\ys+1.5*\ys) {\footnotesize{SPC}};
\node[text width=1cm] at (15.8*\xs*\Xgain,4*\ys+1.5*\ys) {\footnotesize{REP}};
\node[text width=1cm] at (15.5*\xs*\Xgain,1.5*\ys) {\footnotesize{Rate-1}};

\end{tikzpicture}\\
	{\hspace*{20pt} \color{accessblue} \textbf{(a)} \hspace*{95pt} \textbf{(b)}}
	\caption{(a) The full binary tree representation of $\mathcal{P}(16,8)$ illustrated in Fig.~\ref{fig:polar:fg}(a), and (b) the pruned binary tree representation of the same polar code.}
	\label{fig:polar:tree}
\end{figure}
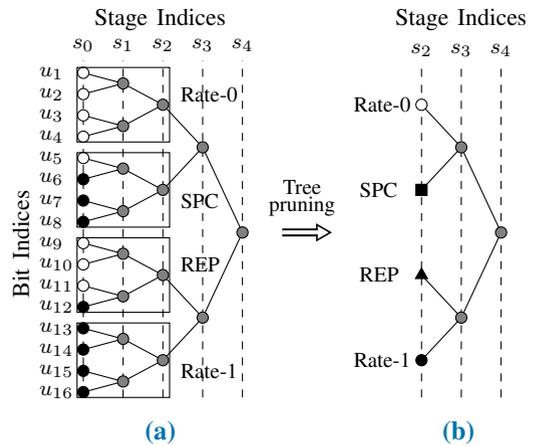

Fig.~\ref{fig:polar:tree}(a) shows a full binary tree representation of $\mathcal{P}(16,8)$, whose factor graph is depicted in Fig.~\ref{fig:polar:fg}(a). The authors in \cite{Ali_FSSCL, Ardakani_TCOM, Hanif_FastSCL} proposed fast decoding operations for various special nodes under SCL decoding, which preserve the error-correction performance of SCL decoding while preventing tree-traversal to the leaf nodes. Thus, the decoding latency of SCL decoding is significantly reduced. Similar to \cite{Ali_FSSCL}, we consider four types of special nodes, namely Rate-0, Rate-1, repetition (REP), and single parity check (SPC), for all the fast SCL-based decoding algorithms in this paper.

Consider a parent node $\nu$ located at the $s$-th stage $(0<s\leq n)$ of the polar code binary tree. There are $N_\nu$ LLR values and $N_\nu$ hard decisions associated with this node, where $N_\nu=2^s$. Let $\bm{\alpha}_{\nu_l}$ and $\bm{\beta}_{\nu_l}$ be the vectors containing the soft and hard values associated with a parent node $\nu$ of the $l$-th decoding path, respectively. $\bm{\alpha}_{\nu_l}$ and $\bm{\beta}_{\nu_l}$ are defined as $\bm{\alpha}_{\nu_l} = \{\alpha_{{s,i_{{\min}_{\nu_l}}}},\ldots,\alpha_{{s,i_{{\max}_{\nu_l}}}}\}$ and $\bm{\beta}_{\nu_l} = \{\beta_{{s,i_{{\min}_{\nu_l}}}},\ldots,\beta_{{s,i_{{\max}_{\nu_l}}}}\}$, respectively, where $i_{{\min}_{\nu_l}}$ and $i_{{\max}_{\nu_l}}$ are the bit indices corresponding to $\nu$ such that $1 \leq i_{{\min}_{\nu_l}} < i_{{\max}_{\nu_l}} \leq N$ and $i_{{\max}_{\nu_l}} - i_{{\min}_{\nu_l}} = N_\nu-1$. For all the FSCL-based decoders considered in this paper, the elements of $\bm{\alpha}_{\nu_l}$ corresponding to the SPC and Rate-1 nodes are considered to be sorted in the following order \cite{Ali_FSSCL}:
\begin{equation}
\label{equ:polar:sorted_LLR}
\abs{\alpha_{{s,i^*_{{\min}_{\nu_l}}}}} \leq \ldots \leq \abs{\alpha_{{s,i^*_{{\max}_{\nu_l}}}}},
\end{equation}
where $i_\text{min} \leq i^*_\text{min}, i^*_\text{max} \leq i_\text{max}$. In addition, let $\tau$ be the minimum number of path splittings occurred at an SPC or a Rate-1 node that allows FSCL decoding to preserve the error-correction performance of the conventional SCL decoding algorithm \cite{Ali_FSSCL}. The definitions and decoding operations of each special node under FSCL decoding are given as follows.

\subsubsection{Rate-0 node}

All the leaf nodes of a Rate-0 node are frozen bits. Therefore, all the hard values associated with the parent node are set to 0 and the path metric of the $l$-th path is given as \cite{Ali_FSSCL}
\begin{equation}
\label{equ:polar:FSSCL:R0}
\text{PM}_l=\text{PM}_l + \sum_{i=i_{\min_{\nu_l}}}^{i_{\max_{\nu_l}}}\frac{\abs{\alpha_{s,i}}-\alpha_{s,i}}{2}.
\end{equation}

\subsubsection{REP node}
All the leaf nodes of a REP node are frozen bits, except for $\beta_{0,i_{{\max}_{\nu_l}}}$. The path metric of the $l$-th decoding path is calculated as \cite{Ali_FSSCL}
\begin{equation}
\label{equ:polar:FSSCL:REP}
\text{PM}_l=\text{PM}_l+\sum_{i=i_{\min_{\nu_l}}}^{i_{\max_{\nu_l}}}\frac{\abs{\alpha_{s,i}}-(1-2\beta_{s,i_{\max_{\nu_l}}})\alpha_{s,i}}{2},
\end{equation}
where $\beta_{s,i_{\max_{\nu_l}}}$ denotes the bit estimate of the information bit of the REP node.

\subsubsection{Rate-1 node} All the leaf nodes of a Rate-1 node are information bits. FSCL decoding performs $\tau$ path splittings, where $\tau=\min(L-1, N_\nu)$ \cite{Ali_FSSCL}. The path metric of the $l$-th decoding path for a Rate-1 node is calculated as \cite{Ali_FSSCL}
\begin{equation}
\label{equ:polar:FSSCL:R1}
\text{PM}_l=\text{PM}_l + \sum_{i=i^*_{\min_{\nu_l}}}^{i^*_{\max_{\nu_l}}}\frac{\abs{\alpha_{s,i}}-(1-2\beta_{s,i})\alpha_{s,i}}{2},
\end{equation}
where $\beta_{s,i}$ denotes the bit estimate of the $i$-th bit of $\nu$.

\subsubsection{SPC node}
All the leaf nodes of an SPC node are information bits, except for $\beta_{0,i_{{\min}_{\nu_l}}}$. The parity check sum of the $l$-th path is first obtained as \cite{Ali_FSSCL}
\begin{equation}
p_l=\bigoplus_{i=i^*_{\min_{\nu_l}}}^{i^*_{\max_{\nu_l}}}\frac{1-\sgn(\alpha_{s,i})}{2}.
\end{equation}
The path metric is then updated as \cite{Ali_FSSCL}
\begin{equation}
\text{PM}_l=\text{PM}_l + p_l\abs{\alpha_{s,i^*_{\min_l}}}.
\end{equation}
The decoding continues with $\tau$ path splittings, where \textcolor{black}{${\tau=\min(L-1,N_\nu-1)}$} \cite{Ali_FSSCL}. In each new path splitting at the $i$-th index, the path metric is updated as \cite{Ali_FSSCL}
\begin{equation}
\label{fig:SPC:PM}
\text{PM}_l=
\begin{cases}
\text{PM}_l + \abs{\alpha_{{s,i}}} +(1-2p_l)\abs{\alpha_{s,i^*_{\min_l}}}\\ 
\hspace*{70pt} \text{if } 1-2\beta_{s,i} \neq \sgn(\alpha_{s,i}),\\
\text{PM}_l \hspace*{53.5pt} \text{otherwise,}\\
\end{cases}
\end{equation}
then the parity check sum is updated as \cite{Ardakani_TCOM}
\begin{equation}
\label{equ:FSSCL-SPC:gamma_update}
p_l=
\begin{cases}
1\oplus p_l &\text{if } 1-2\beta_{s,i} \neq \sgn(\alpha_{s,i}),\\
p_l &\text{otherwise.}\\
\end{cases}
\end{equation}
where $i$ is selected by following the bit indices of the sorted absolute LLR values in (\ref{equ:polar:sorted_LLR}) \cite{Ali_FSSCL}. When all the bits are estimated, the hard decision of the least reliable bit is updated to maintain the parity check condition of the SPC node \cite{Ali_FSSCL}
\begin{equation}
\label{equ:FSSCL-SPC:beta_update}
\beta_{s,i^*_{\min_{\nu_l}}}= \bigoplus_{\substack{\forall i_{\min_{\nu_l}} \le i \le i_{\max_{\nu_l}} \\ i \neq i^*_{\min_{\nu_l}}}} \beta_{s,i}.
\end{equation}

The memory requirements of SCL and FSCL decoding algorithms with list size $L$ in terms of the number of memory bits are given as \cite{Alexios_LLR_SCLD, Ali_FSSCL}
\begin{equation}
\mathcal{M}_\text{SCL} = \mathcal{M}_\text{FSCL} =  N(L+1)b_f+ 2LN,
\end{equation}
where $b_f$ is the number of bits used to quantize a floating-point number.

\subsection{Successive-Cancellation Flip and Dynamic Successive-Cancellation Flip Decoding}
\label{sec:polar:DSCF}

\textcolor{black}{
SC-Flip (SCF) decoding algorithm was proposed in \cite{SCF} to improve the error-correction performance of SC decoding for short to moderate block lengths. Specifically, if the estimated message word $\bm{\hat{u}}$ does not satisfy the CRC test after the initial SC decoding attempt, an additional SC decoding attempt is made by flipping the estimation of an information bit in $\bm{\hat{u}}$ that is most likely to be the first error bit. In the rest of this paper, as we only need to locate the error decision occurred when an information bit is decoded under SC-based decoding, the bit indices are referred to as information bits and are indexed from $1$ to $K+C$. In \cite{SCF}, the most erroneous position is estimated as $\imath=\argmin_{1\leq i \leq K+C} \abs{\alpha_{0,i}}$.}

\textcolor{black}{
A problem associated with SCF decoding is its poor estimation accuracy of the actual error position \cite{DSCF}. To address this issue, the authors in \cite{DSCF} propose the Dynamic SC-Flip (DSCF) decoding algorithm that utilizes a conditional probability model to accurately estimate the error position, which is given as $\imath = \argmin_{1\leq i \leq K+C} Q_i$, where $Q_i$ is the error metric of the $i$-th information bit under SC decoding:
\begin{equation}
	\label{equ:DSCF}
	Q_i = \!\!\abs{\alpha_{0,i}}\!\!+\!\! \sum_{1 \leq j \leq i}\! \frac{1}{\lambda}\ln\left[1\!+\!\exp\left(-\lambda\abs{\alpha_{0,j}}\right)\right]\!.
\end{equation}
The parameter $\lambda \in \mathbb{R}^+$ is a perturbation parameter that is optimized offline \cite{DSCF}. In \cite{9052925, Doan_ICC21, Furkan_TSP20}, fast decoding schemes are proposed to reduce the decoding latency of SCF and DSCF decoding. However, it was observed in \cite{DSCF, doan2020neural} that the maximum number of decoding attempts required by the DSCF-based decoders to obtain a comparable FER performance of SCL decoding with list size 16 is significantly large. This problem prevents the DSCF-based decoders to be practical for applications with a stringent worst-case latency.}

\subsection{Successive-Cancellation List Flip Decoding} 
\label{sec:polar:SCLF}

SCLF decoding also relies on a CRC verification to indicate whether the initial SCL decoding attempt is successful or not. If the first SCL decoding attempt does not satisfy the CRC verification, the SCLF decoding algorithm tries to identify the first information bit index $\imath$, at which the correct path is discarded from the list of the $L$ most probable decoding paths \cite{SCLF}. Given that the $\imath$-th bit index is correctly identified, in the next decoding attempt and after the path splitting occurred at the $\imath$-th bit index, the erroneous path selection is reversed where the $L$ decoding paths that have the highest (worst) path metrics are selected to continue the decoding \cite{SCLF}. This reversed path selection scheme recovers the correct decoding path, which was discarded at the initial SCL decoding at the $\imath$-th bit index, to the list of the active decoding paths. SCLF decoding then performs conventional SCL decoding operations for all the bit indices following $\imath$.

Given that at the $i$-th information bit under SCL decoding, there are $L$ active decoding paths denoted as $l$, $l \in [1, 2L]$. After the path splitting of the current $L$ active paths, the path metrics of the new $2L$ paths are computed and sorted. Let $l'$ be the index of a discarded decoding path after the path metric sorting, i.e., the path metric corresponding to $l'$ is among the $L$ largest path metric values. The probability that the path with index $l'$ is the correct decoding path is \cite{DSCF}
\begin{align}
Pr&(\bm{\hat{u}}_{1_{l'}}^{i_{l'}}=\bm{u}_1^i|\bm{\alpha}_n)=\!\!\prod_{\substack{1 \leq j \leq i\\{\forall j \in \mathbbm{A}_{l'}}}}\!\! Pr(\hat{u}_{j_{l'}}=u_j|\bm{\alpha}_n,\bm{\hat{u}}_{1_{l'}}^{j_{l'}-1}=\bm{u}_1^{j-1})\nonumber\\
&\times \!\!\prod_{\substack{1 \leq j \leq i\\{\forall j \in \mathbbm{A}^c_{l'}}}}\! \left[1-Pr(\hat{u}_{j_{l'}}=u_j|\bm{\alpha}_n,\bm{\hat{u}}_{1_{l'}}^{j_{l'}-1}=\bm{u}_1^{j-1})\right],
\end{align}
where $Pr(\bm{\hat{u}}_{1_{l'}}^{i_{l'}}=\bm{u}_1^i|\bm{\alpha}_n)=Pr(\hat{u}_{1_{l'}}=u_1,\ldots,\hat{u}_{i_{l'}}=u_i|\bm{\alpha}_n)$. $\mathbbm{A}_{l'}$ is the set of information bit indices where their hard decisions follow the sign of the corresponding LLR values, while $\mathbbm{A}^c_{l'}$ is the set of information bit indices whose hard decisions do not follow the sign of the LLR values \cite{SCLF}.

Note that $Pr(\hat{u}_{j_{l'}}=u_j|\bm{\alpha}_n,\bm{\hat{u}}_{1_{l'}}^{j_{l'}-1}=\bm{u}_1^{j-1})$ is not available during the course of decoding as $\bm{u}$ is unknown, thus it is approximated as \cite{DSCF, SCLF}
\begin{equation}
Pr(\hat{u}_{j_{l'}}=u_j|\bm{\alpha}_n,\bm{\hat{u}}_{1_{l'}}^{j_{l'}-1}=\bm{u}_1^{j-1}) \approx  \frac{1}{1+\exp(-\lambda\abs{\alpha_{0,j_{l'}}})},
\label{equ:oldestimate}
\end{equation}
where $\lambda \in \mathbb{R}^+$ is a perturbation parameter that is optimized offline to improve the approximation accuracy of (\ref{equ:oldestimate}). The probability that the correct decoding path is discarded at the information bit with index $i$ is
\begin{equation}
\label{equ:SCLF:Pe}
P_i = \sum_{\forall l'} Pr(\bm{\hat{u}}_{1_{l'}}^{i_{l'}}=\bm{u}_1^i|\bm{\alpha}_n).
\end{equation}
Therefore, the bit index at which the error decision is most likely to take place is $\imath=\argmax_{\log_2 L < i \leq K+C} P_i$ \cite{SCLF}.

Directly computing (\ref{equ:SCLF:Pe}) is not numerically stable \cite{DSCF, SCLF}. Thus, a flipping metric based on the max-log approximation is derived from (\ref{equ:SCLF:Pe}) as \cite{SCLF}
\begin{equation}
\label{equ:SCLF:FM:old}
\begin{split}
Q_i
& = \min_{\forall l'} \left[ \sum_{{\forall j \in \mathbbm{A}^c_{l'}}}\!\!\abs{\alpha_{0,j_{l'}}} \!+\!\! \sum_{1 \leq j \leq i}\! \frac{1}{\lambda}\ln\left[1\!+\!\exp\left(-\lambda\abs{\alpha_{0,j_{l'}}}\right)\right]\! \right].
\end{split}
\end{equation}
\textcolor{black}{With $L=1$, (\ref{equ:SCLF:FM:old}) reverts to (\ref{equ:DSCF}) since $\mathbbm{A}^c_{l'}=\{i\}$ indicates the decoding path forked from the initial SC decoding path at the $i$-th bit. Thus, with $L=1$, SCLF decoding is equivalent to DSCF decoding when only the first error bit of the initial SC decoding is considered.}

The computation of $Q_i$ can be further simplified by using a hardware-friendly approximation introduced in \cite{Furkan_TSP20}:
\begin{equation}
\!f_\lambda(x)=\frac{1}{\lambda}\ln\left[1\!+\!\exp\left(-\lambda\abs{x}\right)\right]\!\approx\begin{cases}
a_\lambda &\text{if } \abs{x} \leq b_\lambda,\\
0 &\text{otherwise,}
\end{cases}
\end{equation}
where $a_\lambda, b_\lambda \in \mathbb{R}^+$ are tunable parameters selected based on a predetermined value of $\lambda$. However, this approach requires the optimizations of the parameters $\{\lambda,a_\lambda,b_\lambda\}$. On the other hand, in \cite{doan2020neural} the authors approximate the term 	${Pr(\hat{u}_{j_{l'}}=u_j|\bm{\alpha}_n,\bm{\hat{u}}_{1_{l'}}^{j_{l'}-1}=\bm{u}_1^{j-1})}$ in (\ref{equ:oldestimate}) as
\begin{equation}
Pr(\hat{u}_{j_{l'}}=u_j|\bm{\alpha}_n,\bm{\hat{u}}_{1_{l'}}^{j_{l'}-1}=\bm{u}_1^{j-1}) \approx  \frac{1}{1+\exp(\theta-\abs{\alpha_{0,j_{l'}}})},
\label{equ:newestimate}
\end{equation}
where similar to $\lambda$, $\theta \in \mathbb{R}^+$ is an additive perturbation parameter used to improve the approximation accuracy. The flipping metric $Q_i$ under the approximation provided in $(\ref{equ:newestimate})$ is then given as \cite{SCLF, doan2020neural}
\begin{equation}
\label{equ:SCLF:FM}
\begin{split}
Q_i \approx \min_{\forall l'} \left[ \sum_{{\forall j \in \mathbbm{A}^c_{l'}}}\!\!\left(\abs{\alpha_{0,j_{l'}}}-\theta\right) \!+\!\! \sum_{1 \leq j \leq i}\!\relu\left(\theta-\abs{\alpha_{0,j_{l'}}}\right)\!\right],
\end{split}
\end{equation}
where \textcolor{black}{$\relu(a)=a$ if $a>0$ and $\relu(a)=0$ otherwise}. The most probable information bit index where the correct path is discarded is then estimated as $\imath=\argmin_{\log_2 L < i \leq K+C} Q_i$ \cite{SCLF}. In this paper, we implement the hardware-friendly SCLF decoder using (\ref{equ:SCLF:FM}) as it only requires the optimization of $\theta$.

\textcolor{black}{Unlike DSCF decoding, SCLF decoding can achieve the FER performance of SCL decoding with a large list size using a reasonable number of maximum decoding attempts \cite{SCLF}. A critical problem associated with SCLF decoding is that it fully traverses the polar binary tree as required by SCL decoding, which results in a high decoding latency. The SSCLF decoder was proposed in \cite{Lee20} to improve the decoding latency of SCLF decoding by introducing a reversed path selection scheme to FSCL decoding. However, the proposed scheme in \cite{Lee20} is only applied to the decoding steps that perform path splitting and path metric sorting, thus it does not apply to the decoding steps that occur after the $\tau$ path-splittings for the SPC and Rate-1 nodes. As a consequent, SSCLF decoding with a small list size ($L\in\{2,4\}$) incurs a significant FER performance degradation when compared to SCLF decoding with the same list size.}

\section{Fast Successive-Cancellation List Flip Decoding}
\label{sec:FSCLF}	

\subsection{Bit-flipping Scheme for FSCL Decoding}
\label{sec:FSCLF:flip_scheme}
We first introduce the bit-flipping scheme tailored to FSCL decoding by illustrating the proposed scheme under various examples. We consider the case where an all-zero codeword of $\mathcal{P}(16,8)$ is transmitted through the channel, whose binary-tree representation is depicted in Fig.~\ref{fig:polar:tree}(a). Similar to SCL-based decoding, under FSCL-based decoding, we denote by $l$ the path index corresponding to the current $L$ active decoding paths, while $\tilde{l}$ is used to indicate the indices of the paths that are forked from $l$. Finally, $l'$ indicates the path indices of the decoding paths that are discarded due to their high path metric values. Note that $l,\tilde{l}, l' \in [1,2L]$. 

\begin{table}[b]
	\centering	
	\footnotesize
	\setlength{\tabcolsep}{3pt}
	\begin{tabular}{c | c | l | l | l | l}		
		\toprule
		\makecell{Node \\type}& \makecell{Path \\ splitting \\ index}&$l'=5$ &$l'=6$&$l'=7$&$l'=8$\\
		\midrule
		\multirow{4}{*}{\textbf{SPC}}&
		- & $\beta_{2,8_5}=\text{n/a}$&$\beta_{2,8_6}=\text{n/a}$&$\beta_{2,8_7}=\text{n/a}$&$\beta_{2,8_8}=\text{n/a}$\\
		&1 &$\beta_{2,5_5}=0$&$\beta_{2,5_6}=0$&$\beta_{2,5_7}=1$&$\beta_{2,5_8}=1$\\
		&2 &$\beta_{2,7_5}=0$&$\beta_{2,7_6}=1$&$\beta_{2,5_7}=0$&$\beta_{2,5_8}=1$\\
		&3 &$\beta_{2,6_5}=0$&$\beta_{2,6_6}=1$&$\beta_{2,6_7}=0$&$\beta_{2,6_8}=1$\\
	\end{tabular}	
	\caption{\textcolor{black}{An example of FSCL decoding applied to an SPC node of size 4 with $L=4$, where the decoding is at the third path splitting. $l'\in \{5,6,7,8\}$ are the indices of the discarded paths.}}
	\label{fig:FSCLF_SPC}
\end{table}

\subsubsection{Bit-Flipping Scheme for SPC Nodes}

Table~\ref{fig:FSCLF_SPC} shows an example of FSCL decoding when applied to the SPC node of $\mathcal{P}(16,8)$ with \textcolor{black}{$L=4$}. The decoding order is first determined by sorting the magnitude of the LLR values associated with the SPC node in the increasing order. In this example, the following decoding order is considered: $\{\beta_{2,8_l}, \beta_{2,5_l}, \beta_{2,7_l}, \beta_{2,6_l}\}$. Thus, $\beta_{2,8_l}$ is selected as the parity bit of the SPC node for all the active decoding paths. The path splittings at \textcolor{black}{$\beta_{2,6_l}$} are considered in this example, and the paths with indices \textcolor{black}{$\tilde{l}\in\{5,6,7,8\}$} are forked from the paths with indices \textcolor{black}{$l\in\{1,2,3,4\}$} at \textcolor{black}{$\beta_{2,6_l}$}, respectively, followed by the path metric sorting operations. \textcolor{black}{The most likely decoding paths with indices $l=\{1,2,3,4\}$ are then selected to continue the decoding, while the paths with indices $l'\in \{5,6,7,8\}$ are discarded as illustrated in Table~\ref{fig:FSCLF_SPC}.}

\textcolor{black}{At this stage, the parity bit $\beta_{2,8_l}$ of the SPC node is not yet decoded. As an all-zero codeword is considered, the correct decoding path is $l'=5$, which is discarded after $\beta_{2,6_l}$ is decoded, i.e., after the third path-splitting index. Given that this erroneous decision in the initial FSCL decoding is detected by a CRC verification, this erroneous path selection is reversed in the next decoding attempt by swapping the path indices of $l'$ and $l$ after the path splittings at $\beta_{2,6_l}$ for all the decoding paths. The decoding continues by setting the values of the parity bit $\beta_{2,8_l}$ with respect to (\ref{equ:FSSCL-SPC:beta_update}) to maintain the parity constraint for all the corrected paths. Similar to the bit-flipping schemes introduced in \cite{Lee20, Doan_ICC21}, in the proposed scheme, the bit-flipping operation is not applicable to the parity bits of the SPC nodes. This is due to the fact that the parity bits are determined after the all the other bits are calculated to ensure the parity check is satisfied.} Therefore, if all the other bits of the SPC node are correctly decoded, the parity bit of this decoding path is also correctly decoded. As a result, the proposed algorithm only considers a maximum of $N_\nu-1$ possibilities to identify a bit flip that occurs in an SPC node. This is significantly smaller than the maximum search space of size ${N_\nu \choose 2}$ required to flip a pair of bits to maintain the parity check constraint, especially as $N_\nu$ increases \cite{Furkan_TSP20}.

\textcolor{black}{Note that in this example, the minimum number of path splittings required by the SPC node to preserve the SCL decoding performance is $\tau=\min\{L-1,N_\nu-1\}=\min\{3,3\}=3$, where $\nu$ indicates the SPC node of size $4$. Under the proposed bit-flipping scheme for SPC nodes, if a decision error occurs at the path-splitting index after the minimum number of $\tau$ path splittings are obtained, in the next decoding attempt, the hard values at the estimated error index are flipped for all the active paths. The path metrics $\text{PM}_l$ of the surviving paths are then updated by following (\ref{fig:SPC:PM}), using the LLR value corresponding to the flipped position and the current parity checksum $p_l$.}

\subsubsection{Bit-Flipping Scheme for REP Nodes}

Since the soft and hard estimate of the information bit associated with a REP node can be directly obtained at the parent node level under FSCL decoding, the path splitting operation under FSCL decoding applied to the information bit of a REP node is similar to that of SCL decoding when applied to an information bit at the leaf node level. Therefore, in this paper, the reversed path selection scheme used in SCLF decoding is directly applied to the information bit associated with a REP node or to an information bit at the leaf-node level under FSCL decoding \cite{Lee20,Doan_ICC21}.

\subsubsection{Bit-Flipping Scheme for Rate-1 Nodes}

Table~\ref{fig:FSCLF_R1} shows an example of FSCL decoding on the Rate-1 node of $\mathcal{P}(16,8)$ at the \textcolor{black}{fifth} path-splitting index with $L=2$. In Table~\ref{fig:FSCLF_R1}, the hard estimates of the discarded paths with indices \textcolor{black}{$l'\in \{2,4\}$} are indicated, while the hard estimates of the surviving paths with indices \textcolor{black}{$l'\in \{1,3\}$} are omitted. It can be observed that the decoding path with index \textcolor{black}{$l'=2$} is the correct path as all the estimated bits are 0, which is discarded after bit \textcolor{black}{$\beta_{2,13_2}$} is decoded. Therefore, in the next decoding attempt, the decoding paths with indices \textcolor{black}{$l'\in\{2,4\}$} will be selected to continue the decoding instead of the paths with indices \textcolor{black}{$l'\in\{1,3\}$} \cite{Lee20,Doan_ICC21}. Similar to the case of SPC nodes, after $\tau$ path splittings, if the hard decision of a bit of the Rate-1 node results in the elimination of the correct path, this erroneous decision is reversed in the next FSCL decoding attempt by flipping the hard estimates of all the active paths at that erroneous index. \textcolor{black}{The path metrics of the active paths are then added with the corresponding absolute LLR values of the flipping indices. In this example, the minimum number of path splittings is $\tau=\min\{L-1,N_\nu\}=\min\{1,4\}=1$, which is obtained at the fifth path-splitting index. Therefore, under FSCL decoding, the hard values of all the active decoding paths following the fifth path-splitting index are set to follow the signs of their LLR values.}

\begin{table}[t]
	\centering
	\setlength{\tabcolsep}{4pt}
	\footnotesize
	\begin{tabular}{c | c | l | l}
		\toprule
		\makecell{Node \\type}& \makecell{Path \\ splitting \\ index}& $l'=2$ & $l'=4$\\
		\midrule
		\multirow{4}{*}{SPC}&
		- & $\beta_{2,8_2}=0$ & $\beta_{2,8_4}=0$\\
		&1 & $\beta_{2,5_2}=0$ & $\beta_{2,5_4}=0$\\
		&2 & $\beta_{2,7_2}=0$ & $\beta_{2,7_4}=0$\\
		&3 & $\beta_{2,6_2}=0$ & $\beta_{2,6_4}=0$\\
		\midrule
		REP &4 & $\beta_{0,12_2}=0$ & $\beta_{0,12_4}=0$\\
		\midrule
		\multirow{4}{*}{\textbf{Rate-1}}&
		5 & $\beta_{2,13_2}=0$ & $\beta_{2,15_4}=1$\\
		&6 & $\beta_{2,15_2}=\text{n/a}$ & $\beta_{2,16_4}=\text{n/a}$\\
		&7 & $\beta_{2,16_2}=\text{n/a}$ & $\beta_{2,14_4}=\text{n/a}$\\
		&8 & $\beta_{2,14_2}=\text{n/a}$ & $\beta_{2,13_4}=\text{n/a}$\\
	\end{tabular}
	\caption{\textcolor{black}{An example of FSCL decoding applied to a Rate-1 node of size 4 with $L=2$, where the decoding is at the 6-th path splitting. $l'\in \{2,4\}$ are the indices of the discarded paths.}}
	\label{fig:FSCLF_R1}
\end{table}

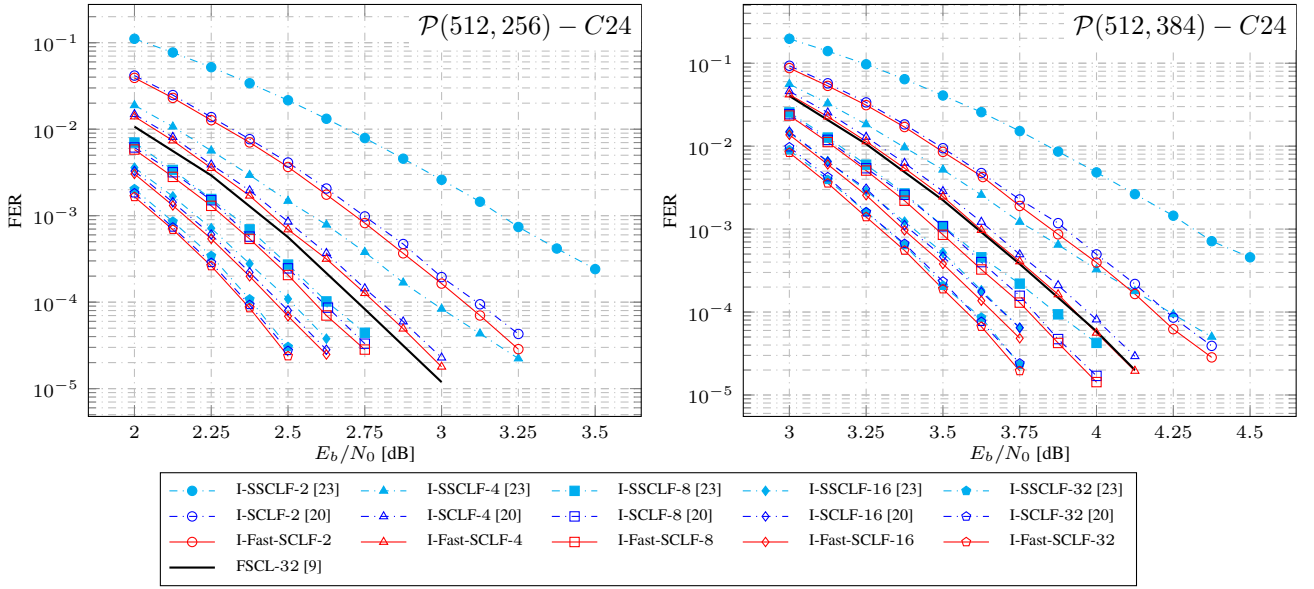
\begin{figure*}[t!]
	\centering
	\begin{tikzpicture}[spy using outlines = {rectangle, magnification=2.0, connect spies}]
	\pgfplotsset{	
		label style = {font=\fontsize{7pt}{7}\selectfont},
		tick label style = {font=\fontsize{7pt}{7}\selectfont}
	}
	
	\begin{axis}[
		scale = 1,
		ymode=log,
		xlabel={$E_b/N_0$ [\text{dB}]}, xlabel style={yshift=0.8em},
		ylabel={FER}, ylabel style={yshift=-0.75em},
		xtick={2,2.25,2.5,2.75,3,3.25,3.5,3.75,4,4.25,4.5,4.75,5},
		ytick={1e-8, 1e-7, 1e-6,1e-5,1e-4,1e-3,1e-2,1e-1,1e-0},
		grid=both,
		ymajorgrids=true,
		xmajorgrids=true,
		grid style=dashdotted,
		width=0.5\linewidth, height=7cm,
		thin,
		mark size=1.75,
		legend cell align={left},
		legend style={
			at={(0,1e-5)},
			anchor=south west,
			column sep= 2mm,
			font=\fontsize{6pt}{7.2}\selectfont,
		},
		legend to name=perf-legend-I-Fast-SCLF,
		legend columns=5,
		]
		
				\addplot[
		color=cyan,
		dashdotted,
		mark=*,
		mark options={solid},
		thin,
		mark size=1.75,
		]
		table {
			2	1.11E-01
			2.125	7.71E-02
			2.25	5.21E-02
			2.375	3.40E-02
			2.5	2.16E-02
			2.625	1.32E-02
			2.75	7.90E-03
			2.875	4.56E-03
			3	2.60E-03
			3.125	1.45E-03
			3.25	7.40E-04
			3.375	4.17E-04
			3.5	2.41E-04
		};
		\addlegendentry{I-SSCLF-$2$ \cite{Lee20}}
		
		\addplot[
		color=cyan,
		dashdotted,
		mark=triangle*,
		mark options={solid},
		thin,
		mark size=1.75,
		]
		table {
			2	1.89E-02
			2.125	1.07E-02
			2.25	5.64E-03
			2.375	2.95E-03
			2.5	1.48E-03
			2.625	7.86E-04
			2.75	3.79E-04
			2.875	1.68E-04
			3	8.37E-05
			3.125	4.33E-05
			3.25	2.23E-05
		};
		\addlegendentry{I-SSCLF-$4$ \cite{Lee20}}
		
		\addplot[
		color=cyan,
		dashdotted,
		mark=square*,
		mark options={solid},
		thin,
		mark size=1.75,
		]
		table {
			2	7.01E-03
			2.125	3.38E-03
			2.25	1.56E-03
			2.375	6.94E-04
			2.5	2.72E-04
			2.625	1.02E-04
			2.75	4.43E-05
		};
		\addlegendentry{I-SSCLF-$8$ \cite{Lee20}}
		
		\addplot[
		color=cyan,
		dashdotted,
		mark=diamond*,
		mark options={solid},
		thin,
		mark size=1.75,
		]
		table {
			2	3.59E-03
			2.125	1.67E-03
			2.25	7.16E-04
			2.375	2.78E-04
			2.5	1.09E-04
			2.625	3.78E-05
		};
		\addlegendentry{I-SSCLF-$16$ \cite{Lee20}}
		
		\addplot[
		color=cyan,
		dashdotted,
		mark=pentagon*,
		mark options={solid},
		thin,
		mark size=1.75,
		]
		table {
			2	2.03E-03
			2.125	8.54E-04
			2.25	3.43E-04
			2.375	1.08E-04
			2.5	3.04E-05
		};
		\addlegendentry{I-SSCLF-$32$ \cite{Lee20}}
		
		\addplot[
			color=blue,
			dashdotted,
			mark=o,
			mark options={solid},
			thin,
			mark size=1.75,
		]
		table {
			2.0000E+00	4.17E-02
			2.1250E+00	2.48E-02
			2.2500E+00	1.38E-02
			2.3750E+00	7.67E-03
			2.5000E+00	4.12E-03
			2.6250E+00	2.06E-03
			2.7500E+00	9.80E-04
			2.8750E+00	4.70E-04
			3.0000E+00	1.95E-04
			3.1250E+00	9.45E-05			
			3.25	4.29E-05
		};
		\addlegendentry{I-SCLF-$2$ \cite{SCLF}}
		
		\addplot[
		color=blue,
		dashdotted,
		mark=triangle,
		mark options={solid},
		thin,
		mark size=1.75,
		]
		table {
			2.0000E+00	1.49E-02
			2.1250E+00	8.11E-03
			2.2500E+00	3.90E-03
			2.3750E+00	1.96E-03
			2.5000E+00	8.52E-04
			2.6250E+00	3.69E-04
			2.7500E+00	1.44E-04
			2.8750E+00	5.96E-05
			3.0000E+00	2.28E-05
		};
		\addlegendentry{I-SCLF-$4$ \cite{SCLF}}
		
		\addplot[
		color=blue,
		dashdotted,
		mark=square,
		mark options={solid},
		thin,
		mark size=1.75,
		]
		table {
			2	6.15E-03
			2.125	3.18E-03
			2.25	1.51E-03
			2.375	5.73E-04
			2.5	2.46E-04
			2.63E+00	8.64E-05
			2.75	3.31E-05
		};
		\addlegendentry{I-SCLF-$8$ \cite{SCLF}}
		
		\addplot[
		color=blue,
		dashdotted,
		mark=diamond,
		mark options={solid},
		thin,
		mark size=1.75,
		]
		table {
			2	3.28E-03
			2.125	1.43E-03
			2.25	6.05E-04
			2.375	2.21E-04
			2.5	8.02E-05
			2.625	2.82E-05
		};
		\addlegendentry{I-SCLF-$16$ \cite{SCLF}}
		
		\addplot[
		color=blue,
		dashdotted,
		mark=pentagon,
		mark options={solid},
		thin,
		mark size=1.75,
		]
		table {
			2	1.82E-03
			2.125	7.32E-04
			2.25	2.85E-04
			2.375	9.31E-05
			2.5	2.74E-05
		};
		\addlegendentry{I-SCLF-$32$ \cite{SCLF}}
		
		\addplot[
			color=red,
			solid,
			mark=o,
			mark options={solid},
			thin,
			mark size=1.75,
		]
		table {
			2.0000E+00	3.92E-02
			2.1250E+00	2.30E-02
			2.2500E+00	1.27E-02
			2.3750E+00	7.03E-03
			2.5000E+00	3.66E-03
			2.6250E+00	1.75E-03
			2.7500E+00	8.21E-04
			2.8750E+00	3.67E-04
			3.0000E+00	1.64E-04
			3.1250E+00	7.01E-05
			3.2500E+00	2.86E-05
			
		};
		\addlegendentry{I-Fast-SCLF-$2$}
		
		\addplot[
		color=red,
		solid,
		mark=triangle,
		mark options={solid},
		thin,
		mark size=1.75,
		]
		table {
			2.0000E+00	1.40E-02
			2.1250E+00	7.38E-03
			2.2500E+00	3.55E-03
			2.3750E+00	1.70E-03
			2.5000E+00	6.92E-04
			2.6250E+00	3.16E-04
			2.7500E+00	1.27E-04
			2.8750E+00	4.93E-05
			3.0000E+00	1.79E-05			
		};
		\addlegendentry{I-Fast-SCLF-$4$}
		
		\addplot[
		color=red,
		solid,
		mark=square,
		mark options={solid},
		thin,
		mark size=1.75,
		]
		table {
			2	5.82E-03
			2.125	2.81E-03
			2.25	1.30E-03
			2.375	5.39E-04
			2.5	2.06E-04
			2.625	7.01E-05
			2.75E+00	2.83E-05
		};
		\addlegendentry{I-Fast-SCLF-$8$}
		
		\addplot[
		color=red,
		solid,
		mark=diamond,
		mark options={solid},
		thin,
		mark size=1.75,
		]
		table {
			2	3.04E-03
			2.125	1.31E-03
			2.25	5.40E-04
			2.375	1.99E-4
			2.5	6.88E-05
			2.625	2.48E-05
		};
		\addlegendentry{I-Fast-SCLF-$16$}
		
		\addplot[
		color=red,
		solid,
		mark=pentagon,
		mark options={solid},
		thin,
		mark size=1.75,
		]
		table {
			2	1.65E-03
			2.125	6.90E-04
			2.25	2.62E-04
			2.375	8.55E-5
			2.5	2.38E-05
		};
		\addlegendentry{I-Fast-SCLF-$32$}
		
		\addplot[
		color=black,
		solid,
		mark=none,
		mark options={solid},
		thick,
		mark size=1.75,
		]
		table {
			2	1.07E-02
			2.25	2.92E-03
			2.5	5.64E-04
			2.75	8.31E-05
			3	1.19E-05
		};
		\addlegendentry{FSCL-$32$ \cite{Ali_FSSCL}}
		
		\node[anchor=north east, fill=white] at (rel axis cs:1,1) {$\mathcal{P}(512,256)-C24$};
	\end{axis}
	
	
\end{tikzpicture}
	\begin{tikzpicture}[spy using outlines = {rectangle, magnification=2.0, connect spies}]
	\pgfplotsset{	
		label style = {font=\fontsize{7pt}{7}\selectfont},
		tick label style = {font=\fontsize{7pt}{7}\selectfont}
	}
	
	\begin{axis}[
		scale = 1,
		ymode=log,
		xlabel={$E_b/N_0$ [\text{dB}]}, xlabel style={yshift=0.8em},
		ylabel={FER}, ylabel style={yshift=-0.75em},
		xtick={2,2.25,2.5,2.75,3,3.25,3.5,3.75,4,4.25,4.5,4.75,5},
		ytick={1e-8, 1e-7, 1e-6,1e-5,1e-4,1e-3,1e-2,1e-1,1e-0},
		grid=both,
		ymajorgrids=true,
		xmajorgrids=true,
		grid style=dashdotted,
		width=0.5\linewidth, height=7cm,
		thin,
		mark size=1.5,
		legend cell align={left},
		legend style={
			at={(0,1e-5)},
			anchor=south west,
			column sep= 2mm,
			font=\fontsize{6pt}{7.2}\selectfont,
		},
		]		
		\addplot[
			color=cyan,
			dashdotted,
			mark=*,
			mark options={solid},
			thin,
			mark size=1.75,
		]
		table {
			3	1.97E-01
			3.125	1.40E-01
			3.25	9.71E-02
			3.375	6.42E-02
			3.5	4.08E-02
			3.625	2.57E-02
			3.75	1.52E-02
			3.875	8.61E-03
			4	4.83E-03
			4.125	2.64E-03
			4.25	1.45E-03
			4.375	7.15E-04
			4.5	4.57E-04
		};
		
		\addplot[
		color=cyan,
		dashdotted,
		mark=triangle*,
		mark options={solid},
		thin,
		mark size=1.75,
		]
		table {
			3	5.60E-02
			3.125	3.28E-02
			3.25	1.83E-02
			3.375	9.65E-03
			3.5	5.20E-03
			3.625	2.58E-03
			3.75	1.22E-03
			3.875	6.46E-04
			4	3.26E-04
			4.125	1.78E-04
			4.25	9.39E-05
			4.375	5.02E-05
		};
		
		\addplot[
		color=cyan,
		dashdotted,
		mark=square*,
		mark options={solid},
		thin,
		mark size=1.75,
		]
		table {
			3	2.55E-02
			3.125	1.27E-02
			3.25	5.98E-03
			3.375	2.69E-03
			3.5	1.09E-03
			3.625	4.57E-04
			3.75	2.20E-04
			3.875	9.38E-05
			4.0	4.250E-05
		};
		
		\addplot[
		color=cyan,
		dashdotted,
		mark=diamond*,
		mark options={solid},
		thin,
		mark size=1.75,
		]
		table {
			3	1.46E-02
			3.125	6.59E-03
			3.25	2.89E-03
			3.375	1.23E-03
			3.5	5.20E-04
			3.625	1.83E-04
			3.75	6.47E-05
		};
		
		\addplot[
		color=cyan,
		dashdotted,
		mark=pentagon*,
		mark options={solid},
		thin,
		mark size=1.75,
		]
		table {
			3	8.94E-03
			3.125	3.93E-03
			3.25	1.61E-03
			3.375	6.44E-04
			3.5	2.08E-04
			3.625	8.58E-05
			3.75	2.32E-5
		};
		
				\addplot[
		color=blue,
		dashdotted,
		mark=o,
		mark options={solid},
		thin,
		mark size=1.75,
		]
		table {
			3	9.32E-02
			3.125	5.73E-02
			3.25	3.37E-02
			3.375	1.83E-02
			3.5	9.45E-03
			3.625	4.74E-03
			3.75E+00	2.28E-03
			3.875	1.18E-03
			4	4.96E-04
			4.125	2.18E-04
			4.25	8.57E-05
			4.375	3.91E-05		
		};
		
		\addplot[
		color=blue,
		dashdotted,
		mark=triangle,
		mark options={solid},
		thin,
		mark size=1.75,
		]
		table {
			3	4.59E-02
			3.125	2.54E-02
			3.25	1.29E-02
			3.375	6.27E-03
			3.5	2.86E-03
			3.625	1.22E-03
			3.75E+00	4.95E-04
			3.875	2.10E-04
			4.00E+00	8.11E-05
			4.125	2.93E-05
		};
		
		\addplot[
		color=blue,
		dashdotted,
		mark=square,
		mark options={solid},
		thin,
		mark size=1.75,
		]
		table {
			3	2.43E-02
			3.125	1.19E-02
			3.25	5.42E-03
			3.375	2.59E-03
			3.5	1.08E-03
			3.625	4.00E-04
			3.75E+00	1.57E-04
			3.875	4.72E-05
			4.00E+00	1.70E-05
		};
		
		\addplot[
		color=blue,
		dashdotted,
		mark=diamond,
		mark options={solid},
		thin,
		mark size=1.75,
		]
		table {
			3	1.50E-02
			3.125 6.47223E-03
			3.25	3.07E-03
			3.375	1.11788E-03
			3.5	4.66E-04
			3.625	1.74657E-04
			3.75	6.46E-05
		};
		
		\addplot[
		color=blue,
		dashdotted,
		mark=pentagon,
		mark options={solid},
		thin,
		mark size=1.75,
		]
		table {
			3	9.72E-03
			3.125	4.24E-03
			3.25	1.60E-03
			3.375	6.66E-04
			3.5	2.34E-04
			3.625	7.64E-05
			3.75E+00	2.41E-05
		};
		
				\addplot[
		color=red,
		solid,
		mark=o,
		mark options={solid},
		thin,
		mark size=1.75,
		]
		table {
			3	8.73E-02
			3.125	5.33E-02
			3.25	3.12E-02
			3.375	1.70E-02
			3.5	8.53E-03
			3.63E+00	4.24E-03
			3.75E+00	1.90E-03
			3.875	8.65E-04
			4.00E+00	3.95E-04
			4.125	1.65E-04
			4.25	6.21E-05
			4.375	2.84E-05
		};
		
		\addplot[
		color=red,
		solid,
		mark=triangle,
		mark options={solid},
		thin,
		mark size=1.75,
		]
		table {
			3	4.19E-02
			3.125	2.27E-02
			3.25	1.14E-02
			3.375	5.42E-03
			3.5	2.46E-03
			3.625	9.73E-04
			3.75	4.07E-04
			3.875	1.63E-04
			4	5.61E-05
			4.125	1.95E-05
		};
		
		\addplot[
		color=red,
		solid,
		mark=square,
		mark options={solid},
		thin,
		mark size=1.75,
		]
		table {
			3	2.34E-02
			3.125	1.12E-02
			3.25	5.08E-03
			3.375	2.20E-03
			3.5	8.55E-04
			3.625	3.26E-04
			3.75	1.30E-04
			3.875	4.24E-05
			4.00E+00	1.43E-05
		};
		
		\addplot[
		color=red,
		solid,
		mark=diamond,
		mark options={solid},
		thin,
		mark size=1.75,
		]
		table {
			3	1.37E-02
			3.125	6.02E-03
			3.25	2.56E-03
			3.375	9.64E-04
			3.5	3.8E-04
			3.625	1.37E-04
			3.75	4.83E-05
		};
		
		\addplot[
		color=red,
		solid,
		mark=pentagon,
		mark options={solid},
		thin,
		mark size=1.75,
		]
		table {
			3	8.35E-03
			3.125	3.57E-03
			3.25	1.40E-03
			3.375	5.53E-04
			3.5	1.89E-04
			3.625	6.68E-05
			3.75E+00	1.95E-05
		};
				
		\addplot[
		color=black,
		solid,
		mark=none,
		mark options={solid},
		thick,
		mark size=1.75,
		]
		table {
			3	4.03E-02
			3.25	1.06E-02
			3.5	2.23E-03
			3.75E+00	3.79E-04
			4	5.86E-05
			4.125	2E-05
		};
		
		\node[anchor=north east, fill=white] at (rel axis cs:1,1) {$\mathcal{P}(512,384)-C24$};	
	\end{axis}
	
	
\end{tikzpicture}
	\ref{perf-legend-I-Fast-SCLF}
	\caption{\textcolor{black}{Ideal error-correction performance in terms of FER of various SCLF-based decoders. The FER values of the FSCL decoder with list size 32 are also plotted for comparison.}}
	\label{fig:fer_IFSCF}
\end{figure*}

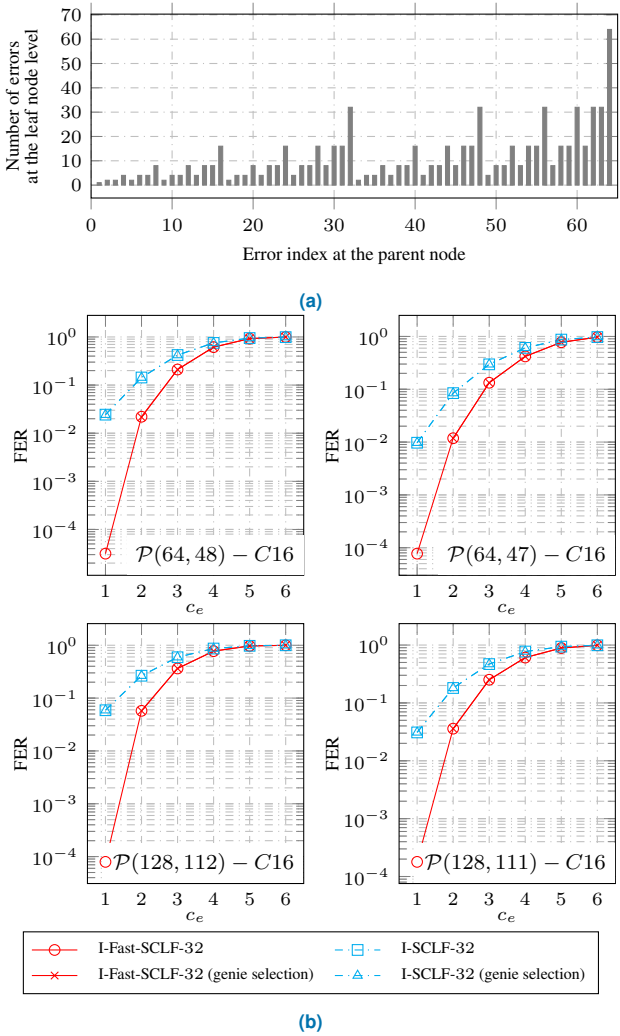
\begin{figure}[t]
	\centering
	\begin{subfigure}{\linewidth}
		\centering
		\begin{tikzpicture}
	\pgfplotsset{	
		label style = {font=\fontsize{7pt}{7}\selectfont},
		tick label style = {font=\fontsize{7pt}{7}\selectfont}
	}
	
	\begin{axis}[
		ybar,
		scale = 1,
		xlabel={Error index at the parent node}, xlabel style={yshift=0.0em},
		ylabel={\makecell{Number of errors\\ at the leaf node level}}, ylabel style={yshift=-1em},
		ytick={0,10,20,30,40,50,60,70},
		grid=both,
		ymajorgrids=true,
		xmajorgrids=true,
		grid style=dashdotted,
		width=1\linewidth, height=4cm,
		thin,
		mark size=2,
		xmin=0, xmax=65,
		legend cell align={left},
		legend style={
			at={(0,1e-5)},
			anchor=south west,
			column sep= 2mm,
			font=\fontsize{6pt}{7.2}\selectfont,
		},
		]
		
		\addplot[
			bar width=1.5pt,
			fill = gray,
			color=gray,
		]
		coordinates {
			(	1	,	1	)
			(	2	,	2	)
			(	3	,	2	)
			(	4	,	4	)
			(	5	,	2	)
			(	6	,	4	)
			(	7	,	4	)
			(	8	,	8	)
			(	9	,	2	)
			(	10	,	4	)
			(	11	,	4	)
			(	12	,	8	)
			(	13	,	4	)
			(	14	,	8	)
			(	15	,	8	)
			(	16	,	16	)
			(	17	,	2	)
			(	18	,	4	)
			(	19	,	4	)
			(	20	,	8	)
			(	21	,	4	)
			(	22	,	8	)
			(	23	,	8	)
			(	24	,	16	)
			(	25	,	4	)
			(	26	,	8	)
			(	27	,	8	)
			(	28	,	16	)
			(	29	,	8	)
			(	30	,	16	)
			(	31	,	16	)
			(	32	,	32	)
			(	33	,	2	)
			(	34	,	4	)
			(	35	,	4	)
			(	36	,	8	)
			(	37	,	4	)
			(	38	,	8	)
			(	39	,	8	)
			(	40	,	16	)
			(	41	,	4	)
			(	42	,	8	)
			(	43	,	8	)
			(	44	,	16	)
			(	45	,	8	)
			(	46	,	16	)
			(	47	,	16	)
			(	48	,	32	)
			(	49	,	4	)
			(	50	,	8	)
			(	51	,	8	)
			(	52	,	16	)
			(	53	,	8	)
			(	54	,	16	)
			(	55	,	16	)
			(	56	,	32	)
			(	57	,	8	)
			(	58	,	16	)
			(	59	,	16	)
			(	60	,	32	)
			(	61	,	16	)
			(	62	,	32	)
			(	63	,	32	)
			(	64	,	64	)			
		};
	\end{axis}	
\end{tikzpicture}
		\caption{}
		\label{fig:example_a}
	\end{subfigure}
	\begin{subfigure}{\linewidth}
		\centering
		\begin{tikzpicture}[spy using outlines = {rectangle, magnification=2.0, connect spies}]
	\pgfplotsset{	
		label style = {font=\fontsize{7pt}{7}\selectfont},
		tick label style = {font=\fontsize{7pt}{7}\selectfont}
	}
	
	\begin{axis}[
		scale = 1,
		ymode=log,
		xlabel={$c_e$}, xlabel style={yshift=0.8em},
		ylabel={FER}, ylabel style={yshift=-1em},
		xtick={1,2,3,4,5,6},
		ytick={1e-5,1e-4,1e-3,1e-2,1e-1,1e0},
		grid=both,
		ymajorgrids=true,
		xmajorgrids=true,
		grid style=dashdotted,
		width=0.52\linewidth, height=5cm,
		thin,
		mark size=2,
		legend cell align={left},
		legend style={
			at={(0,1e-5)},
			anchor=south west,
			column sep= 2mm,
			font=\fontsize{6pt}{7.2}\selectfont,
		},
				legend to name=perf-legend-fer-error,
				legend columns=2,
		]
		
		\addplot[
			color=red,
			solid,
			mark=o,
			mark options={solid},
			thin,
			mark size=2,
		]
		table {
			1	3.1423e-05
			2	2.2002e-02
			3	2.1062e-01
			4	6.1396e-01
			5	9.2449e-01
			6	9.9550e-01
		};
		\addlegendentry{I-Fast-SCLF-$32$}
		
		\addplot[
			color=cyan,
			dashdotted,
			mark=square,
			mark options={solid},
			thin,
			mark size=2,
		]
		table {
			1	2.4302e-02
			2	1.4371e-01
			3	4.2104e-01
			4	7.5128e-01
			5	9.4929e-01
			6	9.9120e-01
		};
		\addlegendentry{I-SCLF-$32$}
		
		\addplot[
		color=red,
		solid,
		mark=x,
		mark options={solid},
		thin,
		mark size=2,
		]
		table {
			2	2.1602e-02
			3	2.1062e-01
			4	6.1396e-01
			5	9.2449e-01
			6	9.9550e-01
		};
		\addlegendentry{I-Fast-SCLF-$32$ (genie selection)}
		
		\addplot[
			color=cyan,
			dashdotted,
			mark=triangle,
			mark options={solid},
			thin,
			mark size=2,
		]
		table {
			1	2.42e-02
			2	1.43e-01
			3	4.20e-01
			4	7.50e-01
			5	9.48e-01
			6	9.90e-01
		};
		\addlegendentry{I-SCLF-$32$ (genie selection)}
		
		
		\node[anchor=south east, fill=white] at (rel axis cs:1,0) {\footnotesize{$\mathcal{P}(64,48)-C16$}};
	\end{axis}
	
\end{tikzpicture}
		\begin{tikzpicture}[spy using outlines = {rectangle, magnification=2.0, connect spies}]
	\pgfplotsset{	
		label style = {font=\fontsize{7pt}{7}\selectfont},
		tick label style = {font=\fontsize{7pt}{7}\selectfont}
	}
	
	\begin{axis}[
		scale = 1,
		ymode=log,
		xlabel={$c_e$}, xlabel style={yshift=0.8em},
		ylabel={FER}, ylabel style={yshift=-1.0em},
		xtick={1,2,3,4,5,6},
		ytick={1E-5, 1E-4,1E-3,1E-2,1E-1,1E0},
		grid=both,
		ymajorgrids=true,
		xmajorgrids=true,
		grid style=dashdotted,
		width=0.52\linewidth, height=5cm,
		thin,
		mark size=1.5,
		legend cell align={left},
		legend style={
			at={(0,1e-5)},
			anchor=south west,
			column sep= 2mm,
			font=\fontsize{6pt}{7.2}\selectfont,
		},
		]
		
		\addplot[
		color=red,
		solid,
		mark=o,
		mark options={solid},
		thin,
		mark size=2,
		]
		table {
			1	7.7294e-05
			2	1.1830e-02
			3	1.3281e-01
			4	4.1864e-01
			5	7.7488e-01
			6	9.6780e-01
		};
		
		\addplot[
		color=red,
		solid,
		mark=x,
		mark options={solid},
		thin,
		mark size=2,
		]
		table {
			2	1.1830e-02
			3	1.3281e-01
			4	4.1864e-01
			5	7.7488e-01
			6	9.6780e-01
		};
						
		\addplot[
			color=cyan,
			dashdotted,
			mark=square,
			mark options={solid},
			thin,
			mark size=2,
		]
		table {
			1	9.7172e-03
			2	8.4508e-02
			3	2.9623e-01
			4	6.0716e-01
			5	8.7119e-01
			6	9.7340e-01
		};
		
		\addplot[
		color=cyan,
		dashdotted,
		mark=triangle,
		mark options={solid},
		thin,
		mark size=2,
		]
		table {
			1	9.7172e-03
			2	8.4508e-02
			3	2.9623e-01
			4	6.0716e-01
			5	8.7119e-01
			6	9.7340e-01
		};
		
		
		\node[anchor=south east, fill=white] at (rel axis cs:1,0) {\footnotesize{$\mathcal{P}(64,47)-C16$ }};
	\end{axis}
	
\end{tikzpicture}\\
		\begin{tikzpicture}[spy using outlines = {rectangle, magnification=2.0, connect spies}]
	\pgfplotsset{	
		label style = {font=\fontsize{7pt}{7}\selectfont},
		tick label style = {font=\fontsize{7pt}{7}\selectfont}
	}
	
	\begin{axis}[
		scale = 1,
		ymode=log,
		xlabel={$c_e$}, xlabel style={yshift=0.8em},
		ylabel={FER}, ylabel style={yshift=-1em},
		xtick={1,2,3,4,5,6},
		ytick={1e-5,1e-4,1e-3,1e-2,1e-1,1e0},
		grid=both,
		ymajorgrids=true,
		xmajorgrids=true,
		grid style=dashdotted,
		width=0.52\linewidth, height=5cm,
		thin,
		mark size=2,
		legend cell align={left},
		legend style={
			at={(0,1e-5)},
			anchor=south west,
			column sep= 2mm,
			font=\fontsize{6pt}{7.2}\selectfont,
		},
		]
		
		\addplot[
			color=red,
			solid,
			mark=o,
			mark options={solid},
			thin,
			mark size=2,
		]
		table {
			1	7.9509e-05
			2	5.7406e-02
			3	3.6384e-01
			4	7.7318e-01
			5	9.7121e-01
			6	9.9660e-01
		};

		\addplot[
		color=red,
		solid,
		mark=x,
		mark options={solid},
		thin,
		mark size=2,
		]
		table {
			2	5.7406e-02
			3	3.6384e-01
			4	7.7318e-01
			5	9.7121e-01
			6	9.9660e-01
		};
		
		\addplot[
			color=cyan,
			dashdotted,
			mark=square,
			mark options={solid},
			thin,
			mark size=2,
		]
		table {
			1	5.8906e-02
			2	2.6512e-01
			3	5.9311e-01
			4	8.6659e-01
			5	9.7574e-01
			6	9.9910e-01
		};

		\addplot[
			color=cyan,
			dashdotted,
			mark=triangle,
			mark options={solid},
			thin,
			mark size=2,
		]
		table {
			1	5.8906e-02
			2	2.6512e-01
			3	5.9311e-01
			4	8.6659e-01
			5	9.7574e-01
			6	9.9910e-01
		};
		
		\node[anchor=south east, fill=white] at (rel axis cs:1,0) {\footnotesize{$\mathcal{P}(128,112)-C16$}};
	\end{axis}
	
\end{tikzpicture}
		\begin{tikzpicture}[spy using outlines = {rectangle, magnification=2.0, connect spies}]
	\pgfplotsset{	
		label style = {font=\fontsize{7pt}{7}\selectfont},
		tick label style = {font=\fontsize{7pt}{7}\selectfont}
	}
	
	\begin{axis}[
		scale = 1,
		ymode=log,
		xlabel={$c_e$}, xlabel style={yshift=0.8em},
		ylabel={FER}, ylabel style={yshift=-1em},
		xtick={1,2,3,4,5,6},
		ytick={1e-5,1e-4,1e-3,1e-2,1e-1,1e0},
		grid=both,
		ymajorgrids=true,
		xmajorgrids=true,
		grid style=dashdotted,
		width=0.52\linewidth, height=5cm,
		thin,
		mark size=2,
		legend cell align={left},
		legend style={
			at={(0,1e-5)},
			anchor=south west,
			column sep= 2mm,
			font=\fontsize{6pt}{7.2}\selectfont,
		},
		]
		
		\addplot[
			color=red,
			solid,
			mark=o,
			mark options={solid},
			thin,
			mark size=2,
		]
		table {
			1	1.7907e-04
			2	3.5898e-02
			3	2.5123e-01
			4	6.0729e-01
			5	8.8896e-01
			6	9.9010e-01
		};

		\addplot[
			color=red,
			solid,
			mark=x,
			mark options={solid},
			thin,
			mark size=2,
		]
		table {
			2	3.5898e-02
			3	2.5123e-01
			4	6.0729e-01
			5	8.8896e-01
			6	9.9010e-01
		};
		
		\addplot[
			color=cyan,
			dashdotted,
			mark=square,
			mark options={solid},
			thin,
			mark size=2,
		]
		table {
			1	3.0903e-02
			2	1.8109e-01
			3	4.6880e-01
			4	7.7271e-01
			5	9.3859e-01
			6	9.9109e-01
		};

		\addplot[
			color=cyan,
			dashdotted,
			mark=triangle,
			mark options={solid},
			thin,
			mark size=2,
		]
		table {
			1	3.0903e-02
			2	1.8109e-01
			3	4.6880e-01
			4	7.7271e-01
			5	9.3859e-01
			6	9.9109e-01
		};
		
		\node[anchor=south east, fill=white] at (rel axis cs:1,0) {\footnotesize{$\mathcal{P}(128,111)-C16$}};
	\end{axis}
	
\end{tikzpicture}\\
		\ref{perf-legend-fer-error}
		\caption{}
		\label{fig:example_b}
	\end{subfigure}
	\caption{\textcolor{black}{(a) The number of translated errors at the leaf node level given a single error at a specific bit index at the parent node level for a sized-64 polar code and (b) the error-correction performance of I-Fast-SCLF-$32$ and I-SCLF-$32$ for the Rate-1 and SPC codes of lengths $N\in\{64,128\}$ with various values of $c_e$.}}
\end{figure}


Similar to SCLF decoding, the proposed scheme only aims at correcting the first erroneous decision in the initial FSCL decoding attempt. Fig~\ref{fig:fer_IFSCF} shows the ideal FER of the proposed bit-flipping scheme where the first erroneous path selection is always accurately corrected. In Fig~\ref{fig:fer_IFSCF}, we use the 5G polar codes $\mathcal{P}(512,256)$ and $\mathcal{P}(512,384)$ concatenated with a $24$-bit CRC\footnote{We use the $24$-bit CRC specified as 24C in the 5G standard.}. Note that the positions of the first erroneous decoding decision can be obtained by comparing the discarded paths with the correct path after each path splitting. \textcolor{black}{The FER performance of the ideal SCLF \cite{SCLF} and SSCLF \cite{Lee20} decoders and the FSCL decoder with list size $32$ \cite{Ali_FSSCL} are also plotted for comparison. In Fig~\ref{fig:fer_IFSCF}, the ideal SCLF, SSCLF, and Fast-SCLF decoders with list size $L$ are denoted as I-SCLF-$L$, I-SSCLF-$L$ and I-Fast-SCLF-$L$, respectively, with $L \in \{2,4,8,16,32\}$.}

\textcolor{black}{As seen from Fig~\ref{fig:fer_IFSCF}, I-Fast-SCLF-$L$ obtains a slight FER performance gain over I-SCLF-$L$. In addition, as the reversed path-selection scheme of \cite{Lee20} is not applied to the decoding steps that occur after the minimum number of $\tau$ path-splittings is obtained for the Rate-1 and SPC nodes, this simplified bit-flipping scheme of \cite{Lee20} introduces FER performance degradation when compared with the ideal SCLF and Fast-SCLF decoders, especially when the list size is small, $(L\in\{2,4\})$. For $L=4$ and at the target FER of $10^{-4}$, the error-correction performance degradations of $0.2$ dB and $0.3$ dB are recorded for the ideal SSCLF decoder when compared to the ideal Fast-SCLF decoder for $\mathcal{P}(512,256)$ and $\mathcal{P}(512,384)$, respectively. Also note that the FER performance of the ideal SSCLF decoder with $L\in\{2,4\}$ degrades quickly as the SNR increases.}

\textcolor{black}{We now explain the slight improvement in the error-correction performance of I-Fast-SCLF-$L$ over that of I-SCLF-$L$ as observed in Fig.~\ref{fig:fer_IFSCF}. The error-correction performance of I-Fast-SCLF-$L$ and I-SCLF-$L$ are identical for REP nodes. Therefore, we empirically show that I-Fast-SCLF-$L$ outperforms I-SCLF-$L$ when applied to the same channel LLR vectors for Rate-1 and SPC nodes, where the LLR vectors contain an exact number of $c_e$ $(c_e>0)$ channel errors. A similar study was conducted in \cite{Ercan8854992} for the case of fast SCF decoding.}

\textcolor{black}{We first consider the error event where only a single error is present in the LLR vector of the Rate-1 and SPC nodes ($c_e=1$), which causes an unsuccessful CRC verification in the first FSCL and SCL decoding attempt. After the correct decoding path is recovered in the second FSCL decoding attempt of I-Fast-SCLF-$L$, FSCL decoding operations, e.g., path forking and path metric sorting, are applied to the $L$ recovered paths. Then, all the hard decisions of the correct path are set to follow the signs of the corresponding LLR values to maintain its path metric. Therefore, the path metric is equal to the absolute LLR value of the flipped bit. Recall that the LLR values of the Rate-1 and SPC nodes are sorted in accordance with (\ref{equ:polar:sorted_LLR}). Thus, at the subsequent decoding steps after the flipped position, any new candidate path with at least a hard decision not following its corresponding LLR value will contain a higher path metric compared to that of the correct decoding path. Therefore, with $c_e=1$, the correct decoding path is always found in the list of the best paths after the second FSCL decoding attempt of I-Fast-SCLF-$L$.}

\textcolor{black}{Note that a single error at the parent node level can translate into multiple errors at the leaf node level. This phenomenon is illustrated in Fig.~\ref{fig:example_a} for an all-zero polar code of length $N=64$, where the number of error bits at the leaf node level is provided with respect to the position of the single error bit at the parent node level. Consequently, there are cases that I-SCLF-$L$ has to perform the reserved path selection schemes multiple times to maintain the correct codeword in the list of the best paths, while I-Fast-SCLF-$L$ only requires a single reserved path selection. Thus with $c_e=1$, the error correction performance of I-Fast-SCLF-$L$ is improved when compared to I-SCLF-$L$. As $c_e$ increases $(c_e\ge2)$, due to the complicated error patterns caused by multiple error bits, we expect that both I-Fast-SCLF-$L$ and I-SCLF-$L$ almost equally likely discard the correct path in the second decoding attempt, causing a wrong estimation of the transmitted codeword. Also note that when the channel reliability is improved at the high SNR regimes and given that $c_e>0$, the performance gain of I-Fast-SCLF-$L$ over I-SCLF-$L$ is mainly obtained from the case of $c_e=1$, as the LLR vectors are more likely to contain a single error than multiple ones ($c_e>1$). On the other hand, at the low SNR regimes, it is more likely to have multiple errors at the parent node level, thus the error-correction performance gain of I-Fast-SCLF-$L$ is incremental when compared to that of I-SCLF-$L$. This phenomenon can also be observed from Fig.~\ref{fig:fer_IFSCF}.}

\textcolor{black}{In Fig.~\ref{fig:example_b}, we plot the FER curves of I-Fast-SCLF-$32$ and I-SCLF-$32$ for the polar codes of lengths $64$ and $128$ concatenated with a $16$-bit CRC used in the 5G standard, the values of $K$ are selected to form the Rate-1 and SPC nodes, respectively. The simulations are carried out at $E_b/N_0=3$ dB and the FER values of the decoders in Fig.~\ref{fig:example_b} are only obtained for the channel LLR vectors that contain exactly $c_e\in\{1,2,3,4,5,6\}$ errors. It can be confirmed from Fig.~\ref{fig:example_b} that with $c_e=1$, I-Fast-SCLF-$32$ has a significant FER performance improvement when compared to I-SCLF-$32$. In addition, the error-correction performance gains of I-Fast-SCLF-$32$ with respect to I-SCLF-$32$ quickly reduce as $c_e$ increases, with the error probabilities of both decoders approaching $1$ when $c_e \ge 4$.}

\textcolor{black}{Note that the error-correction performance degradation of I-Fast-SCLF-$32$ is also caused by the imperfect error detection of the CRC, where a wrong estimate of the correct codeword that satisfies the CRC is selected as the decoding output. As shown in Fig.~\ref{fig:example_b}, if the CRC verification is replaced by a genie selection scheme, the FER values of I-SCLF-$32$ and I-Fast-SCLF-$32$ are relatively unchanged, except for the case of I-Fast-SCLF-$32$ with $c_e=1$, where an FER of $0$ is obtained\footnote{\textcolor{black}{We exclude the FER of I-Fast-SCLF-$32$ with $c_e=1$ from Fig.~\ref{fig:example_b} as an FER of $0$ cannot be plotted in the logarithmic scale.}}. This confirms that after the second FSCL decoding attempt of I-Fast-SCLF-$32$, the correct codeword is always present in the list of the best decoding paths for $c_e=1$.}

In the next section, a path selection error model is derived to accurately estimate the index of the path splitting that causes the elimination of the correct path at the initial FSCL decoding. Therefore, the error-correction performance of the I-Fast-SCLF-$L$ decoder provided in Fig.~\ref{fig:fer_IFSCF} serves as the empirical lower bound of the proposed decoding algorithm.

\subsection{path selection Error Model for FSCL Decoding}
\label{sec:FSCLF:metric}

We use the methods introduced in \cite{DSCF, SCLF} to estimate the erroneous path-splitting index, which predicts the error position using the LLR values associated with each discarded decoding path. Thus, the proposed path selection error model relies on the construction of the LLR vectors obtained at each path splitting under FSCL decoding.

Consider that the first FSCL decoding attempt does not pass the CRC verification and $\nu$ is a node located at the $s$-th stage of the binary tree, that is visited by FSCL decoding. Let $k \in [1, K+C]$ be the path-splitting index that occurs during the decoding of $\nu$. Let $\bm{\gamma}_{l'}=\bm{\gamma}_{1_{l'}}^{k_{l'}}$ be an LLR vector of a discarded decoding path $l'$, which contains $k$ LLR values corresponding to the hard estimates of the discarded path $l'$. After each information bit is decoded, $\bm{\gamma}_{l'}$ is constructed progressively up to the $k$-th path splitting. Formally, $\bm{\gamma}_{l'}$ is obtained using the following procedure:
\begin{itemize}
	\item If $\nu$ is a leaf node that contains an information bit:
		\begin{equation}
		\bm{\gamma}_{l'}= \concat(\bm{\gamma}_{l'},\alpha_{0,i_{{\min}_{\nu_{l'}}}}).
		\label{equ:leaf}
		\end{equation}
	\item If $\nu$ is a REP node:
		\begin{equation}
		\bm{\gamma}_{l'}= \concat(\bm{\gamma}_{l'},\sum_{j=i_{{\min}_{\nu_{l'}}}}^{i_{{\max}_{\nu_{l'}}}} \alpha_{s,j}).
		\end{equation}	
	\item If $\nu$ is a Rate-1 node: Updating $\bm{\gamma}_{l'}$ using the following function after each path splitting at the $j$-th bit:
		\begin{equation}
		\bm{\gamma}_{l'}= \concat(\bm{\gamma}_{l'},\alpha_{s,j}),
		\vspace*{-0.1\baselineskip}
		\end{equation}
		where $j \in \{i^*_{{\min}_{\nu_{l'}}}, \ldots ,i^*_{{\max}_{\nu_{l'}}}\}$ is selected by following the indices of the sorted absolute LLR values in (\ref{equ:polar:sorted_LLR}).
	
	\item If $\nu$ is an SPC node: Updating $\bm{\gamma}_{l'}$ using the following function after each path splitting at the $j$-th bit:
		\begin{equation}
		\label{equ:spc}
		\bm{\gamma}_{l'}= \concat(\bm{\gamma}_{l'},\alpha_{s,j}),
		\end{equation}
		where $j \in \{i^*_{{\min}_{\nu_{l'}}}, \ldots ,i^*_{{\max}_{\nu_{l'}}}\}\setminus i^*_{{\min}_{\nu_{l'}}}$ is selected by following the order of the sorted absolute LLR values in (\ref{equ:polar:sorted_LLR}). Note that $i^*_{{\min}_{\nu_{l'}}}$ is the bit index of the parity bit whose LLR value is ignored when constructing $\bm{\gamma}_{l'}$.
\end{itemize}
$\concat(\bm{\gamma}_{l'},a)$ is a function that concatenates $a \in \mathbb{R}$ to the end of $\bm{\gamma}_{l'}$ and initially $\bm{\gamma}_{l'}=\emptyset$. In addition, $\bm{\gamma}_{l'}$ is not altered if $\nu$ does not satisfy any of the above conditions. \textcolor{black}{For example, the LLR vector $\bm{\gamma}_{l'}$ obtained after the fifth path-splitting index in Table~\ref{fig:FSCLF_R1} for $l'=2$ is $\bm{\gamma}_{2}=\bm{\gamma}_{1_2}^{5_2}=\{\alpha_{2,5_2},\alpha_{2,7_2},\alpha_{2,6_2},\alpha_{0,12_2},\alpha_{2,13_2}\}$. We now define the hard estimates of $\bm{\gamma}_{l'}$ as $\hat{\bm{\eta}}_{l'}=\hat{\bm{\eta}}_{1_{l'}}^{k_{l'}}$, and the correct hard values associated with $\bm{\gamma}_{l'}$ as $\bm{\eta}_{l'}=\bm{\eta}_{1_{l'}}^{k_{l'}}$. For instance,  $\hat{\bm{\eta}}_{2}=\hat{\bm{\eta}}_{1_2}^{5_2}=\{\beta_{2,5_2},\beta_{2,7_2},\beta_{2,6_2},\beta_{0,12_2},\beta_{2,13_2}\}$ is the discarded decoding path obtained after the fifth path-splitting index in Table~\ref{fig:FSCLF_R1} with $l'=2$, and $\bm{\eta}_{2}=\bm{\eta}_{1_{2}}^{5_{2}}=\{0,0,0,0,0\}$. It is worth to note that by not considering the bit-flipping operations for the parity bits of the SPC nodes, the search space of the first error path selection for FSCL decoding contains $K+C$ possible positions, which is equal to that of the SCLF decoder.}

Unlike SCLF decoding, at the same path splitting index the hard estimates and LLR values of $\hat{\bm{\eta}}_{l'}$ and $\bm{\gamma}_{l'}$ of different path indices $l'$ can correspond to different bit indices of the polar binary tree. However, similar to SCLF decoding, each instance of the hard estimates and LLR values of $\hat{\bm{\eta}}_{l'}$ and $\bm{\gamma}_{l'}$ are obtained sequentially by following the course of FSCL decoding. Therefore, in this paper we utilize the conditional error probability model considered in \cite{DSCF, doan2020neural, SCLF} to estimate the erroneous decision occurred at the $k$-th path splitting index of FSCL decoding. Specifically, the probability that the discarded path $l'$ at the $k$-th path splitting index under FSCL decoding is the correct path is
\begin{equation}
\label{equ:FSCLF:error_prob}
\begin{split}
Pr&(\hat{\bm{\eta}}_{1_{l'}}^{k_{l'}}=\bm{\eta}_{1_{l'}}^{k_{l'}}|\bm{\alpha}_n)\\
&=\!\!\prod_{\substack{1 \leq j \leq k\\{\forall j \in \mathbbm{A}_{l'}}}} \!\!Pr(\hat{\eta}_{j_{l'}}=\eta_{j_{l'}}|\bm{\alpha}_n,\hat{\bm{\eta}}_{1_{l'}}^{j_{l'}-1}=\bm{\eta}_{1_{l'}}^{j_{l'}-1})\\
&\times\!\!\prod_{\substack{1 \leq j \leq k\\{\forall j \in \mathbbm{A}^c_{l'}}}} \!\!\left[ 1-Pr(\hat{\eta}_{j_{l'}}=\eta_{j_{l'}}|\bm{\alpha}_n,\hat{\bm{\eta}}_{1_{l'}}^{j_{l'}-1}=\bm{\eta}_{1_{l'}}^{j_{l'}-1}) \right],
\end{split}
\end{equation}
where $\mathbbm{A}_{l'}$ is the set of bit indices $j$ in which the hard estimates $\hat{\eta}_{j_{l'}}$ follow the sign of $\gamma_{j_{l'}}$, and $\mathbbm{A}^c_{l'}$ is the set of bit indices $j$ in which the hard estimates $\hat{\eta}_{j_{l'}}$ do not follow the sign of $\gamma_{j_{l'}}$.

Similar to $\bm{u}$, $\bm{\eta}_{l'}$ is also not available during the decoding process, thus we use the approximation introduced in \cite{doan2020neural} to calculate $Pr(\hat{\eta}_{j_{l'}}=\eta_{j_{l'}}|\bm{\alpha}_n,\hat{\bm{\eta}}_{1_{l'}}^{j_{l'}-1}=\bm{\eta}_{1_{l'}}^{j_{l'}-1})$ as
\begin{equation}
\label{equ:FSCLF:b}
Pr(\hat{\eta}_{j_{l'}}=\eta_{j_{l'}}|\bm{\alpha}_n,\hat{\bm{\eta}}_{1_{l'}}^{j_{l'}-1}=\bm{\eta}_{1_{l'}}^{j_{l'}-1}) \approx \frac{1}{1+\exp(\theta-\abs{\gamma_{j_{l'}}})}.
\end{equation}
The path selection error metric obtained at the $k$-th path splitting based on (\ref{equ:FSCLF:error_prob}) and (\ref{equ:FSCLF:b}) can be obtained as
\begin{align}
Q_k&=- \ln\left[ \sum_{\forall l'} Pr(\hat{\bm{\eta}}_{1_{l'}}^{k_{l'}}=\bm{\eta}_{1_{l'}}^{k_{l'}}|\bm{\alpha}_n) \right] \nonumber\\
& \approx - \ln \left[ \max_{\forall l'} Pr(\hat{\bm{\eta}}_{1_{l'}}^{k_{l'}}=\bm{\eta}_{1_{l'}}^{k_{l'}}|\bm{\alpha}_n) \right] \nonumber\\
& \approx \min_{\forall l'} \left[\sum_{\substack{1 \leq j \leq k\\{\forall j \in \mathbbm{A}^c_{l'}}}} \left(\abs{\gamma_{j_{l'}}}-\theta\right) + \sum_{1 \leq j \leq k}\relu\left(\theta-\abs{\gamma_{j_{l'}}}\right) \right].
\label{equ:FSCLF:FM}
\end{align}
Consequently, the most probable erroneous position $\imath$ is obtained as
\begin{equation}
\label{equ:FSCLF:sel}
\imath = \argmin_{\substack{\log_2 L < k \leq K+C}} Q_k.
\end{equation}

The error metric described in (\ref{equ:FSCLF:FM}) can be progressively calculated during the course of decoding, allowing for an efficient implementation of the proposed decoder. In particular, for each active decoding path $l$ we denote by $q_{{k-1}_{l}}$ the path-error metric at the $(k-1)$-th path splitting index of $l$, which is given as
\begin{equation}
q_{{k-1}_{l}} = \sum_{\substack{1 \leq j \leq k-1\\{\forall j \in \mathbbm{A}^c_{l}}}} \left(\abs{\gamma_{j_{l}}}-\theta\right) + \sum_{1 \leq j \leq k-1}\relu\left(\theta-\abs{\gamma_{j_{l}}}\right)
\end{equation}
if $k > 1$ and $q_{0_{l}}=0$ $\forall l$. Thus, the path-error metric of the path $l$ at the $k$-th path splitting index can be calculated from $q_{{k-1}_{l}}$ as
\begin{equation}
\label{equ:FSCLF:PEM:l}
q_{k_{l}} = q_{{k-1}_{l}} + \relu\left(\theta-\abs{\gamma_{k_{l}}}\right).
\end{equation}
The path-error metric of the forked path with index $\tilde{l}$ originated from $l$, whose hard value at the $k$-th path splitting index does not follow the sign of its LLR value, is calculated as
\begin{equation}
\label{equ:FSCLF:PEM:l_forked}
q_{k_{\tilde{l}}} = q_{k_{l}} + \abs{\gamma_{k_{l}}}-\theta.
\end{equation}
(\ref{equ:FSCLF:PEM:l}) and (\ref{equ:FSCLF:PEM:l_forked}) are used to compute the path-error metrics of all the $2L$ paths associated with the current $L$ active paths and the $L$ forked paths progressively. Next, the path metric sorting is carried out and a list of discarded paths with indices $l'$ is determined. The flipping metric in (\ref{equ:FSCLF:FM}) is obtained as
\begin{equation}
\label{equ:FSCLF:Qi_PEM}
Q_k = q_{k_{l'_\text{min}}},
\end{equation}
where $l'_\text{min} = \argmin_{l'} q_{k_{l'}}$. Therefore, under a practical implementation one only needs to maintain the path-error metrics $q$ corresponding to the $2L$ decoding paths to progressively calculate the path selection error metric $Q_k$.

\textcolor{black}{In this paper, we tackle the disadvantage of Monte-Carlo simulation which optimizes the single parameter $\theta$ offline \cite{DSCF,SCLF,Furkan_TSP20}. This is because in practice, e.g., in the 5G standard, there is a vast number of polar code configurations with different code lengths and rates, and the parameter also requires to be optimized at various SNR values. Thus, optimizing the parameter for each specific configuration is a time-consuming task as adequate training data samples need to be collected for each code configuration. Therefore, we propose an efficient online supervised learning approach to directly optimize the parameter at the operating SNR of the decoder, while obviating the need of pilot signals.}

In particular, let $\mathbb{D}$ be a data batch that contains $B=\abs{\mathbb{D}}$ instances of the path selection error metrics $\bm{Q}=\bm{Q}_1^{K+C}$, where the corresponding message word estimated by the initial FSCL decoding algorithm does not satisfy the CRC test. Under supervised learning, we need to obtain the erroneous path-splitting index $\imath_e$ to train $\theta$. Note that in a practical scenario, the proposed decoder often requires a maximum number of $m$ additional FSCL decoding attempts where a different estimated error index is associated with each additional decoding attempt. By assuming that a correct codeword is obtained if the CRC verification is successful, the error index $\imath_e$ can be obtained when a secondary FSCL decoding attempt passes the CRC verification. Let $\bm{o}$ be a one-hot encoded vector of size $K+C$ that indicates the error bit index $\imath_e$ as
\begin{equation}
\label{equ:softmin}
o_k=
\begin{cases}
1 & \text{ if } k=\imath_e,\\
0 & \text{ otherwise.}\\
\end{cases}
\end{equation}
A data sample $d \in \mathbb{D}$ contains a pair of the input $\bm{Q}$ and its corresponding encoded output $\bm{o}$, i.e., $d \triangleq \{\bm{Q},\bm{o}\}$.

Given a data sample $d$, the path selection error metric introduced in Section~\ref{sec:FSCLF:metric} provides an estimate of $\imath_e$ as $\imath$ by selecting the index corresponding to the smallest element of $\bm{Q}$ (see (\ref{equ:FSCLF:sel})). To enable training, the error metrics are converted to the probability domain using the following softmin conversion:
\begin{equation}
\label{equ:FSCLF:estimate}
\hat{o}_k = \frac{\exp(-Q_k)}{\sum_{j=1}^{K+C} \exp(-Q_j)},
\end{equation}
where $Q_k$ is manually set to $\infty$ for $k \in [1, \log_2 L]$ as the correct decoding path is always present in the first $\log_2 L$ path splittings. It can be seen from (\ref{equ:FSCLF:FM}) and (\ref{equ:FSCLF:estimate}) that the bit index that has the smallest error metric is also the bit index that has the highest probability to be in error. In this paper, we use the binary cross entropy (BCE) loss function to quantify the dissimilarity between the target output $\bm{o}$ and the estimated output $\hat{\bm{o}}$ as
\begin{equation}
\label{equ:BCE:flip}
\text{Loss}= -\sum_{k=1}^{K+C} \left[ o_k\ln{\hat{o}_k} + (1-o_k)\ln(1-{\hat{o}_k})\right].
\end{equation}
The parameter $\theta$ can then be trained to minimize the loss function by using the stochastic gradient descent (SGD) technique or one of its variants. An update step is given as 
\begin{equation}
\label{equ:grad:1}
\theta = \theta - \frac{\textcolor{black}{\mathcal{E}}}{B} \sum_{\forall d \in \mathbb{D}} \frac{\partial \text{Loss}}{\partial \theta},
\end{equation}
where $\textcolor{black}{\mathcal{E}} \in \mathbb{R}^+$  is the learning rate and $\frac{1}{B}\sum_{\forall d \in \mathbb{D}} \frac{\partial \text{Loss}}{\partial \theta}$ is the estimation of the true gradient obtained from a data set that contains an infinite number of data samples. By using the chain rule and simple algebraic manipulations, given an instance $\bm{Q}$ of a data sample $d$, $\frac{\partial \text{Loss}}{\partial \theta}$ can be calculated as
\begin{equation}
\label{equ:grad}
\frac{\partial \text{Loss}}{\partial \theta} = \sum_{k=1}^{K+C} \frac{\hat{o}_k - o_k}{(1-\hat{o}_k)\exp(-Q_k)}\left[ \frac{\partial \phi_k}{\partial \theta}-\hat{o}_k\sum_{j=1}^{K+C} \frac{\partial \phi_j}{\partial \theta} \right],
\end{equation}
where $\phi_k=\exp(-Q_k)$ and $\frac{\partial \phi_k}{\partial \theta}=-\exp(-Q_k)\frac{\partial Q_k}{\partial \theta}$.

It can be observed that the computation of $\frac{\partial \text{Loss}}{\partial \theta}$ requires the computation of $\frac{\partial Q_k}{\partial \theta}$. Similar to $Q_k$, $\frac{\partial Q_k}{\partial \theta}$ can also be progressively calculated during the course of decoding. In particular, from (\ref{equ:FSCLF:PEM:l}) and (\ref{equ:FSCLF:PEM:l_forked}) we obtain
\begin{equation}
\label{equ:FSCLF:dPEM:l}
\begin{split}
\frac{\partial q_{k_{l}}}{\partial \theta}&=\frac{\partial q_{{k-1}_{l}}}{\partial \theta} + \frac{\partial \relu\left(\theta-{\gamma_{k_{l}}}\right)}{\partial \theta}\\
&=\frac{\partial q_{{k-1}_{l}}}{\partial \theta}+\mathds{1}_{\theta>\gamma_{k_{l}}},
\end{split}
\end{equation}
and
\begin{equation}
\label{equ:FSCLF:dPEM:l_forked}
\frac{\partial q_{k_{\tilde{l}}}}{\partial \theta} = \frac{\partial q_{k_{l}}}{\partial \theta} - 1,
\end{equation}
respectively, and $\frac{\partial q_{0_{l}}}{\partial \theta}=0$ $\forall l$. Since the values of $\frac{\partial q_{k_{l}}}{\partial \theta}$ and $\frac{\partial q_{k_{\tilde{l}}}}{\partial \theta}$ are available for all the current active decoding paths with indices $l$ and the forked paths with indices $\tilde{l}$, after the path-metric sorting, $\frac{\partial Q_k}{\partial \theta}$ can be obtained as
\begin{equation}
\label{equ:FSCLF:dQi_PEM}
\frac{\partial Q_k}{\partial \theta} = \frac{\partial q_{k_{l'_\text{min}}}}{\partial \theta}.
\end{equation}
Note that $\frac{\partial Q_k}{\partial \theta}$ contains integer values and $\frac{\partial Q_k}{\partial \theta} \in [-(K+C),K+C]$. \textcolor{black}{To reduce the computational complexity of the training process, we use the method in \cite{7004740} to implement the $\exp(\cdot)$ function as required in (\ref{equ:FSCLF:estimate}) and (\ref{equ:grad}). Specifically, the Taylor series are utilized to approximate the $\exp(\cdot)$ function, which is given as \cite{7004740}
\begin{equation}
	\label{equ:exp:aprx}
	\exp(x) \approx \max\{0,\sum_{t=0}^{T}\frac{x^t}{t!}\},
\end{equation}
where $x\in \mathbb{R}$, $T \ge 0$ is an integer number, and the approximation is exact if $T=\infty$ \cite{taylor1717methodus}.}

\begin{algorithm}[t]
	\DontPrintSemicolon
	\caption{Fast-SCLF Decoding Algorithm}
	\label{alg1}
	\SetKwInOut{Input}{Input}
	\SetKwInOut{Output}{Output}
	\SetKwInput{kwIn}{in}
	\Input{$\bm{y}, L, m, B$}
	\Output{$\hat{\bm{u}}$}
	\vskip 0.2cm

	$\theta \sim (0,1)$ \tcp{Initialize $\theta$}
	$\hat{\bm{u}}_\textup{init}, \bm{Q}, \frac{\partial \bm{Q}}{\partial \theta} \leftarrow \texttt{InitialFSCL}(\bm{y},\theta, L)$\\
	\tcc{Perform FSCL decoding with the reserved path selection scheme}
	\eIf{$\hat{\bm{u}}_\textup{init}$ \textup{passes CRC}}{
		\Return $\hat{\bm{u}}_\textup{init}$\\
	}{
		$\{\imath^*_1,\ldots,\imath^*_m\} \leftarrow \texttt{Sort}(\bm{Q})$\\
		\For{$i \leftarrow 1$ \KwTo $m$}{
			$\hat{\bm{u}}_\textup{flip} \leftarrow \texttt{FSCL}(\bm{y}, \imath^*_i, L)$\\
			\If{$\hat{\bm{u}}$ \textup{passes CRC}}{				
				\textcolor{black}{Construct $\bm{o}$ using (\ref{equ:softmin}) given $\imath_e=\imath^*_i$}\\
				$\theta \leftarrow$ \texttt{OptimizeTheta}($\theta, \textcolor{black}{\bm{o}}, \bm{Q}, \frac{\partial \bm{Q}}{\partial \theta}$)\\
				\Return $\hat{\bm{u}}_\textup{flip}$
			}
		}
	}
	\Return $\hat{\bm{u}}_\textup{init}$
\end{algorithm}

In Algorithm~\ref{alg1}, we outline the proposed Fast-SCLF decoding algorithm integrated with the online training framework. The inputs of Algorithm~\ref{alg1} contain the channel vector $\bm{y}$, the list size $L$, the maximum number of additional FSCL decoding attempts $m$, and the size of the data batch $\mathbb{D}$, denoted as $B$. The parameter $\theta$ is first randomly initialized from $(0,1)$. Given a channel output vector $\bm{y}$, the initial FSCL decoding is carried out in the $\texttt{InitialFSCL}(\cdot)$ function described in Algorithm~\ref{alg2}, which performs the conventional FSCL decoding operations to obtain the estimated message word $\hat{\bm{u}}_\textup{init}$. In addition, at each path splitting with index $k$ of the initial FSCL decoding attempt, the path-error metrics $\{q_{k_l},q_{k_{\tilde{l}}}\}$ and the derivatives $\{\frac{\partial q_{k_{l}}}{\partial \theta},\frac{\partial q_{k_{\tilde{l}}}}{\partial \theta}\}$ of all the paths with indices $l$ and $\tilde{l}$ are progressively calculated (line 3-4, Algorithm 2), followed by the computations of $Q_k$ and $\frac{\partial Q_k}{\partial \theta}$ (line 5-6, Algorithm 2). Note that $\frac{\partial Q_k}{\partial \theta}$ is set to 0 and $Q_k$ is set to $\infty$ for all the path splittings with index $k\in[1,\log_2 L]$. At the end of the $\texttt{InitialFSCL}(\cdot)$ function, the first estimate of the message word $\hat{\bm{u}}_\textup{init}$, the path selection error metrics $\bm{Q}$, and their derivatives $\frac{\partial \bm{Q}}{\partial \theta}$ are returned to the main decoding algorithm.

\begin{algorithm}[t]
	\DontPrintSemicolon
	\caption{Initial FSCL Decoding Algorithm}
	\label{alg2}
	\SetKwInOut{Input}{Input}
	\SetKwInOut{Output}{Output}
	\SetKwInput{kwIn}{in}
	
	\Input{$\bm{y}$,$\theta$,$L$} 
	\Output{$\hat{\bm{u}}_\textup{init}, \bm{Q}, \frac{\partial \bm{Q}}{\partial \theta}$}	
	\vskip 0.2cm
	
	\SetKwFunction{InitialFSCL}{InitialFSCL}
	\SetKwProg{Pn}{Function}{:}{}
	\Pn{\InitialFSCL{$\bm{y}$,$\theta$,$L$}}
	{
		\For{\textup{each path-splitting with index $k \in [1,K+C]$}}{
			Compute $q_{k_l}$ and $q_{k_{\tilde{l}}}$ based on (\ref{equ:FSCLF:PEM:l}) and (\ref{equ:FSCLF:PEM:l_forked}) for all the paths $l$ and $\tilde{l}$\\
			
			Compute $\frac{\partial q_{k_{l}}}{\partial \theta}$ and $\frac{\partial q_{k_{\tilde{l}}}}{\partial \theta}$ based on (\ref{equ:FSCLF:dPEM:l}) and (\ref{equ:FSCLF:dPEM:l_forked}) for all the paths $l$ and $\tilde{l}$\\
			
			Compute $Q_k$ based on (\ref{equ:FSCLF:Qi_PEM})\\
			
			Compute $\frac{\partial Q_k}{\partial \theta}$ based on (\ref{equ:FSCLF:dQi_PEM})\\
		}
		\For{$k \leftarrow 1$ \KwTo $\log_2 L$}{
			$Q_k \leftarrow \infty$\\
			$\frac{\partial Q_k}{\partial \theta} \leftarrow 0$\\
		}
		Obtain $\hat{\bm{u}}_\textup{init}$ from the first FSCL decoding attempt\\
		\Return $\hat{\bm{u}}_\textup{init}, \bm{Q}, \frac{\partial \bm{Q}}{\partial \theta}$
	}
\end{algorithm}

\begin{algorithm}[t]
	\DontPrintSemicolon
	\caption{Parameter optimization}
	\label{alg3}
	\SetKwInOut{Input}{Input}
	\SetKwInOut{Output}{Output}
	\SetKwInput{kwIn}{in}
	
	\Input{$\theta, \textcolor{black}{\bm{o}}, \bm{Q}, \frac{\partial \bm{Q}}{\partial \theta}$} 
	\Output{$\theta$}	
	\vskip 0.2cm
	
	\SetKwFunction{OptimizeTheta}{OptimizeTheta}
	\SetKwProg{Pn}{Function}{:}{}
	
	$c \leftarrow 0$\tcp*{The number of data samples}
	$\Delta \leftarrow 0$\tcp*{The accumulated garadient}
	\Pn{\OptimizeTheta($\theta, \textcolor{black}{\bm{o}}, \bm{Q}, \frac{\partial \bm{Q}}{\partial \theta}$)}
	{
		$c \leftarrow c + 1$\\ 
		Compute $\frac{\partial \text{Loss}}{\partial \theta}$ using (\ref{equ:grad})\\
		$\Delta \leftarrow \Delta + \frac{\partial \text{Loss}}{\partial \theta}$\\
		\tcc{Update $\theta$ and reset the accumulated gradient}
		\If{$c \mod B == 0$} 
		{
			$\theta \leftarrow \theta - \textcolor{black}{\frac{\mathcal{E}}{B}}\Delta$\\
			$\Delta \leftarrow 0$
		}
		\Return $\theta$
	}
\end{algorithm}

In the next step, if $\hat{\bm{u}}_\textup{init}$ satisfies the CRC test, the Fast-SCLF decoder then outputs $\hat{\bm{u}}_\textup{init}$ and terminates. Otherwise, the path selection error metrics $\bm{Q}$ are sorted in the increasing order such that $Q_{i^*_1} \leq \ldots \leq Q_{i^*_{K+C}}$, and the path-splitting indices corresponding to the $m$ smallest elements of $\bm{Q}$ are selected for the secondary FSCL decoding attempts, i.e., $\{\imath^*_1,\ldots,\imath^*_m\}$. The Fast-SCLF decoder then performs a maximum number of $m$ additional FSCL decoding attempts (line 8, Algorithm~\ref{alg1}) with each attempt performs the reversed path selection scheme at a different path-flipping index $\imath^*_i$. If one of the secondary FSCL decoding attempts results in a successful CRC verification, the optimization process of $\theta$ implemented in the $\texttt{OptimizeTheta}(\cdot)$ function is queried, which performs the proposed optimization process based on supervised learning. The details of the function $\texttt{OptimizeTheta}(\cdot)$ are provided in Algorithm~\ref{alg3}. To reduce the memory consumption required to store the data batch $\mathbb{D}$ for each parameter update, we use a variable $\Delta$ in Algorithm~\ref{alg3} to store the accumulated gradients $\sum_{\forall d \in \mathbb{D}} \frac{\partial \text{Loss}}{\partial \theta}$ as shown in (\ref{equ:grad:1}). In addition, each data sample $d$ is completely different from the others due to the presence of channel noise. Therefore, the proposed training framework can prevent overfitting without the need of a separate validation set, which also reduces the memory consumption of the parameter optimization. 

Finally, if the resulting estimated message word $\hat{\bm{u}}_\textup{flip}$ obtained from one of the addtional FSCL decoding attempts satisfies the CRC test, $\hat{\bm{u}}_\textup{flip}$ is returned as the final decoding output. On the other hand, if none of the addtional FSCL decoding attempt can provide a message word that passes the CRC verification, the estimated message word $\hat{\bm{u}}_\textup{init}$ of the initial FSCL decoding is returned as the final output of the decoding process.

\subsection{Quantitative Complexity Analysis}

\textcolor{black}{To quantify the computational complexity of the decoders considered in this paper, we compute a weighted complexity of the performed floating-point additions/subtractions, comparisons, multiplications, and divisions. The complexity of a floating-point addition/subtraction or a floating point comparison is considered to be one unit of complexity, a multiplication requires 3 units of complexity and a division requires 24 units of complexity \cite{guide2011intel}.} In this paper, we use the merge sort algorithm to sort a vector with $N$ elements, which requires a worst case of $N{\lceil\log_2{N}\rceil} - 2^{\lceil\log_2{N}\rceil}+1$ floating-point comparisons if $N$ is not a power of 2, otherwise the number of comparisons needed is $N\log_2{N}$ \cite[Chapter 2]{cormen2009introduction}. We compute the decoding latency of the SCL-based decoders by using the method considered in \cite{Ali_FSSCL, Ardakani_TCOM}. In particular, we count the number of time steps for various decoding operations with the following assumptions. First, the hard decisions obtained from the LLR values and binary operations are computed instantaneously \cite{Ali_FSSCL,Ardakani_TCOM,Alexios_LLR_SCLD}. Second, we consider the time steps required by a merge sort algorithm to sort a vector of size $N$ is $\lceil \log_2 {N} \rceil$ \cite[Chapter 2]{cormen2009introduction}. \textcolor{black}{In addition, we also measure the average runtime in seconds required to decode a frame of all the decoders considered in this paper. The runtime is measured based on a single-core C++ implementation of the considered decoders on a similar Linux system, with an AMD Ryzen 5 CPU and a DRAM memory of 16 GBytes.}

\textcolor{black}{Note that the $\texttt{OptimizeTheta}(\cdot)$ function can be executed in parallel with the decoding process presented in Algorithm~\ref{alg1} and the decoding latency in time steps of the $\texttt{OptimizeTheta}(\cdot)$ function is significantly smaller than the time steps required by an FSCL decoding attempt. Therefore, we do not include the number of time steps needed by the $\texttt{OptimizeTheta}(\cdot)$ function in the time steps of the proposed algorithm. However, to enable a fair comparison with other decoders that do not require parameter optimization during the course of decoding, we include the runtime of the $\texttt{OptimizeTheta}(\cdot)$ function when computing the runtime of the proposed decoder. Furthermore, the computational complexity and memory requirement of the $\texttt{OptimizeTheta}(\cdot)$ function are also considered when computing those of the proposed decoder. The memory consumption of the proposed decoder with list size $L$ can be calculated as}

\textcolor{black}{
\begin{equation}
\label{equ:FSCLF:mem}
\begin{split}
&\mathcal{M}_\text{Fast-SCLF}= \mathcal{M}_\text{FSCL}+\underbrace{b_f}_\text{$\theta$-memory}+\underbrace{Lb_f}_\text{$q$-memory}+\underbrace{Lb_i}_\text{$\frac{\partial q}{\partial \theta}$-memory}\\
&+\underbrace{(K+C)b_f}_\text{$\bm{Q}$-memory} + \underbrace{(K+C)b_i}_\text{$\frac{\partial \bm{Q}}{\partial \theta}$-memory}+\underbrace{(K+C)}_\text{$\bm{o}$-memory}\\
&+\underbrace{5b_f}_\text{$\frac{\partial\text{Loss}}{\partial \theta}$-related memory}+\underbrace{b_f}_\text{$\Delta$-memory}\\
&=\mathcal{M}_\text{FSCL}+\left[K+C+L+7\right]b_f+(K+C+L)b_i\\
&+K+C,
\end{split}
\end{equation}
where $b_i$ is the number of memory bits used to quantize the integer values of $\frac{\partial \bm{Q}}{\partial \theta}$ and $\frac{\partial q}{\partial \theta}$. We consider that $\frac{\partial \text{Loss}}{\partial \theta}$ is progressively calculated, thus $4b_f$ memory bits are used to store the temporal values of $\sum_{j=1}^{K+C} \exp(-Q_j)$, $\exp(-Q_k)$, $\hat{o}_k$, and $\sum_{j=1}^{K+C} \frac{\partial \phi_j}{\partial \theta}$, and $b_f$ memory bits are used to store $\frac{\partial \text{Loss}}{\partial \theta}$, whose value is progressively summed over $K+C$ indices.}

\section{Evaluation}
\label{sec:evaluation}

\subsection{Optimized Parameter and Error-Correction Performance}
\begin{figure*}[t]
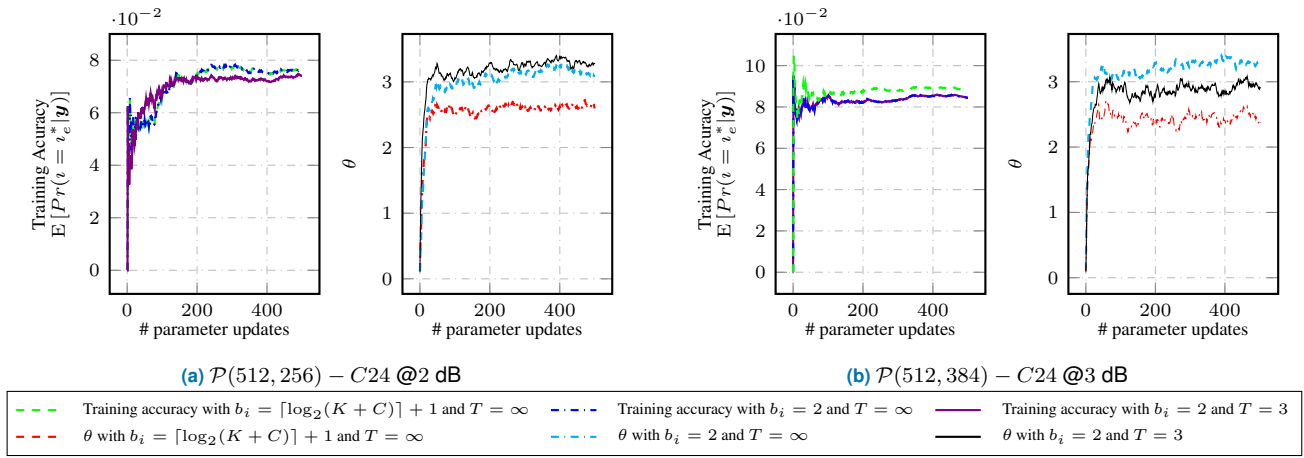

	\centering
	\begin{subfigure}{0.49\linewidth}
		\centering		
		\input{./top1_acc_N512_K256_C24_L32.tikz}
		\input{./theta_N512_K256_C24_L32.tikz}
		\caption{\footnotesize{$\mathcal{P}(512,256)-C24$ @$2$ dB}}
	\end{subfigure}
	\begin{subfigure}{0.49\linewidth}
		\centering
		\input{./top1_acc_N512_K384_C24_L32.tikz}
		\input{./theta_N512_K384_C24_L32.tikz}
		\caption{\footnotesize{$\mathcal{P}(512,384)-C24$ @$3$ dB}}
	\end{subfigure}	
	\ref{perf-legend-accuracy}
	\caption{\textcolor{black}{Training curves of the parameter $\theta$ for $\mathcal{P}(512,256)$ and $\mathcal{P}(512,384)$ with $L=32$ and $m=80$. A 24-bit CRC used in 5G is concatenated with the polar codes.}}
	\label{fig:perf:training}
\end{figure*}

We measure the accuracy of the proposed training framework by calculating the probability that the most probable error index $\imath$ derived from (\ref{equ:FSCLF:sel}) is the actual error index, denoted as $\imath^*_e$, given that the initial FSCL decoding attempt does not satisfy the CRC test. Note that the error index $\imath_e$ used as the training label can be different from $\imath^*_e$. This is because satisfying the CRC test after performing the reserved path selection scheme at the $\imath_e$-th path-splitting index does not warranty that the estimated codeword is the sent codeword. Therefore, the training accuracy is quantified as
\begin{equation}
\operatorname{E}\left[Pr(\imath=\imath^*_e|\bm{y})\right]\approx\frac{\sum_{\text{training samples}} \mathds{1}_{\imath=\imath^*_e}}{\text{Number of training samples}}.
\end{equation}
\textcolor{black}{In this paper, we use the conventional SGD algorithm to optimize $\theta$ with $\mathcal{E}=2^{-4}$ and $B=32$, thus $\frac{\mathcal{E}}{B}$ is fixed to $2^{-9}$ and a multiplication with $\frac{\mathcal{E}}{B}$ can be implemented as a shift operation}. We set $b_f=32$ for both the training and decoding processes as single-precision floating-point format is used to quantize a floating-point number. In addition, an integer number $a$ is quantized using the sign-magnitude representation, which requires $\lceil{\log_2(\abs{a})\rceil}+1$ memory bits.

Fig.~\ref{fig:perf:training} illustrates the learning curves of $\theta$ for $\mathcal{P}(512,256)$ and $\mathcal{P}(512,384)$ with $m=80$, $L=32$, $b_i\in\{\lceil{\log_2(K+C)}\rceil+1, 2\}$, \textcolor{black}{and $T\in\{\infty,3\}$.} With $b_i=\lceil{\log_2(K+C)}\rceil+1$ the maximum and minimum values of the derivatives $\frac{\partial \bm{Q}}{\partial \theta}$ and $\frac{\partial q}{\partial \theta}$ are exactly represented under the sign-magnitude quantization scheme. On the other hand with $b_i=2$, $\frac{\partial \bm{Q}}{\partial \theta}$ and $\frac{\partial q}{\partial \theta}$ are constrained to $\{-1,0,1\}$. As observed from Fig.~\ref{fig:perf:training}, \textcolor{black}{for $T=3$,} constraining $\frac{\partial \bm{Q}}{\partial \theta}$ and $\frac{\partial q}{\partial \theta}$ with the ternary values of $\{-1,0,1\}$ does not significantly degrade the estimation accuracy of the proposed error model \textcolor{black}{compared to the configuration using $T=\infty$ and $b_i=\lceil{\log_2(K+C)}\rceil+1$. Therefore, in the rest of this paper, we set $b_i=2$ and $T=3$ for the proposed decoder as the computational complexity and memory consumption are significantly reduced by using a small value of $T$ and $b_i$, which can be observed from (\ref{equ:exp:aprx}) and (\ref{equ:FSCLF:mem}), respectively.} Note that the spikes in the early part of the training accuracy are caused by the small number of the training samples, which makes the calculation of the training accuracy unreliable at the initial phases of the parameter optimization. As also observed from Fig.~\ref{fig:perf:training}, the value of $\theta$ becomes relatively stable as the number of parameter updates increases. Thus, in practice the function $\texttt{OptimizeTheta}(\cdot)$ can be skipped after a predefined number of parameter updates to further reduce the computational complexity and memory accesses of the proposed framework. \textcolor{black}{In this paper, we stop querying the $\texttt{OptimizeTheta}(\cdot)$ function after 50 parameter updates to further reduce the computational complexity of the proposed decoder.}

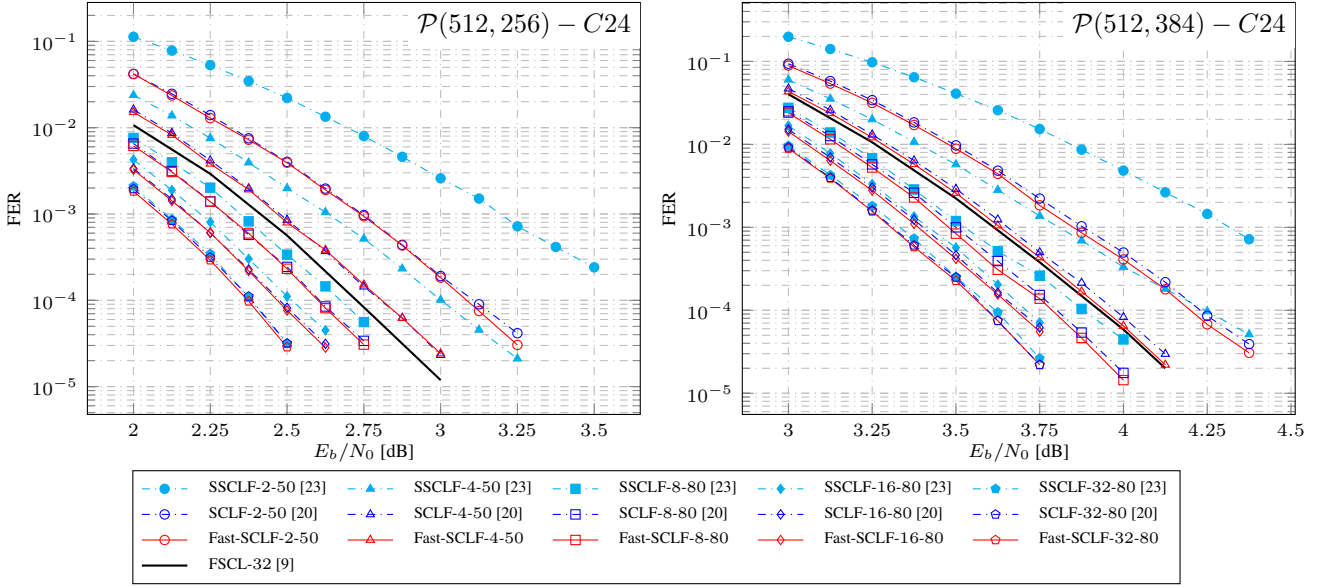
\begin{figure*}[t]
	\centering
	\begin{tikzpicture}[spy using outlines = {rectangle, magnification=2.0, connect spies}]
	\pgfplotsset{	
		label style = {font=\fontsize{7pt}{7}\selectfont},
		tick label style = {font=\fontsize{7pt}{7}\selectfont}
	}
	
	\begin{axis}[
		scale = 1,
		ymode=log,
		xlabel={$E_b/N_0$ [\text{dB}]}, xlabel style={yshift=0.8em},
		ylabel={FER}, ylabel style={yshift=-0.75em},
		xtick={2,2.25,2.5,2.75,3,3.25,3.5,3.75,4,4.25,4.5,4.75,5},
		ytick={1e-8, 1e-7, 1e-6,1e-5,1e-4,1e-3,1e-2,1e-1,1e-0},
		grid=both,
		ymajorgrids=true,
		xmajorgrids=true,
		grid style=dashdotted,
		width=0.5\linewidth, height=7cm,
		thin,
		mark size=1.5,
		legend cell align={left},
		legend style={
			at={(0,1e-5)},
			anchor=south west,
			column sep= 2mm,
			font=\fontsize{6pt}{7.2}\selectfont,
		},
		legend to name=perf-legend-Fast-SCLF,
		legend columns=5,
		]
		
		\addplot[
		color=cyan,
		dashdotted,
		mark=*,
		mark options={solid},
		thin,
		mark size=1.75,
		]
		table {
			2	1.13E-01
			2.125	7.82E-02
			2.25	5.30E-02
			2.375	3.47E-02
			2.5	2.21E-02
			2.625	1.34E-02
			2.75	8.00E-03
			2.875	4.60E-03
			3	2.59E-03
			3.125	1.51E-03
			3.25	7.20E-04
			3.375	4.15E-04
			3.5	2.41E-04
		};
		\addlegendentry{SSCLF-$2$-$50$ \cite{Lee20}}
		
		\addplot[
		color=cyan,
		dashdotted,
		mark=triangle*,
		mark options={solid},
		thin,
		mark size=1.75,
		]
		table {
			2	2.38E-02
			2.125	1.38E-02
			2.25	7.57E-03
			2.375	3.94E-03
			2.5	1.99E-03
			2.625	1.05E-03
			2.75	5.17E-04
			2.875	2.33E-04
			3	1.01E-04
			3.125	4.55E-05
			3.25	2.11E-05
		};
		\addlegendentry{SSCLF-$4$-$50$ \cite{Lee20}}
		
		\addplot[
		color=cyan,
		dashdotted,
		mark=square*,
		mark options={solid},
		thin,
		mark size=1.75,
		]
		table {
			2	7.60E-03
			2.125	3.97E-03
			2.25	2.02E-03
			2.375	8.22E-04
			2.5	3.38E-04
			2.625	1.45E-04
			2.75	5.60E-05
		};
		\addlegendentry{SSCLF-$8$-$80$ \cite{Lee20}}
		
		\addplot[
		color=cyan,
		dashdotted,
		mark=diamond*,
		mark options={solid},
		thin,
		mark size=1.75,
		]
		table {
			2	4.26E-03
			2.125	1.89E-03
			2.25	8.07E-04
			2.375	3.02E-04
			2.5	1.11E-04
			2.625	4.50E-05
		};
		\addlegendentry{SSCLF-$16$-$80$ \cite{Lee20}}
		
		\addplot[
		color=cyan,
		dashdotted,
		mark=pentagon*,
		mark options={solid},
		thin,
		mark size=1.75,
		]
		table {
			2	2.10E-03
			2.125	8.65E-04
			2.25	3.45E-04
			2.375	1.07E-04
			2.5	3.14E-05
		};
		\addlegendentry{SSCLF-$32$-$80$ \cite{Lee20}}
		
		\addplot[
			color=blue,
			dashdotted,
			mark=o,
			mark options={solid},
			thin,
			mark size=1.75,
		]
		table {
			2	4.20E-02
			2.125	2.47E-02
			2.25	1.40E-02
			2.375	7.53E-03
			2.50E+00	4.02E-03
			2.625E+00	1.97E-03
			2.75E+00	9.72E-04
			2.875	4.39E-04
			3.00E+00	1.91E-04
			3.125	8.99E-05
			3.25	4.16E-05
		};
		\addlegendentry{SCLF-$2$-$50$ \cite{SCLF}}
		
		\addplot[
		color=blue,
		dashdotted,
		mark=triangle,
		mark options={solid},
		thin,
		mark size=1.75,
		]
		table {
			2	1.62E-02
			2.125	8.66E-03
			2.25	4.17E-03
			2.375	1.98E-03
			2.50E+00	8.54E-04
			2.625E+00	3.70E-04
			2.75E+00	1.44E-04
			2.875	6.25E-05
			3.00E+00	2.32E-05
		};
		\addlegendentry{SCLF-$4$-$50$ \cite{SCLF}}
		
		\addplot[
		color=blue,
		dashdotted,
		mark=square,
		mark options={solid},
		thin,
		mark size=1.75,
		]
		table {
			2	6.55E-03
			2.125	3.14E-03
			2.25	1.40E-03
			2.375	5.75E-04
			2.50E+00	2.44E-04
			2.625E+00	8.48E-05
			2.75E+00	3.39E-05
		};
		\addlegendentry{SCLF-$8$-$80$ \cite{SCLF}}
		
		\addplot[
		color=blue,
		dashdotted,
		mark=diamond,
		mark options={solid},
		thin,
		mark size=1.75,
		]
		table {
			2	3.36E-03
			2.125	1.49E-03
			2.25	6.04E-04
			2.375	2.28E-04
			2.50E+00	8.20E-05
			2.625E+00	3.13E-05
		};
		\addlegendentry{SCLF-$16$-$80$ \cite{SCLF}}
		
		\addplot[
		color=blue,
		dashdotted,
		mark=pentagon,
		mark options={solid},
		thin,
		mark size=1.75,
		]
		table {
			2	1.95E-03
			2.125	8.37E-04
			2.25	3.14E-04
			2.375	1.11E-04
			2.5	3.18E-05
		};
		\addlegendentry{SCLF-$32$-$80$ \cite{SCLF}}
		
		\addplot[
			color=red,
			solid,
			mark=o,
			mark options={solid},
			thin,
			mark size=1.75,
		]
		table {
			2	4.18E-02
			2.125	2.36E-02
			2.25	1.29E-02
			2.375	7.25E-03
			2.5	3.93E-03
			2.625	1.90E-03
			2.75	9.42E-04
			2.875	4.32E-04
			3	1.83E-04
			3.125	7.52E-05
			3.25	3.04E-05
		};
		\addlegendentry{Fast-SCLF-$2$-$50$}
		
		\addplot[
		color=red,
		solid,
		mark=triangle,
		mark options={solid},
		thin,
		mark size=1.75,
		]
		table {
			2	1.51E-02
			2.125	8.23E-03
			2.25	3.82E-03
			2.375	1.92E-03
			2.50E+00	7.95E-04
			2.625E+00	3.84E-04
			2.75E+00	1.52E-04
			2.875	6.22E-05
			3	2.42E-05		
		};
		\addlegendentry{Fast-SCLF-$4$-$50$}
		
		\addplot[
		color=red,
		solid,
		mark=square,
		mark options={solid},
		thin,
		mark size=1.75,
		]
		table {
			2	6.15E-03
			2.125	3.07E-03
			2.25	1.39E-03
			2.375	5.96E-04
			2.5	2.32E-04
			2.625	8.12E-05
			2.75E+00	3.07E-05
		};
		\addlegendentry{Fast-SCLF-$8$-$80$}
		
		\addplot[
		color=red,
		solid,
		mark=diamond,
		mark options={solid},
		thin,
		mark size=1.75,
		]
		table {		
			2	3.26E-03
			2.125	1.42E-03
			2.25	6.08E-04
			2.375	2.19E-04
			2.5	7.64E-05
			2.625	2.83E-05
		};
		\addlegendentry{Fast-SCLF-$16$-$80$}
		
		\addplot[
		color=red,
		solid,
		mark=pentagon,
		mark options={solid},
		thin,
		mark size=1.75,
		]
		table {
			2	1.83E-03
			2.125	7.625E-04
			2.25	2.92E-04
			2.375	9.75E-05
			2.50E+00	2.87E-05
		};
		\addlegendentry{Fast-SCLF-$32$-$80$}
		
%
%
%
%
		
		\addplot[
		color=black,
		solid,
		mark=none,
		mark options={solid},
		thick,
		mark size=1.75,
		]
		table {
			2	1.07E-02
			2.25	2.92E-03
			2.5	5.64E-04
			2.75	8.31E-05
			3	1.19E-05
		};
		\addlegendentry{FSCL-$32$ \cite{Ali_FSSCL}}	
		
		\node[anchor=north east, fill=white] at (rel axis cs:1,1) {$\mathcal{P}(512,256)-C24$};
	\end{axis}
	
	
\end{tikzpicture}
	\begin{tikzpicture}[spy using outlines = {rectangle, magnification=2.0, connect spies}]
	\pgfplotsset{	
		label style = {font=\fontsize{7pt}{7}\selectfont},
		tick label style = {font=\fontsize{7pt}{7}\selectfont}
	}
	
	\begin{axis}[
		scale = 1,
		ymode=log,
		xlabel={$E_b/N_0$ [\text{dB}]}, xlabel style={yshift=0.8em},
		ylabel={FER}, ylabel style={yshift=-0.75em},
		xtick={2,2.25,2.5,2.75,3,3.25,3.5,3.75,4,4.25,4.5,4.75,5},
		ytick={1e-8, 1e-7, 1e-6,1e-5,1e-4,1e-3,1e-2,1e-1,1e-0},
		grid=both,
		ymajorgrids=true,
		xmajorgrids=true,
		grid style=dashdotted,
		width=0.5\linewidth, height=7cm,
		thin,
		mark size=1.5,
		legend cell align={left},
		legend style={
			at={(0,1e-5)},
			anchor=south west,
			column sep= 2mm,
			font=\fontsize{6pt}{7.2}\selectfont,
		},
		]
		
%
%
%
%
		
		\addplot[
			color=cyan,
			dashdotted,
			mark=*,
			mark options={solid},
			thin,
			mark size=1.75,
		]
		table {
			3	1.98E-01
			3.125	1.41E-01
			3.25	9.76E-02
			3.375	6.45E-02
			3.5	4.09E-02
			3.625	2.58E-02
			3.75	1.53E-02
			3.875	8.63E-03
			4	4.83E-03
			4.125	2.64E-03
			4.25	1.45E-03
			4.375	7.19E-04
		};
		
		\addplot[
		color=cyan,
		dashdotted,
		mark=triangle*,
		mark options={solid},
		thin,
		mark size=1.75,
		]
		table {
			3	6.01E-02
			3.125	3.54E-02
			3.25	2.00E-02
			3.375	1.07E-02
			3.5	5.73E-03
			3.625	2.82E-03
			3.75	1.37E-03
			3.875	6.90E-04
			4	3.33E-04
			4.125	1.81E-04
			4.25	9.72E-05
			4.375 	5.12E-05
		};
		
		\addplot[
		color=cyan,
		dashdotted,
		mark=square*,
		mark options={solid},
		thin,
		mark size=1.75,
		]
		table {
			3	2.76E-02
			3.125	1.39E-02
			3.25	6.80E-03
			3.375	2.86E-03
			3.5	1.19E-03
			3.625	5.14E-04
			3.75	2.61E-04
			3.875	1.04E-04
			4	4.45E-05
		};
		
		\addplot[
		color=cyan,
		dashdotted,
		mark=diamond*,
		mark options={solid},
		thin,
		mark size=1.75,
		]
		table {
			3	1.68E-02
			3.125	7.79E-03
			3.25	3.33E-03
			3.375	1.35E-03
			3.5	5.69E-04
			3.625	2.04E-04
			3.75	7.12E-05
		};
		
		\addplot[
		color=cyan,
		dashdotted,
		mark=pentagon*,
		mark options={solid},
		thin,
		mark size=1.75,
		]
		table {
			3	9.63E-03
			3.125	4.26E-03
			3.25	1.81E-03
			3.375	7.32E-04
			3.5	2.52E-04
			3.625	9.45E-05
			3.75	2.63E-05
		};

		\addplot[
		color=red,
		solid,
		mark=o,
		mark options={solid},
		thin,
		mark size=1.75,
		]
		table {
			3	8.91E-02
			3.125	5.40E-02
			3.25	3.16E-02
			3.375	1.71E-02
			3.50E+00	8.86E-03
			3.625E+00	4.41E-03
			3.75E+00	1.84E-03
			3.875	8.59E-04
			4	4.10E-04
			4.125	1.79E-04
			4.25E+00	6.77E-05
			4.375	3.07E-05
		};
		
		\addplot[
		color=red,
		solid,
		mark=triangle,
		mark options={solid},
		thin,
		mark size=1.75,
		]
		table {
			3	4.35E-02
			3.125	2.37E-02
			3.25	1.21E-02
			3.375	5.76E-03
			3.50E+00	2.59E-03
			3.625E+00	1.05E-03
			3.75E+00	4.35E-04
			3.875	1.68E-04
			4	6.48E-05
			4.125	2.20E-05
		};
		
		\addplot[
		color=red,
		solid,
		mark=square,
		mark options={solid},
		thin,
		mark size=1.75,
		]
		table {
			3	2.40E-02
			3.125	1.15E-02
			3.25	5.26E-03
			3.375	2.29E-03
			3.50E+00	8.40E-04
			3.625E+00	3.08E-04
			3.75E+00	1.38E-04
			3.875	4.64E-05
			4.00E+00	1.44E-05
		};
		
		\addplot[
		color=red,
		solid,
		mark=diamond,
		mark options={solid},
		thin,
		mark size=1.75,
		]
		table {
			3	1.43E-02
			3.125	6.36E-03
			3.25	2.78E-03
			3.375	1.10E-03
			3.50E+00	4.17E-04
			3.625E+00	1.55E-04
			3.75E+00	5.53E-05
		};
		
		\addplot[
		color=red,
		solid,
		mark=pentagon,
		mark options={solid},
		thin,
		mark size=1.75,
		]
		table {
			3	8.91E-03
			3.125	3.89E-03
			3.25	1.56E-03
			3.375	5.81E-04
			3.50E+00	2.27E-04
			3.625E+00	7.47E-05
			3.75E+00	2.20E-05
		};
		
				\addplot[
		color=blue,
		dashdotted,
		mark=o,
		mark options={solid},
		thin,
		mark size=1.75,
		]
		table {
			3	9.37E-02
			3.125	5.79E-02
			3.25	3.41E-02
			3.375	1.84E-02
			3.50E+00	9.72E-03
			3.625E+00	4.82E-03
			3.75E+00	2.22E-03
			3.875	1.03E-03
			4.00E+00	4.97E-04
			4.125	2.18E-04
			4.25	8.57E-05
			4.375	3.91E-05			
		};
		
		\addplot[
		color=blue,
		dashdotted,
		mark=triangle,
		mark options={solid},
		thin,
		mark size=1.75,
		]
		table {
			3	4.70E-02
			3.125	2.60E-02
			3.25	1.31E-02
			3.375	6.39E-03
			3.50E+00	2.90E-03
			3.625E+00	1.24E-03
			3.75E+00	4.99E-04
			3.875	2.14E-04
			4.00E+00	8.25E-05
			4.125	2.97E-05
		};
		
		\addplot[
		color=blue,
		dashdotted,
		mark=square,
		mark options={solid},
		thin,
		mark size=1.75,
		]
		table {
			3	2.50E-02
			3.125	1.26E-02
			3.25	5.74E-03
			3.375	2.61E-03
			3.50E+00	1.00E-03
			3.625E+00	3.92E-04
			3.75E+00	1.53E-04
			3.875	5.40E-05
			4.00E+00	1.76E-05
		};
		
		\addplot[
		color=blue,
		dashdotted,
		mark=diamond,
		mark options={solid},
		thin,
		mark size=1.75,
		]
		table {
			3	1.51E-02
			3.125	6.95E-03
			3.25	3.02E-03
			3.375	1.24E-03
			3.50E+00	4.625E-04
			3.625E+00	1.63E-04
			3.75E+00	6.21E-05
		};
		
		\addplot[
		color=blue,
		dashdotted,
		mark=pentagon,
		mark options={solid},
		thin,
		mark size=1.75,
		]
		table {
			3	9.21E-03
			3.125	3.97E-03
			3.25	1.60E-03
			3.375	6.07E-04
			3.50E+00	2.49E-04
			3.625E+00	7.47E-05
			3.75	2.20E-05
		};
	
		\addplot[
		color=black,
		solid,
		mark=none,
		mark options={solid},
		thick,
		mark size=1.75,
		]
		table {
			3	4.03E-02
			3.25	1.06E-02
			3.5	2.23E-03
			3.75E+00	3.79E-04
			4	5.86E-05
			4.125	2E-05
		};
		
		\node[anchor=north east, fill=white] at (rel axis cs:1,1) {$\mathcal{P}(512,384)-C24$};	
	\end{axis}

\end{tikzpicture}
	\ref{perf-legend-Fast-SCLF}
	\caption{\textcolor{black}{Error-correction performance of all the SCLF-based decoders considered in this paper. The FER values of the FSCL decoder with list size $L=32$ is also plotted for comparison.}}
	\label{fig:fer_FSCLF}
\end{figure*}

\textcolor{black}{Fig.~\ref{fig:fer_FSCLF} shows the error correction performance in terms of FER of various SCLF-based decoders where $m=50$ for $L \in \{2,4\}$ and $m=80$ for $L \in \{8, 16,32\}$. The proposed decoder is denoted as Fast-SCLF-$L$-$m$ while the SCLF and SSCLF decoders are denoted as SCLF-$L$-$m$ and SSCLF-$L$-$m$, respectively.} In addition, the FER values of the FSCL decoder with list size $32$ are also plotted for comparison. The parameter $\theta$ used in (\ref{equ:SCLF:FM}) of the SCLF decoder is optimized offline for each value of $L$ with the Monte-Carlo approach \cite{SCLF}. The $E_b/N_0$ values of the Monte-Carlo simulations are chosen to have an FER of approximately $10^{-4}$ with the selected values of $L$ and $m$. From Fig.~\ref{fig:fer_FSCLF}, it can be observed that under all considered polar codes and list sizes, the SCLF decoder has a relatively similar error-correction performance compared to that of the proposed Fast-SCLF decoder. In some configurations of the polar codes, the Fast-SCLF decoder obtains a slight FER gain over the SCLF decoder with the same list size. \textcolor{black}{This behavior is similar to that of the ideal Fast-SCLF decoder presented in Section~\ref{sec:FSCLF:flip_scheme} when compared with the ideal SCLF decoder. It can also be seen from Fig.~\ref{fig:fer_FSCLF} that with $L\in\{2,4\}$ the simplified path-selection scheme proposed in \cite{Lee20} results in a significant error-correction performance degradation in comparison with those of the Fast-SCLF and SCLF decoders at the target FER of $10^{-4}$, which also degrades quickly as the SNR increases.}

\subsection{Computational Complexity, Decoding Latency, and Memory Requirement}

\begin{table*}[t]
	\def\arraystretch{1.25} 
	
	\caption{\textcolor{black}{Summary of the average computational complexity in terms of weighted complexity of all floating-point operations performed ($\mathcal{C}$) and the average decoding latency in time steps ($\mathcal{L}$) of the SCLF-based decoders considered in Fig.~\ref{fig:fer_FSCLF}.}}
	\footnotesize
	\centering
	\begin{tabular}{c | c  c | c  c | c  c | c  c | c c }
		\toprule
		\multirow{2}{*}{$\mathcal{P}(512,256)$}&\multicolumn{2}{c|}{\makecell{$L=2,m=50$ \\ @$3.125$ dB}}&\multicolumn{2}{c|}{\makecell{$L=4,m=50$ \\ @$2.75$ dB}}&\multicolumn{2}{c|}{\makecell{$L=8,m=80$ \\ @$2.625$ dB}}&\multicolumn{2}{c|}{\makecell{$L=16,m=80$ \\ @$2.5$ dB}}&\multicolumn{2}{c}{\makecell{$L=32,m=80$ \\ @$2.375$ dB}}
		\\
		&$\mathcal{C}$&$\mathcal{L}$&$\mathcal{C}$&$\mathcal{L}$&$\mathcal{C}$&$\mathcal{L}$&$\mathcal{C}$&$\mathcal{L}$&$\mathcal{C}$&$\mathcal{L}$\\
		\midrule
		SCLF \cite{SCLF}&1.46E+4&1.84E+3&2.69E+4&2.17E+3&5.21E+4&2.44E+3&1.13E+5&2.72E+3&2.21E+5&3.02E+3\\
		
		SSCLF \cite{Lee20}&1.32E+4&5.21E+2&2.48E+4&7.87E+2&4.97E+4&1.06E+3&1.06E+5&1.35E+3&2.64E+5&1.62E+3\\
		
		Fast-SCLF&{1.37E+4}&{5.01E+2}&{2.60E+4}&{7.88E+2}&{5.27E+4}&{1.06E+3}&{1.18E+5}&{1.34E+3}&{2.82E+5}&{1.63E+3}\\
		\bottomrule
		
		\toprule
		\multirow{2}{*}{$\mathcal{P}(512,384)$}&\multicolumn{2}{c|}{\makecell{$L=2,m=50$ \\ @$4.25$ dB}}&\multicolumn{2}{c|}{\makecell{$L=4,m=50$ \\ @{$4.0$} dB}}&\multicolumn{2}{c|}{\makecell{$L=8,m=80$ \\ @$3.75$ dB}}&\multicolumn{2}{c|}{\makecell{$L=16,m=80$ \\ @$3.75$ dB}}&\multicolumn{2}{c}{\makecell{$L=32,m=80$ \\ @$3.625$ dB}}
		\\
		
		&$\mathcal{C}$&$\mathcal{L}$&$\mathcal{C}$&$\mathcal{L}$&$\mathcal{C}$&$\mathcal{L}$&$\mathcal{C}$&$\mathcal{L}$&$\mathcal{C}$&$\mathcal{L}$\\
		\midrule
		SCLF \cite{SCLF}&1.71E+4&1.97E+3&{3.18E+4}&{2.40E+3}&6.36E+4&2.87E+3&1.28E+5&3.23E+3&2.68E+5&3.65E+3\\
		
		SSCLF \cite{Lee20}&1.60E+4&5.89E+2&2.94E+4&1.11E+3&5.96E+4&1.43E+3&1.15E+5&1.80E+3&2.60E+5&2.22E+3\\
		
		Fast-SCLF&{1.68E+4}&{5.73E+2}&{3.26E+4}&{9.90E+2}&{6.48E+4}&{1.43E+3}&{1.33E+5}&{1.80E+3}&{2.95E+5}&{2.25E+3}\\
		\bottomrule
	\end{tabular}
	
	\label{tab:compall}
\end{table*}

\textcolor{black}{In Table~\ref{tab:compall}, we summarize the average computational complexities ($\mathcal{C}$) and the average decoding latency in time steps ($\mathcal{L}$) of all the SCLF-based decoders considered in Fig.~\ref{fig:fer_FSCLF}. The $E_b/N_0$ values in Table~\ref{tab:compall} are selected from Fig.~\ref{fig:fer_FSCLF} where the simulated FER values of the proposed decoder are closest to the target FER of $10^{-4}$.}

The effectiveness of the proposed decoder is confirmed in Table~\ref{tab:compall} as the decoding latency of the SCLF decoding algorithm is significantly higher than that of the Fast-SCLF decoder with the same list size. However, with the list size increases the Fast-SCLF decoder imposes a more significant computational complexity overhead when compared to that of the SCLF decoder. This is due to the complexity devoted for sorting the LLR values associated with the special SPC and Rate-1 nodes, which significantly increases with the increase of the list size under FSCL decoding. \textcolor{black}{From Table~\ref{tab:compall}, it is observed that the proposed Fast-SCLF decoder reduces up to $73.4\%$ of the average decoding latency of the SCLF decoder with the same list size at the FER of $10^{-4}$, while incurring a maximum computational overhead of $27.3\%$. As also seen from Table~\ref{tab:compall} and Fig.~\ref{fig:fer_FSCLF}, when compared with the SSCLF decoder with $L\in\{2,4\}$, the proposed decoder with the same list size only incurs negligible overheads in the computational complexity while achieving significantly error-correction performance improvements and maintaining relatively similar decoding latency in time steps. On the other hand, with $L\in\{8,16,32\}$, a maximum complexity overhead of $13.5\%$ is recorded for the proposed decoder when compared with SSCLF decoding with the same list size, while obtaining relatively similar error-correction performance and decoding latency.}

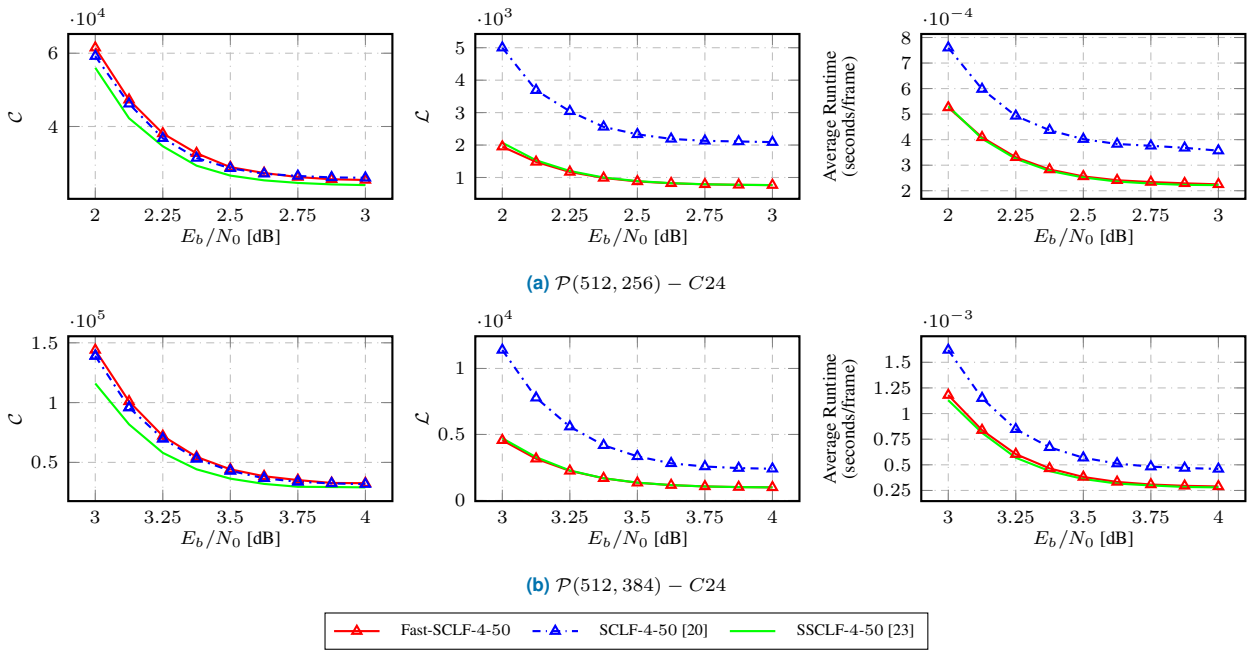
\begin{figure*}
	\centering
	\begin{subfigure}{1.0\linewidth}
		\centering
		\begin{tikzpicture}[spy using outlines = {rectangle, magnification=2.0, connect spies}]
	\pgfplotsset{	
		label style = {font=\fontsize{7pt}{7}\selectfont},
		tick label style = {font=\fontsize{7pt}{7}\selectfont}
	}
	
	\begin{axis}[
		scale = 1,
		xlabel={$E_b/N_0$ [\text{dB}]}, xlabel style={yshift=0.8em},
		ylabel={$\mathcal{C}$}, ylabel style={yshift=-1.5em, rotate=-0},
		xtick={2,2.25,2.5,2.75,3},
		grid=both,
		ymajorgrids=true,
		xmajorgrids=true,
		grid style=dashdotted,
		width=0.33\linewidth, height=3.75cm,
		thick,
		mark size=1.5,
		legend cell align={left},
		legend style={
			at={(0,1e-5)},
			anchor=south west,
			column sep= 2mm,
			font=\fontsize{6pt}{7.2}\selectfont,
		},
		legend to name=perf-legend-compare,
		legend columns=5,
		]

		\addplot[
		color=red,
		solid,
		mark=triangle,
		mark options={solid},
		thick,
		mark size=2,
		]
		table {
			2	6.15E+04
			2.125	4.73E+04
			2.25	3.81E+04
			2.375	3.27E+04
			2.5	2.89E+04
			2.625	2.73E+04
			2.75	2.62E+04
			2.875	2.56E+04
			3	2.54E+04
		};
		\addlegendentry{Fast-SCLF-$4$-$50$}		
		
		\addplot[
		color=blue,
		dashdotted,
		mark=triangle,
		mark options={solid},
		thick,
		mark size=2,
		]
		table {
			2	5.92E+04
			2.125	4.62E+04
			2.25	3.68E+04
			2.375	3.14E+04
			2.5	2.86E+04
			2.625	2.71E+04
			2.75	2.65E+04
			2.875	2.61E+04
			3	2.60E+04
		};
		\addlegendentry{SCLF-$4$-$50$ \cite{SCLF}}
		
		\addplot[
		color=green,
		solid,
		mark=none,
		mark options={solid},
		thick,
		mark size=2,
		]
		table {
			2	5.6037e+04
			2.125	4.2248e+04
			2.25	3.46E+04
			2.375	2.93E+04
			2.5	2.66E+04
			2.625	2.53E+04
			2.75	2.46E+04
			2.875	2.42E+04
			3	2.40E+04
		};
		\addlegendentry{SSCLF-$4$-$50$ \cite{Lee20}}		
		
				
	\end{axis}
	
	
\end{tikzpicture}
		\begin{tikzpicture}[spy using outlines = {rectangle, magnification=2.0, connect spies}]
\pgfplotsset{	
	label style = {font=\fontsize{7pt}{7}\selectfont},
	tick label style = {font=\fontsize{7pt}{7}\selectfont},
	scaled y ticks=base 10:-3,
}

\begin{axis}[
	scale = 1,
	xlabel={$E_b/N_0$ [\text{dB}]},
	xlabel style={yshift=0.8em},
	ylabel={$\mathcal{L}$},
	ylabel style={yshift=-1.5em, rotate=-0},
	xtick={2,2.25,2.5,2.75,3},
	ytick={1E3,2E3,3E3,4E3,5E3,6E3,7E3},
	grid=both,
	ymajorgrids=true,
	xmajorgrids=true,
	grid style=dashdotted,
	width=0.33\linewidth, height=3.75cm,
	thick,
	mark size=1.5,
	legend cell align={left},
	legend style={
		at={(0,1e-5)},
		anchor=south west,
		column sep= 2mm,
		font=\fontsize{6pt}{7.2}\selectfont,
},
]

\addplot[
color=red,
solid,
mark=triangle,
mark options={solid},
thick,
mark size=2,
]
table {
	2	1.95E+03
	2.125	1.48E+03
	2.25	1.17E+03
	2.375	9.86E+02
	2.5	8.78E+02
	2.625	8.21E+02
	2.75	7.89E+02
	2.875	7.73E+02
	3	7.65E+02
};

\addplot[
color=blue,
dashdotted,
mark=triangle,
mark options={solid},
thick,
mark size=2,
]
table {
	2	5.00E+03
	2.125	3.69E+03
	2.25	3.04E+03
	2.375	2.56E+03
	2.5	2.33E+03
	2.625	2.19E+03
	2.75	2.13E+03
	2.875	2.11E+03
	3	2.09E+03
};

\addplot[
color=green,
solid,
mark=none,
mark options={solid},
thick,
mark size=2,
]
table {
	2	2.07E+03
	2.125	1.53E+03
	2.25	1.20E+03
	2.375	9.99E+02
	2.5	8.88E+02
	2.625	8.25E+02
	2.75	7.91E+02
	2.875	7.74E+02
	3	7.66E+02
};


\end{axis}


\end{tikzpicture}
		\begin{tikzpicture}[spy using outlines = {rectangle, magnification=2.0, connect spies}]
\pgfplotsset{	
	label style = {font=\fontsize{7pt}{7}\selectfont},
	tick label style = {font=\fontsize{7pt}{7}\selectfont},
	scaled y ticks=base 10:4,
}

\begin{axis}[
	scale = 1,
	xlabel={$E_b/N_0$ [\text{dB}]},
	xlabel style={yshift=0.8em},
	ylabel={\makecell{Average Runtime\\(seconds/frame)}},
	ylabel style={yshift=-0.5em, rotate=-0},
	xtick={2,2.25,2.5,2.75,3},
	ytick={2E-4,3E-4,4E-4,5E-4,6E-4,7E-4,8E-4},
	grid=both,
	ymajorgrids=true,
	xmajorgrids=true,
	grid style=dashdotted,
	width=0.33\linewidth, height=3.75cm,
	thick,
	mark size=1.5,
	legend cell align={left},
	legend style={
		at={(0,1e-5)},
		anchor=south west,
		column sep= 2mm,
		font=\fontsize{6pt}{7.2}\selectfont,
},
]

\addplot[
color=red,
solid,
mark=triangle,
mark options={solid},
thick,
mark size=2,
]
table {
	2	5.26E-04
	2.125	4.09E-04
	2.25	3.31E-04
	2.375	2.83E-04
	2.5	2.56E-04
	2.625	2.41E-04
	2.75	2.34E-04
	2.875	2.29E-04
	3	2.25E-04
};

\addplot[
color=blue,
dashdotted,
mark=triangle,
mark options={solid},
thick,
mark size=2,
]
table {
	2	7.60E-04
	2.125	5.98E-04
	2.25	4.93E-04
	2.375	4.37E-04
	2.5	4.02E-04
	2.625	3.83E-04
	2.75	3.76E-04
	2.875	3.68E-04
	3	3.57E-04
};

\addplot[
color=green,
solid,
mark=none,
mark options={solid},
thick,
mark size=2,
]
table {
	2	5.31E-04
	2.125	4.04E-04
	2.25	3.25E-04
	2.375	2.78E-04
	2.5	2.52E-04
	2.625	2.36E-04
	2.75	2.28E-04
	2.875	2.23E-04
	3	2.22E-04
};


\end{axis}


\end{tikzpicture}
		\caption{$\mathcal{P}(512,256)-C24$}
		\vspace*{5pt}
	\end{subfigure}
	\begin{subfigure}{1.0\linewidth}
		\centering
		\begin{tikzpicture}[spy using outlines = {rectangle, magnification=2.0, connect spies}]
	\pgfplotsset{	
		label style = {font=\fontsize{7pt}{7}\selectfont},
		tick label style = {font=\fontsize{7pt}{7}\selectfont}
	}
	
	\begin{axis}[
		scale = 1,
		xlabel={$E_b/N_0$ [\text{dB}]}, xlabel style={yshift=0.8em},
		ylabel={$\mathcal{C}$}, ylabel style={yshift=-1.5em, rotate=-0},
		xtick={3,3.25,3.5,3.75,4,4.25},
		grid=both,
		ymajorgrids=true,
		xmajorgrids=true,
		grid style=dashdotted,
		width=0.33\linewidth, height=3.75cm,
		thick,
		mark size=1.5,
		legend cell align={left},
		legend style={
			at={(0,1e-5)},
			anchor=south west,
			column sep= 2mm,
			font=\fontsize{6pt}{7.2}\selectfont,
		},
		]
		
		\addplot[
		color=red,
		solid,
		mark=triangle,
		mark options={solid},
		thick,
		mark size=2,
		]
		table {
			3	1.44E+05
			3.125	1.01E+05
			3.25	7.21E+04
			3.375	5.46E+04
			3.5	4.42E+04
			3.625	3.84E+04
			3.75	3.52E+04
			3.875	3.28E+04
			4	3.26E+04
		};

		\addplot[
		color=blue,
		dashdotted,
		mark=triangle,
		mark options={solid},
		thick,
		mark size=2,
		]
		table {
			3	1.39E+05
			3	1.39E+05
			3.125	9.59E+04
			3.125	9.59E+04
			3.25	6.98E+04
			3.25	6.98E+04
			3.375	5.31E+04
			3.375	5.30E+04
			3.5	4.29E+04
			3.5	4.30E+04
			3.625	3.67E+04
			3.625	3.67E+04
			3.75	3.38E+04
			3.75	3.38E+04
			3.875	3.24E+04
			4		3.18E+04
		};
		
		\addplot[
		color=green,
		solid,
		mark=none,
		mark options={solid},
		thick,
		mark size=2,
		]
		table {
			3	1.16E+05
			3.125	8.18E+04
			3.25	5.80E+04
			3.375	4.42E+04
			3.5	3.64E+04
			3.625	3.20E+04
			3.75	2.96E+04
			3.875	2.94E+04
			4	2.91E+04
		};
		
		
	\end{axis}
	
	
\end{tikzpicture}
		\begin{tikzpicture}[spy using outlines = {rectangle, magnification=2.0, connect spies}]
	\pgfplotsset{	
		label style = {font=\fontsize{7pt}{7}\selectfont},
		tick label style = {font=\fontsize{7pt}{7}\selectfont}
	}
	
	\begin{axis}[
		scale = 1,
		log ticks with fixed point,
		xlabel={$E_b/N_0$ [\text{dB}]}, xlabel style={yshift=0.8em},
		ylabel={$\mathcal{L}$},
		ylabel style={yshift=-1.5em, rotate=-0},
		xtick={3,3.25,3.5,3.75,4,4.25},
		grid=both,
		ymajorgrids=true,
		xmajorgrids=true,
		grid style=dashdotted,
		width=0.33\linewidth, height=3.75cm,
		thick,
		mark size=1.5,
		legend cell align={left},
		legend style={
			at={(0,1e-5)},
			anchor=south west,
			column sep= 2mm,
			font=\fontsize{6pt}{7.2}\selectfont,
		},
		]
		
		\addplot[
		color=red,
		solid,
		mark=triangle,
		mark options={solid},
		thick,
		mark size=2,
		]
		table {
			3	4.56E+03
			3.125	3.16E+03
			3.25	2.24E+03
			3.375	1.68E+03
			3.5	1.34E+03
			3.625	1.16E+03
			3.75	1.06E+03
			3.875	1.01E+03
			4	9.89E+02
		};
		
		\addplot[
		color=blue,
		dashdotted,
		mark=triangle,
		mark options={solid},
		thick,
		mark size=2,
		]
		table {
			3	1.14E+04
			3.125	7.78E+03
			3.25	5.59E+03
			3.375	4.18E+03
			3.5	3.34E+03
			3.625	2.81E+03
			3.75	2.57E+03
			3.875	2.45E+03
			4.0	2.40E+03
		};
		
		\addplot[
		color=green,
		solid,
		mark=none,
		mark options={solid},
		thick,
		mark size=2,
		]
		table {
			3	4.71E+03
			3.125	3.28E+03
			3.25	2.27E+03
			3.375	1.68E+03
			3.5	1.35E+03
			3.625	1.16E+03
			3.75	1.06E+03
			3.875	1.01E+03
			4	9.91E+02
		};
		

	\end{axis}
	
	
\end{tikzpicture}
		\begin{tikzpicture}[spy using outlines = {rectangle, magnification=2.0, connect spies}]
\pgfplotsset{	
	label style = {font=\fontsize{7pt}{7}\selectfont},
	tick label style = {font=\fontsize{7pt}{7}\selectfont}
}

\begin{axis}[
	scale = 1,
	scaled y ticks=base 10:3,
	xlabel={$E_b/N_0$ [\text{dB}]},
	xlabel style={yshift=0.8em},
	ylabel={\makecell{Average Runtime\\(seconds/frame)}},
	ylabel style={yshift=-0.5em, rotate=-0},
	xtick={3,3.25,3.5,3.75,4,4.25},
	ytick={2.5E-4,5E-4,7.5E-4,10E-4,12.5E-4,15E-4},
	grid=both,
	ymajorgrids=true,
	xmajorgrids=true,
	grid style=dashdotted,
	width=0.33\linewidth, height=3.75cm,
	thick,
	mark size=1.5,
	legend cell align={left},
	legend style={
		at={(0,1e-5)},
		anchor=south west,
		column sep= 2mm,
		font=\fontsize{6pt}{7.2}\selectfont,
},
]

\addplot[
color=red,
solid,
mark=triangle,
mark options={solid},
thick,
mark size=2,
]
table {
	3	1.18E-03
	3.125	8.38E-04
	3.25	6.04E-04
	3.375	4.65E-04
	3.5	3.81E-04
	3.625	3.33E-04
	3.75	3.08E-04
	3.875	2.95E-04
	4	2.89E-04
};

\addplot[
color=blue,
dashdotted,
mark=triangle,
mark options={solid},
thick,
mark size=2,
]
table {
	3	1.62E-03
	3.125	1.15E-03
	3.25	8.47E-04
	3.375	6.71E-04
	3.5	5.69E-04
	3.625	5.12E-04
	3.75	4.82E-04
	3.875	4.69E-04
	4	4.60E-04
};

\addplot[
color=green,
solid,
mark=none,
mark options={solid},
thick,
mark size=2,
]
table {
	3	1.13E-03
	3.125	8.12E-04
	3.25	5.69E-04
	3.375	4.39E-04
	3.5	3.62E-04
	3.625	3.17E-04
	3.75	2.99E-04
	3.875	2.84E-04
	4	2.80E-04
};


\end{axis}


\end{tikzpicture}
		\caption{$\mathcal{P}(512,384)-C24$}
		\vspace*{5pt}
	\end{subfigure}
	\ref{perf-legend-compare}
	\caption{\textcolor{black}{Average computational complexity and latency in terms of time steps and runtime of the SCLF-based decoders with list size 4.}}
	\label{fig:comp}
\end{figure*}

Note that the path selection error metric of SCLF decoding can be progressively calculated using a similar approach as described in (\ref{equ:FSCLF:PEM:l}) and (\ref{equ:FSCLF:PEM:l_forked}). Therefore, the memory consumption of the SCLF decoder with list size $L$ is calculated as
\begin{equation}
\begin{split}
\mathcal{M}_\text{SCLF} &= \mathcal{M}_\text{SCL} + \underbrace{b_f}_\text{$\theta$-memory}+\underbrace{Lb_f}_\text{$q$-memory}+\underbrace{(K+C)b_f}_\text{$\bm{Q}$-memory}\\
&= \mathcal{M}_\text{SCL} + (K+C+L+1)b_f.
\end{split}
\end{equation}
\textcolor{black}{In addition, the memory consumption of the SSCLF decoder only requires an addition of $(K+C)b_f$ memory bits to store the path-selection error metric when compared with that of the FSCL decoder with the same list size \cite{Lee20}. The memory consumption of the SSCLF decoder with list size $L$ is given as \cite{Lee20}
\begin{equation}
	\mathcal{M}_\text{SSCLF} = \mathcal{M}_\text{FSCL} + \underbrace{(K+C)b_f}_\text{$\bm{Q}$-memory}.
\end{equation}
In Table~\ref{tab:comp}, we summary the memory consumption in KBits of all the SCL-based decoders considered in this paper.}

\begin{table}
	\centering
	\caption{\textcolor{black}{Memory requirement in KBits of all the SCL-based decoders considered in this paper.}}
	\footnotesize
	\begin{tabular}{cccccc}
		\toprule
		\multirow{2}{*}{$\mathcal{P}(512,256)$} & \multicolumn{5}{c}{$L$}\\
		\cmidrule{2-6}&2 & 4 & 8 & 16 & 32\\		
		\midrule
		FSCL\cite{Ali_FSSCL} &50.0&84.0&152.0&288.0&560.0\\
		SCLF\cite{SCLF}&58.8&92.9&161.0&297.3&569.8\\
		SSCLF\cite{Lee20}&58.7&92.7&160.7&295.7&568.7\\
		Fast-SCLF&{60.1}&{94.2}&{162.3}&{298.6}&{571.1}\\
		\bottomrule
		
		\toprule		
		\multirow{2}{*}{$\mathcal{P}(512,384)$} & \multicolumn{5}{c}{$L$}\\
		\cmidrule{2-6}&2 & 4 & 8 & 16 & 32\\
		\midrule
		FSCL\cite{Ali_FSSCL}&50.0&84.0&152.0&288.0&560.0\\
		SCLF\cite{SCLF}&62.8&97.0&165.0&301.3&573.8\\
		SSCLF\cite{Lee20}&62.7&96.7&164.7&300.7&572.7\\
		Fast-SCLF&{64.6}&{98.7}&{166.8}&{303.1}&{575.6}\\
		\bottomrule
	\end{tabular}
	\label{tab:comp}
\end{table}

\begin{table*}
	\centering
	\caption{\textcolor{black}{The average computational complexity, average decoding latency, memory consumption, and error-correction performance degradation of the Fast-SCLF, SCLF, and SSCLF decoders with $L=4$ and $m=50$ in comparison with those of the FSCL-32 decoder.}}
	\footnotesize
	\begin{tabular}{r|rrrr|rrrr}
		&\multicolumn{4}{c|}{$\mathcal{P}(512,256)-C24$}&\multicolumn{4}{c}{$\mathcal{P}(512,384)-C24$}\\
		\cmidrule{2-9}
		&\makecell{Fast-SCLF-4\\@2.75 dB}&\makecell{SCLF-4\\@2.75 dB}& \makecell{SSCLF-4\\@3.0 dB}&FSCL-32&\makecell{Fast-SCLF-4\\@4.0 dB}&\makecell{SCLF-4\\@4.0 dB}& \makecell{SSCLF-4\\@4.25 dB}&FSCL-32\\
		\midrule		
		$\mathcal{C}$ (weighted complexity)& 2.60E+4&2.69E+4&2.40E+04&2.42E+5&3.26E+4&3.18E+4&2.91E+04&2.12E+5\\
		$\mathcal{L}$ (time steps) & 7.88E+2&2.17E+3&7.66E+02&1.40E+3&9.90E+2&2.40E+3&9.78E+02&1.36E+3\\
		$\mathcal{M}$ (KBits)&94.2&92.9&92.7&560.0&98.7&97.0&96.7&560.0\\		
		Avg. Runtime (seconds/frame) &2.34E-4&3.76E-4&2.22E-4&1.92E-3&2.95E-4&4.69E-4&2.85E-4&2.05E-3\\
		FER Degradation (dB)&{0.07}&0.07&0.27&-&{0.02}&0.05&0.31&-\\
	\end{tabular}
	\label{tab:FSCL_comp}
\end{table*}

\textcolor{black}{We illustrate the average complexity, average decoding latency in time steps, and average runtime of the Fast-SCLF-$4$-$50$, SCLF-$4$-$50$, and SSCLF-$4$-$50$ decoders in Fig.~\ref{fig:comp}. As seen from Fig.~\ref{fig:comp}, the proposed decoder requires a relatively similar decoding complexity when compared with the SCLF, while the SSCLF decoder has the lowest average computational complexity among all the SCLF-based decoders. In addition, the SSCLF and Fast-SCLF decoding algorithms require significantly smaller average decoding latency both in terms of time steps and runtime when compared with those of the SCLF decoder.}

\textcolor{black}{In Table~\ref{tab:FSCL_comp}, we summarize the average computational complexity, memory requirement, and average decoding latency in terms of time steps and runtime of the SCLF-based decoders with $L=4, m=50$, and those of the FSCL-32 decoder. The error-correction performance degradation of the SCLF-based decoders when compared to FSCL-32 is also provided in Table~\ref{tab:FSCL_comp}. The $E_b/N_0$ values are selected to allow an FER performance close to the target FER of $10^{-4}$ for all the considered decoders. In particular, for $\mathcal{P}(512,256)$, the average complexity and average latency in time steps of Fast-SCLF-$4$-$50$ account for approximately $10.7\%$ and $56.3\%$ of the complexity and time steps of FSCL-32, respectively. For $\mathcal{P}(512,384)$, Fast-SCLF-$4$-$50$ reduces $84.6\%$ of the average complexity and $27.2\%$ of the average time steps in comparison with FSCL-32. In addition, the proposed decoder with list size 4 requires around $17\%$ of the memory requirement of FSCL-32, while having an FER degradation of less than 0.07 dB. When compared with the SSCLF decoder, the proposed decoder obtains the FER performance gains of 0.2 dB and 0.3 dB at the cost of $8.3\%$ and $12.0\%$ computational complexity overhead for $\mathcal{P}(512,256)$ and $\mathcal{P}(512,384)$, respectively, while the average decoding latency and memory consumption are relatively preserved at the target FER of $10^{-4}$. Note that due to its high complexity, the average runtime of FSCL-32 is significantly higher than those of all the SCLF-based decoders with list size 4.}

\begin{figure}
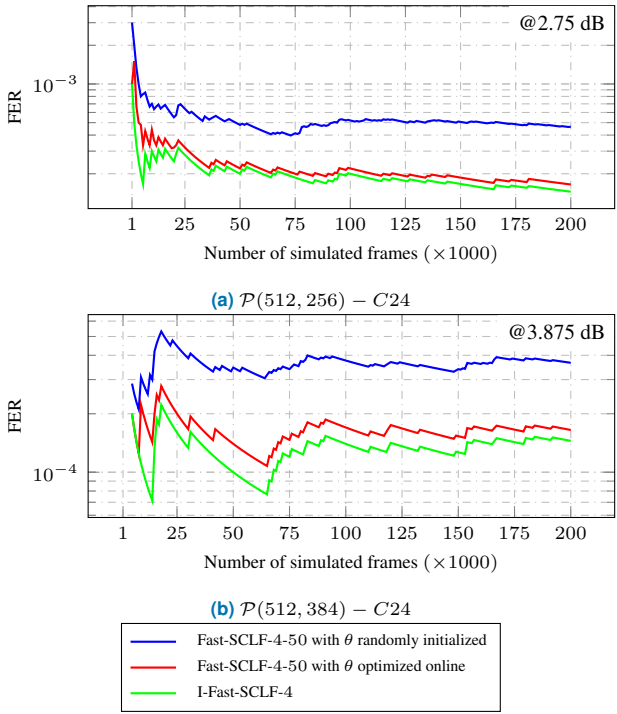

	\centering
	\begin{subfigure}{1\linewidth}
		\centering
		\input{./progress_FER_N512_K256.tikz}
		\caption{$\mathcal{P}(512,256)-C24$}
	\end{subfigure}	
	\begin{subfigure}{1\linewidth}
		\centering
		\input{./progress_FER_N512_K384.tikz}
		\caption{$\mathcal{P}(512,384)-C24$}
	\end{subfigure}	
	\ref{perf-legend-FER-evolution}
	\caption{\textcolor{black}{Effects of online training on the error-correction performance of the Fast-SCLF-$4$-$50$ decoder.}}
	\label{fig:thetavsFER}
\end{figure}

\textcolor{black}{In Fig.~\ref{fig:thetavsFER} we study the effects of the $\theta$ parameter on the error-correction performance of the proposed decoder when online training is considered. Specifically, we illustrate the FER values obtained at the first $200000$ frames of Fast-SCLF-$4$-$50$ with and without online learning. In addition, the FER values of the ideal Fast-SCLF-$4$ decoder are also plotted for reference. It can be observed that the proposed online learning scheme effectively optimizes the $\theta$ parameter, allowing the FER of the proposed decoder to quickly approach its ideal FER performance. On the other hand, when online training is not considered, using the proposed decoder with the initialized value of $\theta$ results in a poor error-correction performance.}

\section{Conclusion}
\label{sec:conclude}

In this paper, we proposed a bit-flipping scheme tailored to the state-of-the-art fast successive-cancellation list (FSCL) decoding, forming the fast successive-cancellation list flip decoder (Fast-SCLF). We then derived a parameterized path selection error metric that estimates the erroneous path-splitting index at which the correct decoding path is eliminated from the initial FSCL decoding. The trainable parameter of the proposed error model is optimized using online supervised learning, which directly trains the parameter at the operating signal-to-noise ratio of the decoder without the need of pilot signals. We numerically evaluated the proposed decoding algorithm and compared its error-correction performance, average computational complexity, average decoding latency, and memory requirement with those of the state-of-the-art FSCL decoder, the successive-cancellation list flip (SCLF) decoder, \textcolor{black}{and the simplified SCLF (SSCLF) decoder}. The simulation results confirm the effectiveness of the proposed decoder when compared with the FSCL and the SCLF decoders for different polar codes and various list sizes. \textcolor{black}{As also observed from the simulation results, the error-correction performance of the Fast-SCLF decoder significantly outperforms that of the SSCLF decoder with small list sizes (2 and 4), at the cost of negligible computational complexity overhead, while maintaining relatively similar memory consumption and decoding latency also compared to SSCLF decoding. Future research includes designing and implementing a hardware architecture of the proposed decoder, where the bit-flipping scheme is extended to other special nodes of polar codes.}


\begin{IEEEbiography}[{\includegraphics[width=1in,height=1.25in,clip,keepaspectratio]{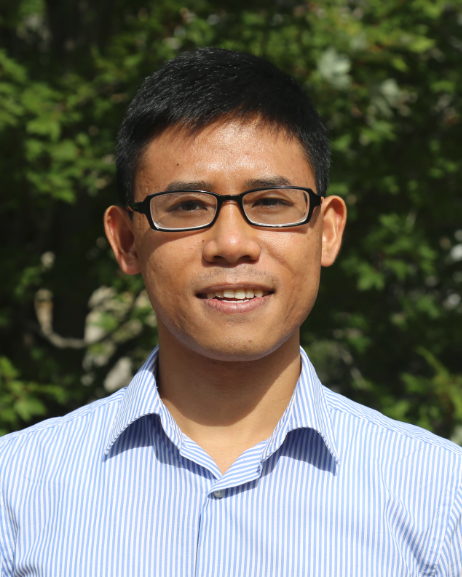}}]{Nghia~Doan} (Student Member, IEEE) received the B.Sc. degree from Posts and Telecommunications Institute of Technology, Hanoi, Vietnam, in 2014 and the M.Sc. degree from Seoul National University, Seoul, South Korea, in 2017, both in electrical and computer engineering. He is currently working toward the Ph.D. degree in electrical and computer engineering at McGill University, Montreal, QC, Canada. His research interests include channel coding, machine learning for communications, and hardware-aware algorithm optimization of digital signal processing applications.
\end{IEEEbiography}

\begin{IEEEbiography}[{\includegraphics[width=1in,height=1.25in,clip,keepaspectratio]{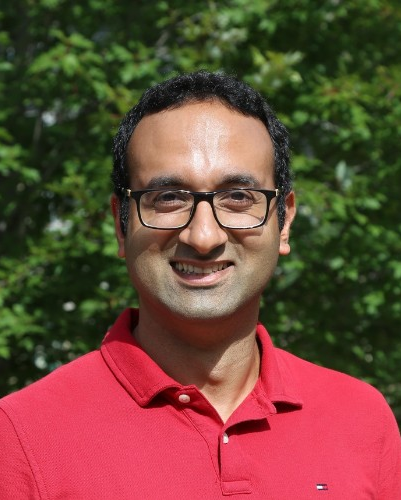}}]{Seyyed~Ali~Hashemi} (Member, IEEE) was born in Qaemshahr, Iran. He received the B.S. degree from Sharif University of Technology, Iran, the M.S. degree from the University of Alberta, Canada, and the Ph.D. degree from McGill University, Canada, all in Electrical Engineering. He is currently a Senior Engineer at Qualcomm, USA. Prior to that, he was a Postdoctoral Fellow with the Department of Electrical Engineering, Stanford University, USA. His research interests include machine learning for communications, error-correcting codes, and VLSI implementation of digital signal processing systems. He was the recipient of a Best Student Paper Award at ISCAS 2016, and a Postdoctoral Fellowship from Natural Sciences and Engineering Research Council of Canada (NSERC) in 2018.
\end{IEEEbiography}

\begin{IEEEbiography}[{\includegraphics[width=1in,height=1.25in,clip,keepaspectratio]{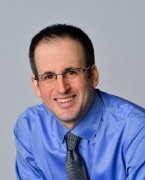}}]{Warren~J.~Gross} (Senior Member, IEEE) received the B.A.Sc. degree in electrical engineering from the University of Waterloo, Waterloo, ON, Canada, in 1996, and the M.A.Sc. and Ph.D. degrees from the University of Toronto, Toronto, ON, Canada, in 1999 and 2003, respectively. He is currently a Professor and the Chair of the Department of Electrical and Computer Engineering, McGill University, Montreal, QC, Canada. His research interests are in the design and implementation of signal processing systems and custom computer architectures. Dr. Gross served as the Chair for the IEEE Signal Processing Society Technical Committee on Design and Implementation of Signal Processing Systems. He served as the General Co-Chair for the IEEE GlobalSIP 2017 and the IEEE SiPS 2017 and the Technical Program Co-Chair for SiPS 2012. He also served as an Organizer for the Workshop on Polar Coding in Wireless Communications at WCNC 2017, the Symposium on Data Flow Algorithms and Architecture for Signal Processing Systems (GlobalSIP 2014), and the IEEE ICC 2012 Workshop on Emerging Data Storage Technologies. He served as an Associate Editor for the IEEE TRANSACTIONS ON SIGNAL PROCESSING and as a Senior Area Editor. He is a Licensed Professional Engineer in the Province of Ontario.
\end{IEEEbiography}

\EOD

\end{document}